\documentclass[%
nofootinbib,
 amsmath,amssymb,
 aps,
]{revtex4-1}\usepackage{graphicx}% Include figure files
\usepackage{bm}% bold math
\usepackage{hyperref}
\usepackage{amssymb}
\usepackage{amsmath}
\usepackage{epsfig}
\usepackage{slashed}
\usepackage{comment}
\usepackage{mathtools}
\usepackage{datetime}
\usepackage{rotating}
\usepackage{relsize}
\def\vecsign{\mathchar"017E}
\def\dvecsign{\smash{\stackon[-1.95pt]{\vecsign}{\rotatebox{180}{$\vecsign$}}}}
\def\dvec#1{\def\useanchorwidth{T}\stackon[-4.2pt]{#1}{\,\dvecsign}}
\usepackage{stackengine}
\stackMath
\usepackage[english]{babel}
\usepackage[utf8]{inputenc}
\usepackage{fancyhdr}
\newcommand{\Dwii}{{\hspace{.015in}{}^3\hspace{-.015in}D_2}}
\newcommand{\Dwiii}{{\hspace{.015in}{}^3\hspace{-.015in}D_3}}
\newcommand{\Pw}{{\hspace{.015in}{}^3\hspace{-.015in}P_0}}
\newcommand{\Pwi}{{\hspace{.015in}{}^3\hspace{-.015in}P_1}}
\newcommand{\Pwii}{{\hspace{.015in}{}^3\hspace{-.015in}P_2}}

\newcommand{\half}{\frac{1}{2}}
\newcommand{\be}{\begin{equation}}
\newcommand{\ee}{\end{equation}}
\newcommand\beq{\begin{eqnarray}}
\newcommand\eeq{\end{eqnarray}} 
\newcommand\eqn[1]{\label{eq:#1}} 
\newcommand\eq[1]{eq.~(\ref{eq:#1})}

\newcommand{\GeV}{{\rm ~GeV }}

\newcommand{\MeV}{{\rm ~MeV }}
\newcommand{\CA}{{\cal A}}

\newcommand{\CC}{{\cal C}}

\newcommand{\CI}{{\cal I}}

\newcommand{\CV}{{\cal V}}

\newcommand{\MJS}{{$ {\text{MJS}}$}}
\newcommand{\MJSb}{{$\overline{\text{MJS}}$}}
\newcommand{\mjs}{ {\text{MJS}}}
\newcommand{\mjsb}{\overline{\text{MJS}}}

\newcommand\bra[1]{\langle #1 |}
\newcommand\ket[1]{\,| #1 \rangle}

\newcommand\expect[3]{\langle #1|#2|#3\rangle}
\newcommand{\mybar}[1]%
        {\kern 0.6pt\overline{\kern -0.6pt#1\kern -0.6pt}\kern 0.6pt}

\makeatletter
\renewcommand*\env@matrix[1][\arraystretch]{%
  \edef\arraystretch{#1}%
  \hskip -\arraycolsep
  \let\@ifnextchar\new@ifnextchar
  \array{*\c@MaxMatrixCols c}}
\makeatother
\begin{document}

\preprint{INT-PUB-19-015}

\title[title]{On the convergence of  nuclear effective field theory with perturbative pions}
\author{David B. Kaplan}
 \email{dbkaplan@uw.edu}
\affiliation{Institute for Nuclear Theory, Box 351550, University of Washington, Seattle, WA 98195-1550}

%\date{}%
%\dedicatory{}%
%\commby{}%
% ----------------------------------------------------------------

%\cite{Ohnishi:2014uea}
\begin{abstract}
 The classic paper by Fleming, Mehen and Stewart (FMS)  \cite{Fleming:1999ee} cast doubts on the  convergence of spin-triplet nucleon-nucleon partial wave scattering amplitudes when following the proposal of Kaplan, Savage and Wise to construct nuclear effective field theory around the unitary fermion limit with perturbative pion exchange. FMS identified the subclass of iterated one-pion exchange potential graphs as the cause of this poor convergence, which they showed persisted in the chiral limit.  Theoretical tools are developed here to compute these Feynman graphs analytically to high order in all angular momentum channels simultaneously,  examining the amplitudes computed to seven loops in the $L=J$ channels, and three loops in the coupled $L=J\pm1$ channels.  One finds that there is nothing pathological about the perturbative expansion of a $1/r^3$ potential in general, and that the expansion converges satisfactorily in all partial waves except those with the lowest angular momentum, particularly the ${}^3P_0$ and the coupled ${}^3S_1-{}^3D_1$ channels. The results corroborate work by Birse \cite{birse2005renormalization}, which suggests possible avenues to explore for improving the range of validity of the EFT expansion.

\end{abstract}

\maketitle

\section{Introduction}

The modern application of effective field theory to multi-nucleon interactions originated in work by Weinberg \cite{Weinberg:1990rz,Weinberg:1991um} and serves as a bridge between QCD and nuclear physics. It was initially developed in Ref.~\cite{Ordonez:1992xp}, while  for more modern reviews see  \cite{machleidt2011chiral,epelbaum2009modern}. Low energy interactions between nucleons are computed in an expansion in powers of  $k/\Lambda$, where $k$ is the momentum scale of the process and $\Lambda$ is a scale characterizing short distance interactions.  What makes this approach particularly powerful is that long range interactions due to pion exchange can be easily incorporated in a way consistent with the approximate chiral symmetry of QCD, while our ignorance about short distance physics can be parametrized in a model-independent way by a relatively small number of coupling constants.  The result is a predictive framework with a systematic path toward reduction of theoretical errors that is consistent with  chiral perturbation theory, a theory which has proved quite successful in the meson and single baryon sectors.  Unlike chiral interactions between mesons, however,  nucleon interactions are not perturbative; the breakdown of a perturbative expansion arises due to infrared singularities which enhance amplitudes by factors proportional to the nucleon mass, reflecting the familiar fact from quantum mechanics that  heavy particles can be bound by weak potentials.     Weinberg's proposal was to perform a chiral expansion of the nucleon potential to a desired order, and then to  iterate insertions of this potential in multi-nucleon amplitudes -- essentially solving the Schr\"odinger equation with the approximate potential.  The value of this approach is that nucleon interactions can be described by relatively few parameters at low orders in the expansion, and the observed hierarchy between 2-, 3-, and 4-nucleon interactions can be explained.

There are drawbacks  to Weinberg's theory as well.  For one thing, the power counting scheme does not account for the large anomalous dimensions of some operators; for example, it is seen   that the 3-body contact interaction can be more important than the 2-body effective range term, in contrast to what one expects from the engineering dimensions of these operators used in the Weinberg scheme \cite{bedaque1999three}.   A violation of Weinberg's dimensional analysis is similarly found in two nucleon scattering in the spin-triplet channels \cite{Nogga:2005hy}.  Another issue is that at any given order in the chiral expansion of the nucleon potential,  short-distance singularities  give rise to an infinite number of   ultraviolet (UV) divergences in scattering amplitudes which cannot be renormalized at the order one is working; this necessitates keeping a finite cutoff in the theory \cite{Kaplan:1996xu}.  If the cutoff is chosen too high,  results are spuriously sensitive to the cutoff, while if it is too low, the energy range of applicability of the theory is greatly restricted.
Ideally one would like to find a large window for the cutoff for which physical results are insensitive to its precise value; in practice, however, one never finds   a broad plateau, and since the procedure is carried out numerically, the boundary is blurred between having a well defined expansion with predictable errors, and the traditional potential approach involving a model for short-range interactions with numerous parameters that can be fit to experimental data  but with little control over systematic errors.

 The alternative nuclear effective field theory of Refs.~\cite{Kaplan:1998tg,Kaplan:1998we} introduced the unitary Fermi gas to nuclear physics as the starting point for a low energy expansion, which also naturally incorporates chiral perturbation theory.  The KSW expansion, as it is generally referred to,  was defined in the framework of the renormalization group, where operators enter the expansion at an order dictated by their effective scaling  dimension in the presence of the strong interactions between nucleons; this leads to a straightforward power counting scheme in the two body sector, but has some rather surprising results for three-body interactions where   nontrivial anomalous dimensions arise which govern the relative importance of different operators \cite{Bedaque:1998kg,birse1999renormalisation,Bedaque:2002yg,barford2003renormalization,Griesshammer:2005ga,birse2005renormalization,Nishida:2007pJ}. An advantage of the KSW approach is that the scattering amplitude is expanded consistently to a given order, and therefore can be completely renormalized, eliminating all dependence on a UV cutoff.  The KSW expansion has been extremely successful at very low energy in its ``pionless" form \cite{chen1999nucleon}, with applications varying from nuclear astrophysics \cite{rupak2000precision}  to neutrinoless $\beta\beta$ decay \cite{cirigliano2018neutrinoless}.    It also holds potential promise for the simulation of nuclear matter, since the starting point of the unitary Fermi gas is relatively simple and without a sign problem, while deviations from this limit due to pion exchange and shorter range interactions are incorporated perturbatively.  Despite these advantages, the  KSW   program stalled when  Fleming, Mehen and Stewart (FMS) gave evidence in NNLO calculations  that the expansion fails to converge at relatively low  momenta in numerous spin triplet partial waves \cite{Fleming:1999ee}, casting doubt on whether the inclusion of propagating pions in the KSW expansion could extend its range of applicability beyond the relatively low energies required for the validity of the pionless theory. Since then there have been several suggestions for how to improve upon the KSW expansion (e.g. \cite{Beane:2001bc,Beane:2008bt}), but none have become widespread tools for doing nuclear physics.
 
\begin{figure}[t]
\includegraphics[width=10 cm]{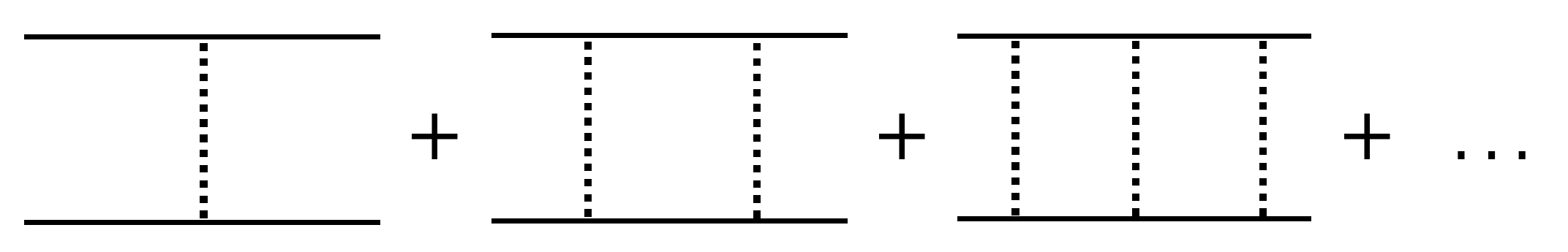}
\caption{ {\it Feynman graphs  corresponding to contributions to the two-nucleon scattering amplitudes $ \CA_{0}, \CA_{1}, \CA_{2},\ldots  $ via one-pion exchange in the spin-triplet channel, where the solid lines are nucleons and  the dashed lines are pions.}}
\label{fig:ladders}
\end{figure}

An interesting feature of the  FMS paper which I pursue here  is their observation that  the problem they had uncovered in the spin-triplet channels could be attributed to graphs corresponding to the iterated contribution to nucleon scattering from the one pion exchange (OPE) potential (i.e. the ladder diagrams shown in Fig.~\ref{fig:ladders}), and that the problems with convergence could be seen even in the chiral ($m_\pi\to 0$) limit, where the potential scales as $1/r^3$.  It is well known that such a potential has pathologies that require renormalization (no ground state exists for attractive channels, for example) and it is worth asking whether the lack of convergence suspected by FMS is a result of this pathology, or a result of the potential simply being too strong, or whether their conclusion was in fact  just an artifact of computing just the first two orders in the OPE expansion, and  in only a few partial waves.   

Since the NNLO calculation in the FMS paper was already heroic, these speculations might seem academic; however, pursuing their calculation to higher orders and for many partial waves is not unreasonable thanks to their observation that (i) one can focus on the graphs in Fig.~\ref{fig:ladders}, and  (ii)  one can compute them in the chiral limit.  In this paper I devise techniques for turning the computation of these graphs   into an algebraic recursion relation, allowing one to compute the diagrams to relatively high order with ease in all angular momentum channels at once, with a novel regularization and renormalization scheme that involves analytic continuation of angular momentum to render the diagrams finite.  This gives us the tools to look more deeply into the convergence problems encountered by FMS.  This work has been preceded by a number of excellent papers which have examined the convergence of the perturbative expansion as well as nonperturbative behavior in various partial waves, including  \cite{Nogga:2005hy,Birse:2005um,PavonValderrama:2016lqn,Wu:2018lai}.  I  reach the same general conclusion as this prior work, which is that perturbation theory works poorly in the ${}^3S_1$, ${}^3D_1$, $\epsilon_1$, and ${}^3P_0$ partial waves, is borderline in ${}^3P_1$, and converges well for other partial waves.   What this paper brings to the discussion that is new is the machinery for carrying out high order analytic calculations in the chiral limit, allowing one to make somewhat more precise statements about the convergence than found in the FMS paper, including especially for asymptotically large angular momentum.  

A feature of the techniques developed here is the special role played by angular momentum.  Not only are divergences regulated by analytically continuing to non-integer angular momentum, but one finds in the perturbative expansion that the spherical Bessel functions for the solution of the scattering wave function are systematically shifted away from integer value in an energy-dependent manner. This latter feature creates a natural connection to the 2005 work by Birse \cite{Birse:2005um} and the atomic physics literature he refers to, notably by Cavagnero and Gao \cite{cavagnero1994secular,gao1999repulsive}. Those earlier papers found that wave function solutions to the $1/r^3$ potential could be represented as a sum over Bessel functions with shifted order, and that at least in the repulsive interaction, could be solved for exactly. The shifted order can be interpreted as a wave function corresponding to non-integer angular momentum.  Furthermore, the shift is energy dependent, and is consistent with what I find here using field theoretic techniques \footnote{References \cite{cavagnero1994secular,gao1999repulsive} and their consistency with calculations presented here were brought to my attention by M. Birse.}.  Fully developing these connections is outside the scope of this paper, but suggest possibly fruitful directions of research for advancing the understanding of nuclear effective theory.

The layout of this paper is as follows.  In \S\ref{sec:KSW} I briefly review the KSW expansion and state the findings of FMS that cast doubt on the convergence of the KSW expansion.  In \S\ref{sec:computation} I set up  the  problem of computing the ladder diagrams in of Fig.~\ref{fig:ladders} in the spin-triplet channel and chiral limit, turning it into a recursion problem.  I do this first for the $L=J$ partial waves, computing the scattering amplitudes up to seven loops, and then for the $L=J\pm 1$ coupled channels, where the calculation is somewhat more involved,  up to three loops.  The actual amplitudes I compute are rather complicated expressions and are relegated to the Appendix.  A feature of the amplitudes is that divergences appear for a given number of loops only for angular momenta $L\le \half n_\text{loops}$.  Thus at any fixed order in the EFT expansion, the amplitudes for large enough $L$ are finite and parameter-free, and one can directly study their convergence as I do in \S\ref{sec:asymptotic}.  In order to study  the convergence of the expansion for low $L$, the amplitudes have to be renormalized, and in   \S\ref{sec:renormalization} I introduce the Minimal Angular Momentum Subtraction Scheme (\MJS) which allows us to do so by the subtraction of  poles in angular momentum.  I then proceed to renormalize the amplitudes with insertions of counterterms at tree level and one-loop, extending their utility by two orders in the perturbative expansion.  This is a somewhat technical section and can be skipped if the reader's main interest is in the discussion of the convergence behavior for low angular momentum, the topic of \S\ref{sec:lowJ}, where I consider scattering for $J= 1,\ldots,4$.  
%Up to this point I focused on the perturbative expansion sown in Fig.~\ref{fig:ladders}; in \S\ref{sec:unitary} I show that the nonperturbative physics of unitary fermions in the ${}^3S_1-{}^3D_1$ coupled channels can be simply incorporated into our calculation by analytically continuing the angular momentum of the $S$-wave to $L=-1$.  
I conclude with a discussion about my results in the final section, along with a discussion of connections with the work of references \cite{Birse:2005um,cavagnero1994secular,gao1999repulsive}.

 \section{The KSW expansion}
 \label{sec:KSW}
 
 The point of the KSW expansion for nuclear EFT is to have a consistent power-counting scheme which allows one to  compute scattering amplitudes to a consistent order, with a small expansion parameter $\hat k$ which enters to a higher power at each subsequent order \cite{Kaplan:1998tg,Kaplan:1998we}.  This parameter $\hat k$ is the ratio of nucleon momentum $k$ to a fundamental parameter of nature,  $\Lambda_{NN}$: 
 \beq
 \hat k= \frac{k}{\Lambda_{NN}}\ ,\qquad \Lambda_{NN}   = \frac{8\pi f^2}{g_A^2 M} = 285\MeV\ .
  \eqn{Lambda}
 \eqn{khatdef}\eeq 
 In this expression, $g_A = 1.27$ is the axial coupling, $M=939\MeV$ is the nucleon mass, $f = 131\MeV$ is the pion decay constant.  The value of $\Lambda_{NN}$, which arises when computing OPE scattering amplitudes,  can be interpreted as the parametric value of the pion mass for which the OPE Yukawa potential would support a 2-nucleon bound state, obtained by equating the Bohr radius for the Coulomb potential with identical short distance strength to the Compton wavelength of the pion.  As such it provides a natural scale for where a perturbative chiral expansion of the scattering amplitude  should break down  --- one that is unfortunately low, comparable to the Fermi momentum of matter at nuclear density.   Furthermore, this naive derivation leaves open the question of whether the expansion suffers from pathologies not parametrized by $\Lambda_{NN}$ due to the more singular parts of the OPE potential, a question this paper addresses explicitly and finds is not the case.
 Central to the KSW expansion is its power counting, where the nucleon-nucleon scattering length is treated as $O(1/\hat k)$; the leading operator in the low-energy theory is then a four-nucleon contact interaction whose coefficient scales as $O(1/\hat k)$.  This large large anomalous dimension  explains why the low energy effective theory for the strong interactions differs markedly from Fermi's effective theory for the weak interactions, even though both are described by four-fermion operators.  With every nucleon loop contributing a compensating factor of $\hat k$,  one finds that to leading order,   $O(1/\hat k)$, one must sum an infinite set of bubble diagrams to describe nucleon-nucleon scattering.  Thus, although the expansion is perturbative in $\hat k$ starting with $O(\hat k^{-1})$, an  infinite number of Feynman diagrams are computed at each order, explaining how the theory can give rise to nuclear bound states  and large scattering lengths.  These infinite sums take the form of a geometric series and  can be performed analytically, giving rise to a nontrivial pole  in the leading order amplitude, fit to the deuteron pole and and the nearly bound di-neutron.   Since amplitudes are computed to a consistent order in $\hat k$, they can be renormalized at each order with a finite number of counterterms.  This eliminates all cutoff dependence in the answer, allowing $S$-matrix elements at each order in the expansion to be expressed entirely in terms of a finite number of low energy constants fit to nature.
 
 The pion field must be included in the theory for scattering at $k\gtrsim m_\pi/2$, and the KSW expansion is consistent with the chiral expansion so long as one considers $m_\pi \sim O(\hat k)$.  Since the pions are derivatively coupled, one-pion exchange contributes to the amplitude a factor 
 \beq
 \frac{4\pi  i}{M \Lambda_{NN}} \vec\tau_1\cdot\vec\tau_2 \frac{(\vec q\cdot \vec \sigma_1)(\vec q\cdot \vec \sigma_2)}{\vec q^2+m_\pi^2} \ .\eqn{OPE} \eeq
With $q\sim m_\pi \sim k$ one  sees that this expression is $O(1)$ in the $\hat k$ expansion; therefore the diagrams in Fig.~\ref{fig:ladders} scale as $\hat k^\ell$, where $\ell$ is the number of loops, since  each loop integration  contributes a factor of $\hat k$   Ref.~\cite{Kaplan:1998tg}. The loop expansion  in Fig.~\ref{fig:ladders} is thus equivalent to the KSW expansion of the amplitude in powers of $\hat k$.  These ladder graphs are only a subset of what must be summed to a given order in the realistic problem.

   The crucial observation of FMS was that there appeared to be a failure with the convergence of the KSW expansion for physically relevant momenta in some spin-triplet channels \cite{Fleming:1999ee}.  Their NNLO computation (to $O(\hat k)$) included the first two OPE ladder diagrams in Fig.~\ref{fig:ladders}, tree level and one-loop, and the authors identified these contributions   to be large even in the chiral limit $m_\pi\to 0$.       The results FMS found in the chiral limit included \cite{Fleming:1999ee}
   \beq
   \hat\CA_{\Pw} &=& 1 +\hat k\left(\frac{2\pi}{5}+i\right)\cr
   \hat\CA_{\Pwi} &=& -\frac{1}{2} + \hat k\left(\frac{\pi}{10} + \frac{i}{4}\right)\cr
   \hat\CA_{\Pwii} &=& \frac{1}{10} + \hat k\left(\frac{3\pi}{50} + \frac{i}{60}\right)\cr
   \hat\CA_{\Dwii} &=& \frac{1}{2} + \hat k\left(\frac{3\pi}{70} + \frac{i}{4}\right)\cr
   \hat\CA_{\Dwiii} &=& -\frac{1}{7} + \hat k\left(\frac{3\pi}{49} + \frac{i}{28}\right)\ .
 \eqn{FMSamps}
   \eeq
   where I have defined the dimensionless scattering amplitude $\hat \CA$ in terms of the S matrix as
   \beq
   S=1+\frac{i M k}{2\pi}\CA =1+ 2i\hat k\hat\CA  \ ,\qquad \hat\CA \equiv \frac{M\Lambda_{NN}}{4\pi}\,\CA\ .
   \eeq

Each of these amplitudes is exact to  $O(\hat k)$ and   parameter-free.   FMS noted that the real nonanalytic terms linear in $\hat k$ are  accompanied by a factor of $\pi$ (nonanalytic because $k=\sqrt{ME}$), greatly reducing the range where the  $O(\hat k)$ term is small compared to the leading $O(1)$ contribution.  For example, in the $\Dwiii$ partial wave one sees  that the ratio of leading to subleading contributions to the real part of $\hat\CA$ is $(3\pi/7) \hat k$ implying a   correction of more than  50\%    for $k \gtrsim 105 \MeV$; in the $\Pwii$ channel that ratio is $(3\pi/5)\hat k$, and 50\% corrections are encountered for $k\gtrsim 75\MeV$.
     This suggests that the expansion breaks down for nucleon momenta much less than those of interest in nuclei, where one would like a theory valid up to   $k\sim \Lambda_{NN}$, and expect it to be valid well above $k\sim m_\pi$.  The authors speculated that the nonanalytic contributions to the ladder diagrams, bringing with them powers of $\pi$, were responsible for the precocious breakdown of the KSW expansion.   However, with just two orders in the expansion to compare, it is hard to make a definitive statement. To address this question I pursue a calculation of the ladder diagrams to higher order in the KSW expansion.

 \section{A recursion relation for scattering amplitudes for the chiral  $1/r^3$ potential}
 \label{sec:computation}
 
 In this section I develop the technology for transforming the ladder diagrams of Fig.~\ref{fig:ladders} with massless pion exchange into a set of recursive algebraic relations, starting with the spin-triplet $L=J$ partial waves, and then the coupled $L=J\pm 1$ channels.  I do not consider the spin-singlet channels where the OPE potential is just an uninteresting Dirac $\delta$-function in the chiral limit.  In this paper do  not incorporate a hallmark feature of the KSW expansion, namely the resummation of contact interactions to all orders in the $S$-wave channels which accounts for the nonperturbative physics of the large scattering lengths; in my conclusion I discuss how one might do so within the framework of this paper by simply taking the angular momentum of the ${}^3S_1$ partial wave to be $\ell=-1$ instead of $\ell=0$.
  
 \subsection{Spin-triplet   amplitudes   for $L=J$}
 \label{sec:LeqJcomputation}
 
 I first consider the  $L=J$ spin-triplet partial wave scattering amplitudes due to OPE in the chiral limit.  As mentioned, the diagrams in Fig.~\ref{fig:ladders} form only a subset of the diagrams to sum in the KSW expansion, but they form a self consistent subset in their own right, so long as they are augmented by contact interactions as needed for renormalization.  After all, formally they are just the solution to the nonrelativistic (and not chiral invariant) Schr\"odinger equation for two nucleons interacting vie the OPE potential \eq{OPE}. Thus, rather than computing the diagrams of  Fig.~\ref{fig:ladders} by using conventional momentum space Feynman rules, I will work in coordinate space, a technique used to advantage in Ref.~\cite{Fleming:1999ee}.
 
  My starting point is the radial Schr\"odinger equation with a general rescaled potential $\CV$ in the form
 \beq
\left(\partial_\rho^2+ \frac{2}{\rho} \partial_\rho+1- \frac{L(L+1)}{\rho^2} \right)u(\rho) =\CV(\rho) \,u(\rho)  \ . 
 \eqn{SeqLeqU}
 \eeq
  where  $L$ is the orbital angular momentum, $\rho = k r$ with $k=\sqrt{ME}$. To solve this perturbatively in powers of $\CV$ I introduce the Green function
 \beq
 g_L(\rho,\rho' )  = \theta(\rho'-\rho)j_L(\rho) h^{(1)}_L(\rho') + \theta(\rho-\rho') j_L(\rho') h^{(1)}_L(\rho)\ ,
 \eqn{gedef}
 \eeq
 where $j_L$ and $h_L^{(1)}$ are spherical Bessel and Hankel functions respectively,  with $g_L$ satisfying the equation
 \beq
\left(\partial_\rho^2+ \frac{2}{\rho} \partial_\rho+1- \frac{L(L+1)}{\rho^2} \right) g_L(\rho,\rho' )  =  \frac{i}{\rho^2} \delta(\rho-\rho')\ .
\eeq
This definition incorporates the boundary conditions that the scattered wave function be regular at the origin, and an outgoing spherical wave at infinity. Then at $O(n)$ in the perturbative expansion,  the wave function is given by
\beq
u^{(n)} = -i\int_0^\infty d\rho' \,\rho^{\prime\, 2} \, g_L(\rho,\rho') \,\CV(\rho')\, u_L^{(n-1)}(\rho')\ ,\qquad u_L^{(0)}(\rho) = j_L(\rho)\ ,
\eqn{upert}
\eeq
while the amplitude is
\beq
\hat \CA^{(n)} = -\frac{1}{\hat k} \int_0^\infty d\rho' \,\rho^{\prime\, 2} \, u^{(0)}(\rho) \CV(\rho) u^{(n)}(\rho)\ .
\eqn{apert}
\eeq

In the present case of interest, $L=J$ nucleon-nucleon scattering in the spin-triplet channel by the OPE potential,  one has $\CV = -2\left(1+2(-1)^L\right) \hat k/\rho^3$, so that
 \beq
\left(\partial_\rho^2+ \frac{2}{\rho} \partial_\rho+1- \frac{L(L+1)}{\rho^2} \right)u(\rho) =-2\left(1+2(-1)^L\right)\frac{\hat k}{\rho^3}\,u(\rho)  \ , 
 \eqn{SeqLeqJL}\eeq
where the factor $- \left(1+2(-1)^L\right)$ arises from the $ \vec\tau_1\cdot\vec\tau_2$ isospin factor in \eq{OPE}.  This equation is quite singular, as an attractive $1/r^3$ potential has no ground state, and positive energy scattering solutions will all have an infinite number of nodes in any neighborhood around the origin, with vanishing amplitude.  A perturbative expansion for scattering solutions is still possible, but will require renormalization.  A simple power counting of the lowest dimension contact interactions possible in the $L$ partial wave reveals that one would expect divergences to appear at order $\hat k^{2L}$.
With this potential, the above equations yield  
\beq
u_L^{(n)}(\rho) = 2i\left(1+2(-1)^L\right)\hat k  \int_0^\infty \frac{d\rho'}{\rho'} g_L(\rho,\rho') u_L^{(n-1)}(\rho')\ ,\qquad u_L^{(0)} = j_L(\rho)\ .
\eqn{urecur}
\eeq
and
\beq
\hat\CA_L^{(n)} =
2\left(1+2(-1)^L\right) \int_0^\infty \frac{d\rho}{\rho} j_L(\rho) u_L^{(n)}(\rho)\ .
\eqn{Arecur}
\eeq
Note that the superscript ``$n$" starts at $n=0$ and refers to the power of $\hat k$; for example, while $\hat\CA_L^{(0)} $ entails one insertion of the pion potential (representing the tree diagram in Fig.~\ref{fig:ladders}) but is independent of $\hat k$.

The scattering amplitude and wave function can be computed recursively to any order in $\hat k$ by using the following two integrals, which converge for sufficiently large $\ell_1,\ell_2$:
\beq
%X_{(\ell,0)}^{\ell_1,\ell_2}& =& 
\int_0^\infty \frac{d\rho}{\rho} j_{\ell_1}(\rho)   j_{\ell_2}(\rho) 
=
-\frac{2 \cos \left(\frac{1}{2} \pi  (\ell_1 -\ell_2)\right)}{(\ell_1 -\ell_2 -1) (\ell_1-\ell_2 +1) (\ell_1+\ell_2) (\ell_1+\ell_2 +2)}\ .
 \eqn{jjint} 
 \eeq
and
\beq
 \int_0^\infty \frac{d\rho'}{\rho'} g_{\ell_1}(\rho,\rho') j_{\ell_2}(\rho')= \alpha_{\ell_1 \ell_2}\, j_{\ell_1}(\rho) + \beta_{\ell_1 \ell_2}\, j_{\ell_2-1}(\rho) + \gamma_{\ell_1\ell_2} \,  j_{\ell_2+1}(\rho)\ .
 \eqn{jgjint}
 \eeq
with  $\rho$-independent coefficients given by 
\beq
\alpha_{\ell_1 \ell_2} &=& -\frac{2 e^{i\pi(\ell_2-\ell_1)/2}}{(\ell_2-\ell_1-1)(\ell_2-\ell_1+1)(\ell_1+\ell_2)(2 + \ell_1+\ell_2)}\cr &&\cr
\beta_{\ell_1 \ell_2} &=& \frac{i}{(\ell_2-\ell_1-1)(\ell_1+\ell_2)(1+2\ell_2)}\cr&&\cr
\gamma_{\ell_1 \ell_2} &=& \frac{i}{(\ell_2-\ell_1+1)(2 + \ell_1+\ell_2)(1 +2\ell_2)}\ .
\eqn{abc}
\eeq
The derivation of this formula is given in Appendix~\ref{sec:Bessel}.
The poles in $\alpha, \beta, \gamma$ at $\ell_2=\ell_1\pm 1$ do not  imply ill-defined integrals for those values, but for those cases the expression in \eq{jgjint} must be evaluated as the limit $\ell_2\to \ell_1\pm 1$, which yields the finite results
\beq
 \int_0^\infty \frac{d\rho'}{\rho'} g_{\ell_1}(\rho,\rho') j_{\ell_2}(\rho')\Biggl\vert_{\ell_2=\ell_1\pm 1}= \eta_{\ell_1\ell_2}\, j_{\ell_1}(\rho) + \omega_{\ell_1\ell_2} \, j_{2\ell_2-\ell_1}(\rho) + \zeta_{\ell_1,\ell_2} \,\partial_\nu j_\nu(\rho)\biggl\vert_{\nu=\ell_1}\ ,
 \eeq
 where
 \beq
 \eta_{\ell_1,\ell_2} = \frac{\pi  \left(2 \ell_2+1\right)+i \left(2 \ell_1+1\right) \left(\ell_2-\ell_1\right)}{2 \left(2 \ell_1+1\right) \left(2
   \ell_2+1\right)^2}\ ,\qquad
   \omega_{\ell_1,\ell_2} = \frac{i^{\ell_2-\ell_1}}{2 \left(2 \ell_2+1\right)^2}\ ,\qquad
     \zeta_{\ell_1,\ell_2} =\frac{i}{\left(\ell_1+\ell_2\right) \left(\ell_1+\ell_2+2\right)}\ .
\eqn{limex}     \eeq	
When such an expression arises as an intermediate result, the derivative with respect to order can be taken after the sequence of integrations has been performed.
Therefore the integrals \eq{urecur} and \eq{Arecur} can  in principle be solved recursively to any order $n$ using the two integrals in \eq{jjint} and \eq{jgjint}, and computing the pion ladder graphs in Fig.~\ref{fig:ladders} reduces to a recursive algebraic problem.

As  expected, one encounters infinities at sufficiently high order in $n$ for any angular momentum $L$, the first generically appearing at $n=2L$, a logarithmic singularity at the origin for $\hat\CA^{(n)}$ arising in the integral $\int (d\rho/\rho)\,j_L(\rho) j_{-L}(\rho)$.  Since I am performing these integrals in coordinate space with the aid of the very particular integrals \eq{jjint} and \eq{jgjint}, it is not feasible to regulate the integrals via dimensional regularization; however a simple opportunity presents itself through angular momentum regularization: I simply analytically continue the calculation to non-integer angular momenta $\ell$, defining the integrals in \eq{jjint} and \eq{jgjint} by the expressions on the right for values of $\ell_{1,2}$ outside their regions of convergence.  As one will see, divergences in the amplitudes then appear as poles in $(\ell-L)$, where $L$ is integer, in a similar way to how divergences appear as poles at integer dimension in the dimensional regularization scheme; later in this paper I discuss how such amplitudes can be renormalized by subtraction of those poles.  For the rest of this paper $L$ will denote integer angular momentum, while $\ell$ will correspond to its continuation to non-integer values.

Following this program, it is relatively easy to compute the diagrams in Fig.~\ref{fig:ladders} up to seven loops for the spin-triplet, $L=J$ partial waves; the seven-loop diagram takes about 0.01 seconds to compute using Mathematica on a laptop.  My results are given in Appendix~\ref{sec:LeqJamps}, \eq{ampsLeqJ}. Two nontrivial checks of the result are (i) the $n=0,1$ results agree with the FMS results in the chiral limit; (ii) the amplitudes exhibit unitarity, a highly nontrivial constraint.  The latter property is most easily seen by computing the phase shifts and checking that they are real to the order one is working. A curious feature of the result is that the amplitudes can be written as a sum of poles at both positive and negative integer and half-integer values of angular momentum; one only has an explanation for the poles at positive integer $\ell$, which are associated with the expected divergences for physical scattering.
  
The  phase shifts $\delta$ in each partial wave  may be computed perturbatively by writing them as an expansion
$\delta = \sum_{n=1}^\infty \delta^{(n)}$, 
where $\delta^{(n)} = O(\hat k^n)$,  and solving the equation
\beq
S  =1 + 2 i \hat k\,\hat \CA = e^{2i\delta} 
\eqn{smat1}
\eeq
order by order in the momentum expansion.  One finds
\beq
\delta^{(1)} &=&\hat k \hat \CA^{(0)} \cr
\delta^{(2)} &=& \hat k\hat \CA^{(1)} - i \hat k^2 \left(\hat\CA^{(0)}\right)^2 \cr
\delta^{(3)} &=& \hat k\hat\CA^{(2)} -2i \hat k^2 \hat \CA^{(0)}\hat\CA^{(1)}-\frac{4}{3}\hat k^3\left(\hat\CA^{(0)}\right)^3 \cr
\delta^{(4)} &=& \hat k\hat\CA^{(3)}-2i \hat k^2 \hat\CA^{(0)}\hat\CA^{(2)}-i\hat k^2\left(\hat\CA^{(1)}\right)^2-4\hat k^3 \left(\hat\CA^{(0)}\right)^2\hat\CA^{(1)}+2i\hat k^4\left(\hat\CA^{(0)}\right)^4\ ,
\eeq
and so on. I adopt the notation that $\delta^{(n)}$ is a contribution to the phase shift at $O(\hat k^n)$, even though it arises from computing  $\hat\CA^{(n-1)}$ so that the superscript always describes the power of $\hat k$ to expect. Substitution of these expressions for $\hat\CA^{(n)} $ in \eq{ampsLeqJ}   into these equations yields the phase shifts shown in \eq{LeqJphase}, and they are all manifestly real as required by unitarity.

 \subsection{Spin triplet amplitude for the ${}^3P_0$ channel}
 
I next turn to the $L=J\pm 1$ partial waves. A special case  is for $L=1$, $J=0$, the ${}^3P_0$ channel, since it is not coupled to any other angular momentum.  For this special partial wave the Schr\"odinger equation reads
 \beq
\left(\partial_\rho^2+ \frac{2}{\rho} \partial_\rho+1- \frac{\ell(\ell+1)}{\rho^2} \right)u(\rho) =4\left(1+2(-1)^\ell\right)\frac{\hat k}{\rho^3}\,u(\rho)  \ , 
 \eqn{SeqLeqJ}\eeq
with $\ell\to 1$.  Comparing with \eq{SeqLeqJL}, one sees that the equation for ${}^3P_0$ scattering is identical up to a factor of $-2$ in the interaction strength to that for ${}^3P_1$, and so the  scattering amplitudes $\hat\CA^{(n)}$ for this channel are trivially given by those in \eq{ampsLeqJ} with an additional factor of $(-2)^{n+1}$.

\subsection{Spin-triplet   amplitudes   for  $L=J\pm1$ coupled channels}

The radial Schr\"odinger equation for scattering in the coupled $L=J\pm 1$ channels, analytically continued to non-integer $\ell=j\pm 1$,  is (see, for example, Ref.~\cite{wu2014quantum})
 \beq
\left[\left(\partial_\rho^2+ \frac{2}{\rho} \partial_\rho+1\right) - \frac{\ell_a(\ell_a+1)}{\rho^2} \right]u_{\ell_a}(\rho) =  \left(1-2(-1)^j\right) \,\hat k \sum_b \frac{v_{ab}}{\rho^3}\,u_{\ell_b}(\rho)  \ , 
\eqn{Seqmix}
 \eeq
 where $a,b$ run over $1,2$ with angular momentum $\ell_{1,2} = j\mp 1$ respectively and
 \beq
%\bfeps_{ab}=\frac{k}{\Lambda_{NN}}\frac{ \left(1-2(-1)^j\right)}{{2j+1}}\begin{pmatrix}
 %2(j-1)  & -6 \sqrt{j(j+1)}  \\ -6 \sqrt{j(j+1)} & 2(j+2) \end{pmatrix}_{ab}  \ ,
v = \frac{ 1}{{2j+1}}\begin{pmatrix}
 2(j-1)  & -6 \sqrt{j(j+1)}  \\ -6 \sqrt{j(j+1)} & 2(j+2) \end{pmatrix}   \ .
 \eqn{bfepsdef}\eeq
 Repeated indices are not automatically summed.
 The analog of \eq{urecur} and \eq{Arecur} are 
 \beq
u_{\ell_a}^{(n)}(\rho) &=& -i\left(1-2(-1)^j\right)\hat k \, \sum_b v_{ab}\int_0^\infty \frac{d\rho'}{\rho'} g_{\ell_a}(\rho,\rho')  u_{\ell_b}^{(n-1)}(\rho')\ ,\qquad u_{\ell_a}^{(0)} = j_{\ell_a}(\rho)\ ,\cr &&\cr
\hat\CA^{(n)}_{\ell_a \ell_b} &=& -\left(1-2(-1)^j\right) v_{ab} \int\frac{d\rho}{\rho} j_{\ell_a}(\rho) u^{(n)}_{\ell_b}(\rho)\ .
\eqn{urecurCoupled}
\eeq
 
  %%%%%%

One can proceed as for the $L=J$ case,  calculating the scattering amplitude at successively higher loops by means of the integrals \eq{jjint} and \eq{jgjint}, which reduce the problem to solving recursive algebraic relations.  For the coupled channels I have computed the amplitudes to three loops,  $O(\hat k^3)$, and my results are given in Appendix~\ref{sec:LJcoupledJg2},  \eq{ampsLJCoupled}.  One then computes the phase shifts $\delta_\pm$ and $\epsilon$ by equating at each order in $\hat k$ an expansion  of the equation 
 \beq
S = 1 + \frac{i M k}{2\pi}\CA = 1 + 2 i \hat k \hat\CA =
\begin{pmatrix}
e^{2i\delta_-}\cos2\epsilon & i e^{i(  \delta_-+  \delta_+)}\sin2  \epsilon\\ i e^{i( \delta_- +  \delta_+)}\sin2 \epsilon &  e^{2i  \delta_+}\cos2  \epsilon\end{pmatrix}\ ,
\eeq
where phase shifts are given in the ``barred" convention of Stapp et al. \cite{stapp1957phase}. All of the phase shifts I find are real, providing a nontrivial check on the calculation showing that unitarity is preserved.   

 \section{Asymptotic behavior for large $L$ and convergence of the expansion}
\label{sec:asymptotic}

I now turn to the central question of interest, whether the results I have derived display convergence of the perturbative KSW expansion or not.  To address that question for the low $L$ partial waves one will have to deal with renormalization; here I sidestep that issue by considering first the behavior of the phase shifts for asymptotically large $L$.  One expects the EFT expansion to work better at large $L$ due to the angular momentum barrier which makes scattering events less sensitive to short distance physics, and that is indeed what one finds.  This physics was previously explored for spin-singlet scattering at finite pion mass in Ref.~\cite{PavonValderrama:2016lqn}.

\subsection{$L=J$ partial wave amplitudes for large $L$}

 %%%%%%%%%
 %%%%%%%%%

\begin{figure}[t]
\includegraphics[width=16cm]{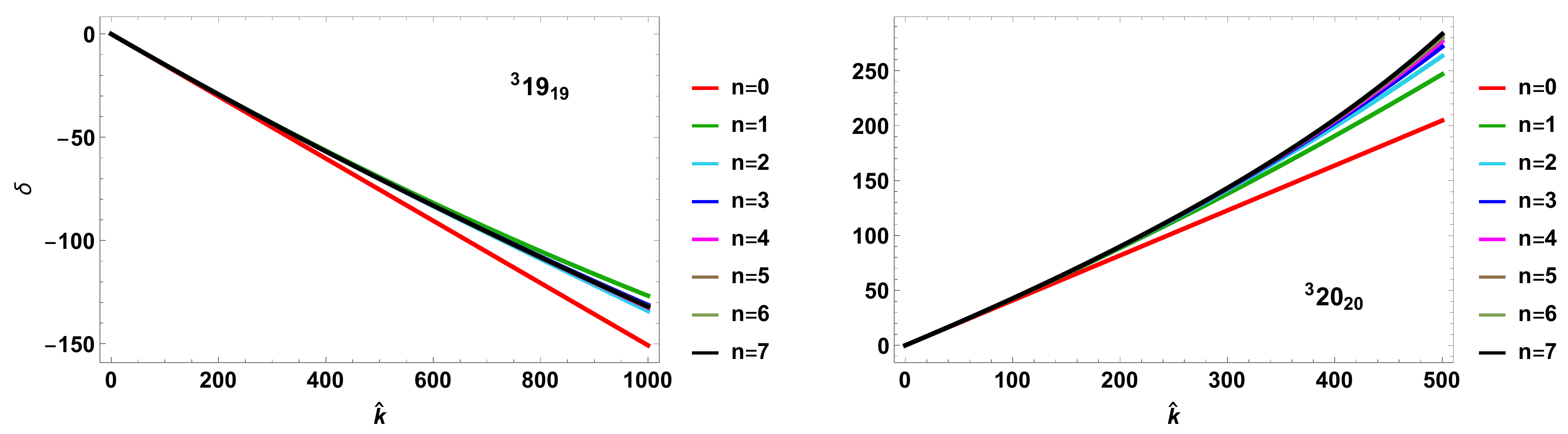}
\caption{ {\it Illustration of the large-$L$ behavior of spin-triplet $L=J$ phase shifts   for the  $L=19$ and $L= 20$ partial waves, with the cumulative phase shift to $O(n)$ in degrees  plotted   versus $\hat k $, derived from the expressions for $\hat \CA^{(n)}$ for $n=0,\ldots,7$.  Note the scale of the horizontal axes; $\hat k$ is defined in \eq{khatdef}, and $\hat k = 1000$ corresponds to $k = 285\GeV$. }}
\label{fig:LeqJasymp}
\end{figure}

%%%%%%%%%
 %%%%%%%%%

\begin{figure}[t]
\includegraphics[width=16cm]{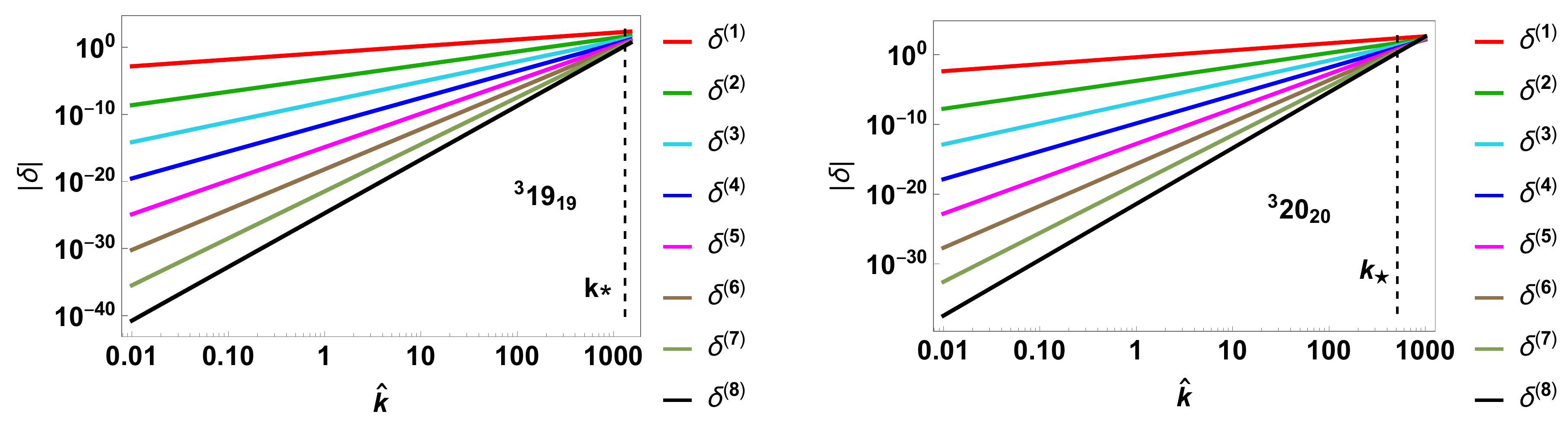}
\caption{ {\it The Lepage plot of the contributions $|\delta^{(n)}|$ in radians from \eq{LeqJphase} at each order for $n=1,\ldots,8$ versus $\hat k$ on a log-log plot for spin triplet, $L=J$ partial waves with $L=19$ and $L=20$.   The estimated radius of convergence $k\le k_\star(L)$ from  \eq{kstar} is marked in each case, which appears to work well.}}
\label{fig:LeqJasympLepage}
\end{figure}

%%%%%%%%%
 %%%%%%%%%

   For the $L=J$ partial waves one finds from \eq{ampsLeqJ}
\beq
\hat\CA^{(n)} \sim \frac{\left(1+2 (-1)^L\right)}{L^2\,(n+1)!}  
\left(\frac{2i \hat k\left(1+2 (-1)^L\right)}{L^2}\right)^n
  \, \,\left[ 1 -\frac{n+1}{L}\left(1 + \frac{3i\pi n}{16}\right) + O(L^{-2})\right]\ ,
%\qquad L=J;\quad L\to\infty
\eqn{nval}\eeq
so one sees that the leading term for large $L$ is  proportional  to the combination
\beq
  \frac{(2\hat k)^n}{L^{2n+2} (n+1)!}\ ,
 \eeq
 and thus should be expected to have a radius of convergence much larger than $\hat k = 1$.  A caveat however: this expansion is performed for large  $L$ at fixed $n$, while  the subleading terms in \eq{nval} suggest that the large-$n$ and large-$L$ expansions do not commute, and that it would not be surprising if the perturbative expansion were asymptotic, eventually diverging for large $n$ at fixed $L$.  
 
Instead of examining the large-$L$ behavior of the perturbative amplitudes, it is more   instructive to look at  the phase shifts given  in \eq{LeqJphase},  whose asymptotic behavior is given by:
 \beq
 \left\{\frac{\delta^{(1)}}{\left(1+2(-1)^L\right)},\ldots,\frac{\delta^{(8)}}{\left(1+2(-1)^L\right)^8}\right\}
 & \sim &
 \left\{  \frac{\hat k}{L^2},
 \frac{3 \pi  \,\hat k^2 }{8 L^5}, 
 \frac{8 \,\hat k^3  }{3 L^8},
 \frac{315 \pi  \,\hat k^4  }{128 L^{11}},
 \frac{128  \,\hat k^5 }{5 L^{14}},
 \frac{15015 \pi   \,\hat k^6 }{512 L^{17}},
\frac{12288  \,\hat k^7 }{35 L^{20}},
\frac{14549535 \pi \,\hat k^8    }{32768 L^{23}}
 \right\}\ ,\cr &&
 \eqn{LeqJasympphase}\eeq
 a sequence that can be exactly reproduced by the formula
 \beq
 \delta^{(n)} \sim \left(1+2(-1)^L\right)^n \frac{\hat k^n }{L^{3 n-1}}\frac{\sqrt{\pi }\, 2^{n-2}  \Gamma \left(\frac{3 n}{2}-\frac{1}{2}\right)}{\Gamma \left(\frac{n}{2}+1\right) \Gamma (n+1)}\ ,
  \eqn{Gam} \eeq
where the $\Gamma$ functions depending on $n/2$ account for the peculiar behavior of a factor of $\pi$ appearing at every other order in the expansion.  While this formula was derived empirically from the eight orders I have computed, it seems likely to be correct to all orders.  

Although I have pointed out that there is reason to doubt that this large-$L$ expansion is valid for $n\gg L$, if one naively sums  the expression in \eq{Gam} to all orders in $n$ one arrives at an expression involving hypergeometric functions
 \beq
\delta&=& \sum_{n=1}^\infty \delta^{(n)} \sim\sum_{n=1}^\infty \left(1+2(-1)^L\right)^n \frac{\hat k^n }{L^{3 n-1}}\frac{\sqrt{\pi }\, 2^{n-2}  \Gamma \left(\frac{3 n}{2}-\frac{1}{2}\right)}{\Gamma \left(\frac{n}{2}+1\right) \Gamma (n+1)}
\cr &&
 =\frac{\left(2 (-1)^L+1\right) \hat k \  _3F_2\left(\frac{1}{3},\frac{2}{3},1;\frac{3}{2},\frac{3}{2};\frac{27 \hat k^2\left(2 (-1)^L
   +1\right)^2}{L^6}\right)}{L^2}-\frac{1}{2} \pi  L \left(\, _2F_1\left(-\frac{1}{6},\frac{1}{6};1;\frac{27 \hat k^2\left(2 (-1)^L
   +1\right)^2}{L^6}\right)-1\right)\ ,\cr &&
 \eeq
   which exhibits a breakdown of the $\hat k$ expansion when the argument of the hypergeometric functions exceeds one, namely for
   \beq
 \frac{27 \hat k^2\left(2 (-1)^L
   +1\right)^2}{L^6} \gtrsim 1\ ,\qquad \Longrightarrow\qquad  \hat k \gtrsim k_\star(L)\equiv \frac{L^3}{\sqrt{27}\, \left\vert 2 (-1)^L
   +1\right\vert}\ ,
  \eqn{kstar} \eeq
   which suggests that the KSW expansion should converge well in the region $\hat k\lesssim  \hat k_\star(L)$.  Note that $ \hat k_\star(L)>1$  for  $L\ge 3$.  The $L^3$ dependence  of $k_\star$ agrees with the conclusion in eq. (29) of Ref.~\cite{,Birse:2005um}, derived from semi-classical arguments.

A plot of the cumulative phase shifts computed  from \eq{LeqJphase} is presented in Fig.~\ref{fig:LeqJasymp} for the $L=J=19$ and $L=J=20$ partial waves  for which $\hat k_\star(19) = 1320$ and  $\hat k_\star(20) = 513$ respectively.  The horizontal axes extend to $\hat k=1000$ in each case,  corresponding to the very high momentum $k=285\GeV$.   
 In both one sees the sort of convergence expected from the above argument; this is made even more clear by the Lepage plots in  Fig.~\ref{fig:LeqJasympLepage} where the individual contributions $\delta^{(n)}$ are plotted versus $\hat k$ on a log-log plot.  With each successive order one sees a lower and steeper line, all of which converge in the vicinity of $\hat k \simeq \hat k_\star(L)$, providing visual  confirmation that the expansion is well understood and convergent.

It is worth noting that   in each case in  Fig.~\ref{fig:LeqJasymp}, if one were to only look at the leading two orders of the expansion (the red and green lines) one would erroneously underestimate the radius of convergence  revealed by the Lepage plots in Fig.~\ref{fig:LJcoupledLepage}, perhaps by a factor of three or so.  This is the order to which the FMS paper computed  \cite{Fleming:1999ee}, and one sees that at least in these examples such a calculation can give an unreliably pessimistic conclusion about the convergence of the expansion.

\subsection{$L=J\pm1$ partial wave amplitudes for large $L$}

%%%%%%%%%
 %%%%%%%%%

\begin{figure}[t]
\includegraphics[width=18cm]{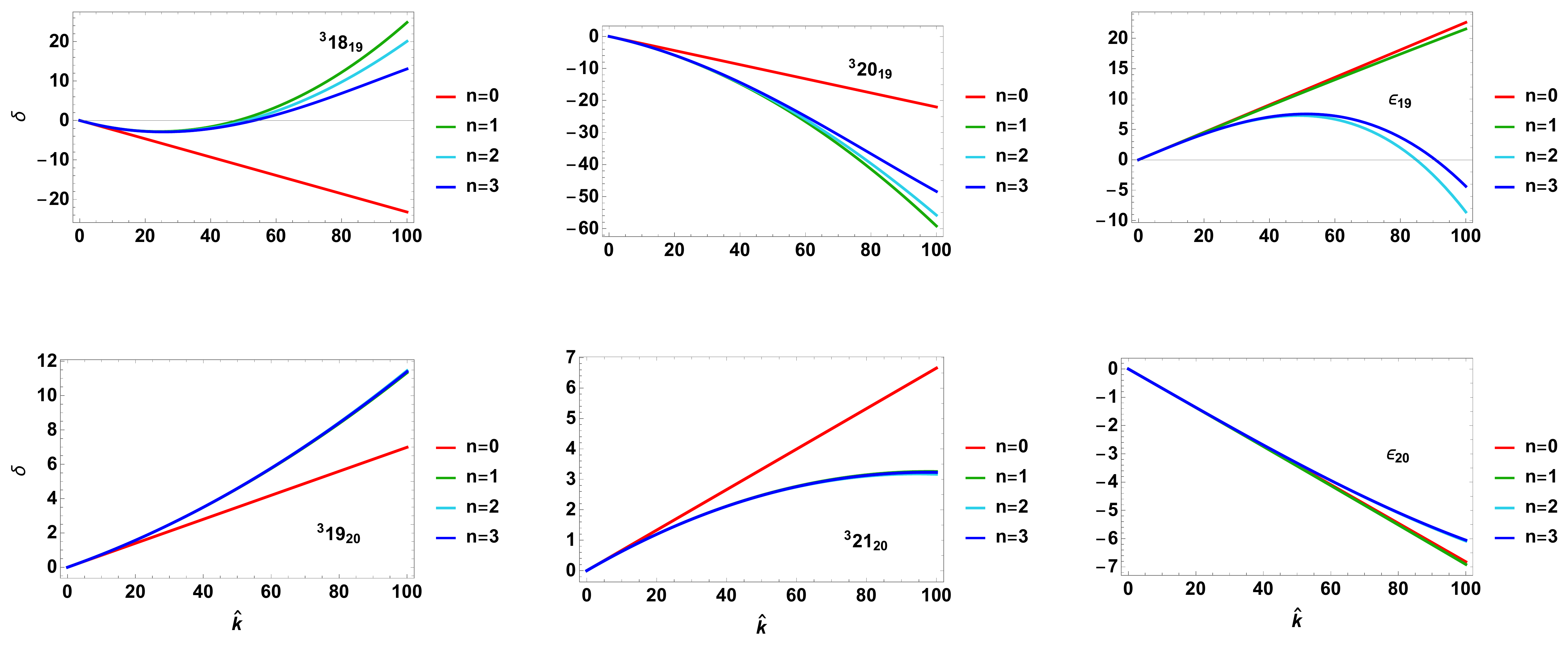}
\caption{ {\it  Phase shifts (in degrees) for the $L=J\pm1$ partial waves with the $J=19$ and $J=20$,  plotted   versus $\hat k $.  }}
\label{fig:LJcoupledasymp}
\end{figure}

%%%%%%%%%
 %%%%%%%%%

\begin{figure}[t]
\includegraphics[width=18cm]{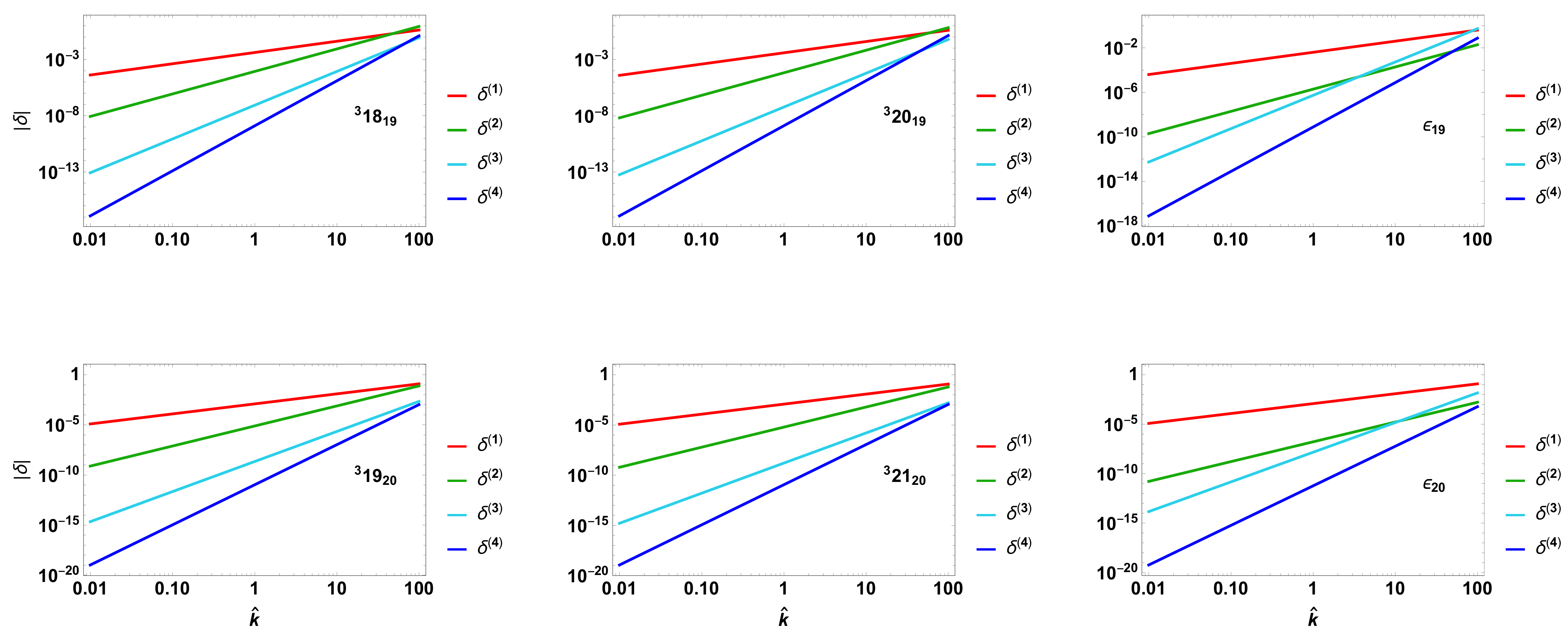}
\caption{ {\it Lepage plots  for the $L=J\pm1$ partial waves with the $J=19$ and $J=20$,  plotted   versus $\hat k $. }}
\label{fig:LJcoupledLepage}
\end{figure}

%%%%%%%%%
 %%%%%%%%%

 The limiting expressions for the coupled channel amplitudes in \eq{ampsLJCoupled} for large $J$ are given by
\beq
\hat\CA^{(0)} &\sim &\frac{\left(2(-1)^J-1\right)}{ J^2}\,
\half \begin{pmatrix*}[r]
1 & -1 \\
 -1& 1 \\
\end{pmatrix*}
\ ,
\cr &&\cr
\hat\CA^{(1)} &\sim &\hat k\frac{\left(2(-1)^J-1\right)^2}{J^4}\,\left[ \frac{i}{2}\begin{pmatrix*}[r]1 & -1\\-1&1\end{pmatrix*}+\frac{3 \pi}{8 }\begin{pmatrix*}[r]1 & 0\\0 & -1\end{pmatrix*} \right]\ ,
\cr &&\cr
\hat\CA^{(2)} &\sim &\hat k^2\frac{\left(2(-1)^J-1\right)^3}{J^6}
\begingroup
\left[
\renewcommand*{\arraystretch}{1.3}
\begin{pmatrix*}[r]
-\frac{1}{3}& \frac{7}{5}\\ \frac{7}{5}&-\frac{1}{3}\end{pmatrix*}
+\frac{3\pi i}{8}\begin{pmatrix*}[r]1 & 0\\0 & -1
\end{pmatrix*} 
 \right]\ ,
\endgroup
\cr &&\cr
\hat\CA^{(3)} &\sim &\hat k^3\frac{\left(2(-1)^J-1\right)^4}{J^8}\left[
-\frac{37i}{30}\begin{pmatrix*}[r]1 & -1\\-1&1\end{pmatrix*}-\frac{63\pi}{128}\begin{pmatrix*}[r]1 & 0\\0 & -1\end{pmatrix*}  \ 
+\frac{9\pi^2i}{64}\begin{pmatrix*}[r]1 & 0\\0 & 1\end{pmatrix*} \right]
\eqn{nvalCoupled}
\eeq
Unlike for the $L=J$ scattering amplitudes, here one sees the factors of $\pi$ appearing at leading order in the expansion, as observed by FMS.  
However, if one looks instead at the large-$J$ expansion of the phase shifts from  \eq{deltamJneq1}, \eq{deltapJneq1}, \eq{epsJneq1}, one obtains the asymptotic behavior
\begin{equation}
\begin{aligned}
&\delta_-^{(1)} \sim \hat k \left( 2(-1)^J-1\right)\frac{1} {2J^2}& \qquad & 
\delta_+^{(1)} \sim \hat k \left( 2(-1)^J-1\right)\frac{ 1}{2J^2}& \qquad & 
\epsilon^{(1)} \sim  -\hat k \left( 2(-1)^J-1\right)\frac{1}{2J^2}\\
&\delta_-^{(2)} \sim \hat k^2 \left( 2(-1)^J-1\right)^2\frac{3\pi} {8J^4}& \qquad & 
\delta_+^{(2)} \sim -\hat k^2 \left( 2(-1)^J-1\right)^2\frac{ 3\pi}{8J^4}& \qquad & 
\epsilon^{(2)} \sim  -\hat k^2 \left( 2(-1)^J-1\right)^2\frac{3\pi}{16J^5}\\
&\delta_-^{(3)} \sim \hat k^3 \left( 2(-1)^J-1\right)^3\frac{11} {4J^7}& \qquad & 
\delta_+^{(3)} \sim -\hat k^3 \left( 2(-1)^J-1\right)^3\frac{ 11}{4J^7}& \qquad & 
\epsilon^{(3)} \sim  \hat k^3 \left( 2(-1)^J-1\right)^3\frac{16}{15J^6}\\
&\delta_-^{(4)} \sim -\hat k^4 \left( 2(-1)^J-1\right)^4\frac{15\pi} {128J^8}& \qquad & 
\delta_+^{(4)} \sim \hat k^4 \left( 2(-1)^J-1\right)^4\frac{ 15\pi}{128J^8}& \qquad & 
\epsilon^{(4)} \sim  \hat k^4 \left( 2(-1)^J-1\right)^4\frac{75\pi}{64J^9}
\end{aligned}
\eqn{LneqJasympphase}
\end{equation}
One sees that despite the factors of $\pi$ appearing in the expansion of the amplitude in \eq{nvalCoupled}, the above phase shifts exhibit the same  pattern of having a simple factor of $\pi$ appear at every other order for large $J$  as they do for the $L=J$ phase shifts in \eq{LeqJasympphase}.     Note that overall these expressions  do not fall off as fast for large $J$, however.  That suggests that one  should find a significantly smaller radius of convergence in $\hat k$ for the same large $J$ than what I found for the $L=J$ channels, although the factors of $\pi$ are not implicated as being the source of  the  problem,   as had been surmised in Ref.~\cite{Fleming:1999ee}.  Indeed,   plots of the phase shifts for coupled channels   in Fig.~\ref{fig:LJcoupledasymp} and Fig.~\ref{fig:LJcoupledLepage} with $J=19$ and $J=20$ confirm a smaller radius of convergence, as well as several striking features:

\begin{itemize}
\item While the radius of convergence is   smaller  than seen in the $L=J$ channels, it is still much greater than $\hat k\sim 1$ for large $J$;
\item For the cumulative phase shifts $\delta_\pm$ in Fig.~\ref{fig:LJcoupledasymp}, in every case there is a  large correction to the phase shifts between leading and subleading orders (as observed by FMS \cite{Fleming:1999ee}), while the next three orders in the expansion converge fairly well, albeit only up to a  lower value of $\hat k$ than in the $L=J$ channels;
\item The Lepage plots for  $\delta^{(n)}_\pm$ in Fig.~\ref{fig:LJcoupledLepage} are identical and exhibit a pairwise clustering of corrections;
\item The mixing angles $\epsilon_J$  in Fig.~\ref{fig:LJcoupledasymp} exhibit a surprising pattern where the first two orders give similar results, which jumps at third order, and then changes little at fourth order.  This  nonuniform convergence manifests itself in the Lepage plots Fig.~\ref{fig:LJcoupledLepage} in the crossing of the   $\delta^{(2)}$ line with those for $\delta^{(3)}$  and $\delta^{(4)}$ at values of $\hat k$. 
\end{itemize}
It would be interesting if this analysis could be extended to higher orders to see if a clear  pattern can be recognized, but that is beyond the scope of this paper.

 \section{Renormalization: the \MJS~subtraction scheme}
 \label{sec:renormalization}
   
    While establishing how the perturbative pion expansion converges for large angular momentum is interesting, and establishes that there is nothing inherently pathological about performing a perturbative expansion with a $1/r^3$ potential,  the convergence properties of the expansion at low angular momentum are of more practical relevance for nuclear physics.  
Before one can examine the phase shifts for low values of $L$ one has to confront the question of renormalization, which is the topic of this section.  Here I show how to define a regularization and renormalization scheme sufficient to remove the poles in the amplitudes in $(\ell-L)$, where $L$ is an integer, from the first two orders where they appear.  This involves insertion of a counterterm at tree level and in one-loop diagrams. The calculations are somewhat technical, and a reader mainly interested in the discussion of convergence can skip to the next section.

The amplitudes I have computed for partial waves with angular momentum $L$ are divergent at orders $n\ge 2L$, the only exception being the ${}^3S_1$ channel, where the divergence first appears at $n=2$.  In a Lagrangian approach these divergences coincide with the existence of 4-nucleon contact interactions with projection operators onto the relevant angular momentum state (see Ref.~\cite{Fleming:1999ee} for a lucid discussion). The projection operators require $2L$ derivatives and so will enter the EFT expansion at order $n=2L$, with higher dimension contact interactions allowed with additional pairs of derivatives.  The existence of such operators implies that one should expect divergences in the OPE scattering amplitudes at that order, where the coefficients of contact terms can be chosen to renormalize the amplitude; these divergences will therefore appear at order $n=2L+2m$, where $m $ is any positive integer.  In the Lagrangian approach it is convenient to  regulate the theory using dimensional regularization, use the coefficients of the contact operators to absorb the $1/(d-4)$ poles in the amplitude (where $d$ is the space-time dimension), and then continue back to $d=4$. Care must be taken with how the partial wave projection operators are treated in arbitrary dimension, since the irreducible representations of $SO(3)$ cannot be analytically continued to representations of $SO(d-1)$ \cite{Fleming:1999ee}.  

In the Schr\"odinger approach in coordinate space that I have adopted, dimensional regularization is not practical.  Instead, the fact that one is able to derive amplitudes as rational functions of angular momentum $L$  suggests that one develop a more exotic subtraction scheme where one analytically continues the amplitudes to arbitrary angular momentum  $\ell$, subtract poles in $(\ell-L)$, and then take the limit $\ell\to L$.   The poles can be absorbed into a singular $\ell$-dependent potential to the Schr\"odinger equation, adjusting its coefficient to render all the amplitudes finite.  

%\subsection{Contact interactions for renormalizing the $L=J$ amplitudes}

\begin{figure}[t]
\includegraphics[width=15cm]{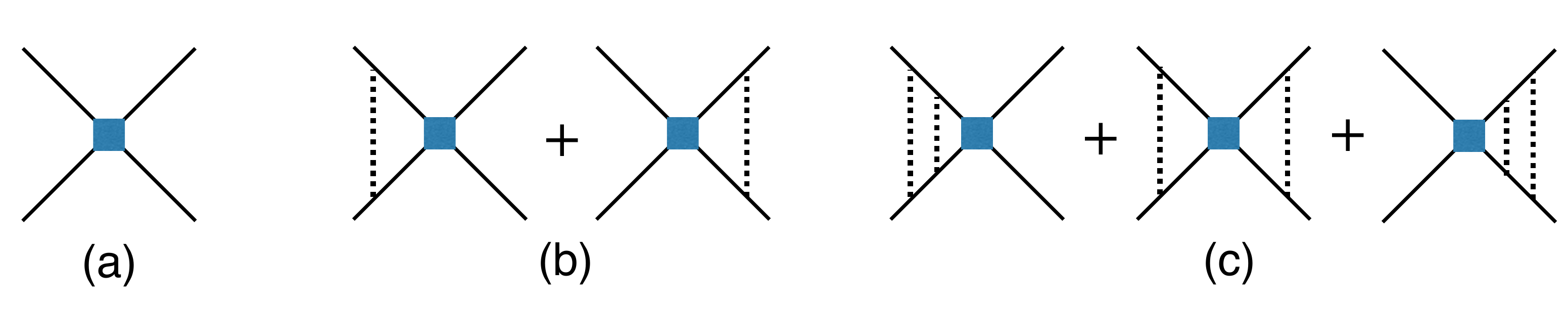}
\caption{ {\it Perturbative insertions of the $\CC_L$ counterterm from \eq{SeqLeqJren} involving (a) no pions, entering at order $n=2L$, (b) one pion at order $n=2L+1$, or (c) two pions at order $n=2L+2$. In addition one will encounter a new contact interaction with an additional pair of derivatives at order $n=2L+2$.  }}
\label{fig:cts}
\end{figure}

 At integer $L$, contact terms with   derivatives in the Lagrangian correspond to potentials in coordinate space which are spatial derivatives acting on Dirac $\delta$-functions.  One should be able to renormalize this theory by adding such potentials to the Schr\"odinger equation to cancel the infinities observed in the OPE amplitudes in \eq{ampsLeqJ} and \eq{ampsLJCoupled}.  For example, for the ${}^3P_1$ channel, the leading contact interaction in an EFT would look like
 \beq
%\frac{C}{M\Lambda_{NN}^3}
 \left(\epsilon_{ijp} N^T \sigma_2 \sigma_i\dvec \partial_j N\right)
 \left(\epsilon_{mnp} N^T \sigma_2 \sigma_m\dvec 
 \partial_nN\right)^\dagger
 \eeq
 which corresponds to a radial potential  $\hat V_L$ in this channel with 
 \beq
 \expect{r}{\hat V_1}{r'} \propto \frac{\partial_r\delta(r)}{r^2}\frac{ \partial_{r'}\delta(r')}{{r'}^2}\ ,\qquad\Longrightarrow\qquad \expect{\psi}{\hat V_1}{\chi} \propto \psi^{\prime *}(0) \chi'(0)\ ,
 \eeq
 where $\int dr\, r^2\, \ket{r}\bra{r}$ is the unit operator.  More generally, the leading counterterm in each channel with angular momentum $L$ would be of the form 
   \beq
 \expect{r}{\hat V_L}{r'} \propto \frac{\partial^L_r \delta(r)}{r^2}\frac{\partial^L_{r'}\delta(r')}{{r'}^2}\ ,
 \qquad\Longrightarrow\qquad \expect{\psi}{\hat V_L}{\chi} \propto \psi^{(L)*}(0) \chi^{(L)}(0)\ ,
 \eeq
where the $(L)$ superscripts denote the $L^{th}$ radial derivative.  By dimensional analysis the coefficient of this potential   has mass dimension $-2(L+1)$, which I write as $\CC_L/M\Lambda_{NN}^{2L+1}$ with dimensionless coupling $\CC_L$.   For coupled channels there are three types of counterterms to consider, which can take the form
 \beq
 \expect{r}{\hat V^\pm_J}{r'}& \propto&  \frac{\partial^{J\pm 1}_r \delta(r)}{r^2}\frac{\partial^{J\pm 1}_{r'}\delta(r')}{{r'}^2} \ ,\cr
\expect{r}{\hat V^\epsilon_J}{r'}& \propto& \left( \frac{\partial^{J-1}_r \delta(r)}{r^2}\frac{\partial^{J+1}_{r'}\delta(r')}{{r'}^2}+ \frac{\partial^{J+1}_r \delta(r)}{r^2}\frac{\partial^{J-1}_{r'}\delta(r')}{{r'}^2}\right)\ .
 \eeq
%and the coefficients of these potentials have coefficients $\CC^\pm_J/M\Lambda_{NN}^{2(J\pm1)+1}$  $\CC^\epsilon_J/M\Lambda_{NN}^{2 J+1}$.    
I incorporate the counterterm into the Schr\"odinger equation \eq{SeqLeqJ} for the $L=J$ channel as
  \beq
\left(\partial_\rho^2+ \frac{2}{\rho} \partial_\rho+1- \frac{L(L+1)}{\rho^2} \right)u(\rho) = \frac{\hat k}{\rho^3} \,u(\rho) + \CC_L  \hat k^{2L+1}  \, \left[\frac{ 1}{\rho^2}\,\left(\partial_\rho^L \right)^\dagger\delta(\rho)\right]\left[\partial_{\rho'}^L u(\rho')\right]_{\rho'=0}\ , 
\eqn{contactpotential}
\eeq
 where  the dagger on $\partial_\rho^L$ simply  indicates that there are no signs when integrating by parts to remove the derivatives from the $\delta$-function, i.e. that the signs have been absorbed into the definition of $\CC_L$.  With this additional contribution to the potential, \eq{urecur} and \eq{Arecur} have to be modified appropriately to include contributions to $u^{(n)}$ and $\hat \CA^{(n)}$, modifications which inform one how to compute the diagrams in Fig.~\ref{fig:cts}.

 In order to regulate the theory I now analytically continue the problem to non-integer angular momentum $\ell$.  Following the conventional procedure used in dimensional regularization, I leave unchanged the interaction strength $\CC_L/M\Lambda_{NN}^{2L+1}$, which means that one must introduce a renormalization scale $\mu$ to absorb the operator dimension change when one replaces $\partial^L\to \partial^\ell$.  When I do this, the modified Schr\"odinger equation which I need to solve is
   \beq
\left(\partial_\rho^2+ \frac{2}{\rho} \partial_\rho+1- \frac{\ell(\ell+1)}{\rho^2} \right)u(\rho) &=& -2\left(1+2(-1)^L\right)\frac{\hat k}{\rho^3} \,u(\rho) + \CC_L  \hat k^{2L+1}\left(\frac{k}{\mu}\right)^{2(\ell-L)} \,\left[\frac{ 1}{\rho^2}\,\left(\partial_\rho^\ell \right)^\dagger\delta(\rho)\right]\left[\partial_{\rho'}^\ell u(\rho')\right]_{\rho'=0}\ .  
 \eqn{SeqLeqJren}\eeq
 The strategy then is to expand the counterterm $\CC_L$ as
 \beq
 \CC_L =  \xi_{L,0} + \frac{\xi_{L,1}}{\ell-L} +  \frac{\xi_{L,2}}{(\ell-L)^2} +\ldots
 \eeq
 where the values of the $\xi_{L,m}$ for $m\ge 1$ are fixed by the requirement that the scattering amplitudes at each order in $\hat k$ are finite at $\ell\to L$. I call this the minimal angular subtraction scheme, \MJS.  A related subtraction scheme calls for subtracting all extraneous constants generated in the $\ell\to L$ limit, which simply shifts the leading term, $\xi_{L,0}\to \tilde \xi_{L,0}$, and I call this the \MJSb~scheme, which is what I will use here.   The finite contribution due to $\tilde \xi_{L,0}$ at a given value of $\mu$ would be fixed to data, if this were a theory of the real world; as it is not, and one is only interested in convergence properties of the perturbative expansion, I will simply set all the $\tilde \xi_{L,0} = -1$ at $\mu=\Lambda_{NN}$ when plotting renormalized phase shifts.

 To make sense of \eq{SeqLeqJren} one needs to supply a definition for what one means by a non-integer valued derivative, which I take to be
\beq
\partial_x^\ell x^k = \frac{\Gamma(k+1)}{\Gamma(k-\ell+1)} x^{k-\ell}\ ,
\eqn{lderiv}\eeq
a definition which agrees with conventional differentiation for integer $\ell$.
Fractional derivatives acting on spherical Bessel functions are then defined by applying them to  the Bessel functions' Taylor series, using the definition \eq{lderiv}.  Since such a Taylor expansion takes the form
\beq
j_\ell(x) = 2^\ell x^\ell\sum_{k=0}^\infty (-1)^k\frac{\Gamma(k+\ell+1)}{\Gamma(k+1)\Gamma(2k+2\ell+2)} \, x^{2k}\ ,
\eeq
it follows that   for non-negative integer $m$ one has
\beq
\delta(x)\, \partial_x^{\ell+m}j_\ell(x)& \equiv& \delta(x) \lim_{x\to 0} 2^\ell  \sum_{k=0}^\infty (-1)^k\frac{\Gamma(k+\ell+1)}{\Gamma(k+1)\Gamma(2k+2\ell+2)} \,\frac{\Gamma(2k+\ell+1)}{\Gamma(2k-m+1)}\, x^{2k-m}\cr&&\cr
&=&
 \delta(x)\times  \begin{cases}
2^\ell (-1)^{m/2}\frac{\Gamma(m/2+\ell+1) \Gamma(m+\ell+1)}{\Gamma(m/2+1)\Gamma(m+2\ell+2)} & m\text{ even}\\
0 & m \text{ odd}
\end{cases}\cr &&\cr
&\equiv& \Delta_{\ell,m}\,\delta(x)\ ,
\eqn{Deltadef}\eeq
where terms in the sum with $2k<m$ vanished before taking the $x\to 0$ limit due to the divergence of $\Gamma(2k-m+1)$ in the denominator. The values I will be using in this section are 
\beq
\Delta_{0,0} = 1\ ,\qquad\Delta_{1,0} = \frac{1}{3}\ ,\qquad \Delta_{2,0} = \frac{2}{15}\ . 
\eeq
Note that \eq{Deltadef} only holds for integer $m$, which will limit the applicability of this simple renormalization strategy.  
  For non-integer $m$ one obtains $x$ to a non-integer power times $\delta(x)$ which  either vanishes or is infinite.  For integrals where one must take a limit with respect to order of the Bessel function, as in the example \eq{limex}, there will be graphs which involve fractional derivatives  acting on Bessel functions of arbitrary order and the above machinery is insufficient for regulating the infinities in such cases.    In practice this will allow us to compute insertions of the contact interactions in the graphs (a) and (b) in Fig.~\ref{fig:cts}, but not the graphs (c). This is sufficient, however, to extend the applicability of the computed amplitudes by two orders in the perturbative expansion.

  \subsection{Renormalization of the   ${}^3P_1$ and ${}^3P_0$ amplitudes at $n=2,3$}

I start by renormalizing the ${}^3P_1$ amplitude, for which  one sees from \eq{ampsLeqJ}  that  poles in the amplitude at $\ell=1$  appear at orders $n=2$ and higher; I will include the graph in Fig.~\ref{fig:cts}(a) to renormalize $\CA^{(2)}$, and the graphs in Fig.~\ref{fig:cts}(b) to render $\CA^{(3)}$ finite.   

 I start with Fig.~\ref{fig:cts}(a), the tree level insertion of $\CC_L$ which contributes at order $n=2$ in the $L=1$ partial wave, and use the notation $\partial_x^p j_\ell(x) = j_\ell^{(p)}(x)$.    This graph corresponds to the integral, with $\ell\sim 1$,
\beq
 \delta \hat\CA_{(6a)} &=& 
 - \CC_p
\hat k^{2} \left(\frac{k}{\mu} \right)^{2(\ell-1)}\, \int_0^\infty  d\rho\,\int_0^\infty  d\rho'\,  j^{(\ell)}_\ell(\rho)  j^{(\ell)}_\ell(\rho') \,  \delta(\rho)\, \delta(\rho')
\cr &=&
  -  \CC_p \hat k^{2} \left(\frac{k}{\mu} \right)^{2(\ell-1)}
  \Delta_{\ell,0}^2  \ ,
 %  \frac{ \pi  \Gamma \left( \ell+1\right)^2}{4^{ \ell+1}\Gamma \left(\ell+\frac{3}{2}\right)^2}\ ,
\eqn{tree1}\eeq
a contribution at order $n= 2$.  One can therefore write $\CC_p = \xi_{p,0} + \xi_{p,1}/(\ell-1)$ and choose  $\xi_{p,1}$ to cancel the $1/(\ell-1)$ pole in $\CA^{(2)}$ (the subscript ``$p$" indicating $p$-wave).  The residue  of that pole at $\ell=1$ is easily read off from \eq{ampsLeqJ}:
\beq
\hat\CA^{(2)} = \hat k^2\left[ -\frac{1}{9(\ell-1)} + O\left((\ell-1)^0\right) \right]\ ,
\eeq
and therefore since $ \Delta_{1,0}^2=1/9$, the renormalization condition for the $L=J=1$ channel is:
\beq
- \xi_{p,1} \times \lim_{\ell\to 1}\left(  \left(\frac{k}{\mu} \right)^{2(\ell-1)}
    \Delta_{\ell,0}^2\right) = \frac{1}{9} \qquad \Longrightarrow\qquad \xi_{1,1}= -1
\ ,\
\eqn{gL0renorm}\eeq
where I made use of \eq{Deltadef}. With $\xi_{p,1}$ fixed,  the renormalized amplitude at $n=2$ is then given by  
\begin{equation}
\begin{aligned}
\hat \CA_\text{ren}^{(2)}  &= \lim_{\ell\to 1} \left(\hat\CA^{(2)} + \delta\hat\CA_{(6a)}\right) 
 &\cr
 &=
  \lim_{\ell\to 1} \left(\hat\CA^{(2)} - \hat k^2\left(\frac{k}{\mu} \right)^{2(\ell-1)}\left[\xi_{p,0}-\frac{1}{\ell-1} \right]
  \Delta_{\ell,0}^2\right)& \cr
    &=
\hat k^2\left(\frac{1}{9} \left[\ln \left(\frac{ k^2}{\mu^2 }\right) - \xi_{p,0} \right] + \frac{2\ln 2}{9}-\frac{i \pi }{10}+\frac{1}{4}\right)&\hfill (\mjs)\ \cr
     &\equiv 
   \hat k^2\left(\frac{1}{9} \left[\ln \left(\frac{ k^2}{\mu^2 }\right) - \tilde \xi_{p,0} \right] -\frac{i \pi }{10}\right)\ &\hfill (\mjsb).
\eqn{C10}
\end{aligned}
\end{equation}
 One sees from the above answer that $\mu$ independence of physical amplitudes implies logarithmic running for $ \tilde \xi_{p,0} $.     Note that the expressions for $\hat\CA^{(2)}$  in \eq{ampsLeqJ}  are unchanged for any of the angular momenta other than $L=2$, as those channels are all finite at order $n=2$ and there are no counterterms in those partial waves at $O(\hat k^2)$.

 %%%%%%%%%%%%%%%%

At the next order, the expression for $\CA^{(3)}$ in \eq{ampsLeqJ}   also has a singularity at $\ell=1$, and is finite for $\ell\ge 2$.  This singularity is in the imaginary part, and is expected from unitarity since $\CA^{(2)}$ had a pole at $\ell=1$.  Thus one should find a finite result in the $\ell\to 1$ limit when one adds the 1-loop contributions from the graphs in Fig.~\ref{fig:cts}b to $\CA^{(3)}$,  without any further subtractions required.      These graphs yield
\beq
 \delta \hat\CA_{(6b)} &=& 
 -4i \left(1+2(-1)^L\right) \CC_p\hat k^3\left(\frac{k}{\mu}\right)^{2(\ell-1)} \left[\partial^\ell_\rho  j_\ell(\rho)\right]_{\rho=0}\left[\partial^\ell_\rho \int \frac{d\rho' }{\rho'} g_\ell(\rho,\rho')  j_\ell(\rho')\right]_{\rho=0}\cr
&=&
-4i \left(1+2(-1)^\ell\right) \CC_p\hat k^3\left(\frac{k}{\mu}\right)^{2(\ell-1)} \Delta_{\ell,0}^2\, \alpha_{\ell,\ell}   \cr
 &=&
 -2i \frac{\left(1+2(-1)^\ell\right)}{\ell(\ell+1)} \hat k^3\left(\xi_{p,0}-\frac{1}{\ell-1} \right)\left(\frac{k}{\mu}\right)^{2(\ell-1)}  \Delta_{\ell,0}^2\ .
\eqn{b63p1}\eeq
where $\alpha_{\ell_1,\ell_2}$ is given in \eq{abc}.  
 
I now add this to  the unrenormalized amplitude $\CA^{(3)}$  in \eq{ampsLeqJ} or  \eq{ampsLeqJ} in the limit $\ell\to1$, and obtain    the renormalized amplitude
\begin{equation}
\begin{aligned}
\hat \CA_\text{ren}^{(3)} \Biggr\vert_{L=1} &=\lim_{\ell\to 1} \left(\hat\CA^{(3)} + \delta\hat\CA_{6(b)}\right)& \cr
 &=
\hat k^3\,\left[-\frac{2341
   \pi }{21000}+i\left( \frac{   \xi_{p,0}}{9}-\frac{1}{9}  \ln  \frac{k^2}{\mu^2 } +\frac{ \pi ^2}{100}-\frac{3 }{16}-\frac{2}{9}  \ln 2\right)\right]&\hfill(\mjs)\ 
   \cr &=
\hat k^3\left[ -\frac{2341 \pi }{21000} + i\left( \frac{1}{9} \left[\tilde \xi_{p,0}-\ln \frac{k^2}{\mu^2 } \right]+\frac{ \pi
   ^2}{100}+\frac{1}{16}\right)\right]&\hfill(\mjsb)
   \end{aligned}
   \end{equation}

The renormalization of the   ${}^3P_0$ amplitude at $n=2,3$ follows trivially from the above analysis since the OPE potential for the ${}^3P_0$ channel equals $(-2)$ times the OPE potential for  ${}^3P_1$, and they both have $L(L+1) = 2$.  Therefore one need only multiply the renormalized amplitude $\CA^{(n)}$ for  ${}^3P_1$ scattering by $(-2)^{n+1}$, and replace the finite $\tilde \xi_{1,0}$ coupling   for ${}^3P_1$ scattering by an independent one  for the ${}^3P_0$ channel.

  \subsection{Renormalization of the   ${}^3D_2$ amplitude at $n=4,5$}

The ${}^3D_2$ amplitude exhibits a pole at order $n=4$, and is renormalized by a contact interaction in much the same way as the ${}^3P_1$ amplitude was renormalized at order $n=2$. The residue of the $\ell=2$ pole in $\hat\CA^{(4)}$ in   \eq{ampsLeqJ} is readily found to equal $3/100$, and so repeating the 
 procedure used in the ${}^3P_1$ channel yields the counterterm
\beq
\xi_{d,1} = \frac{3/100}{\Delta_{2,0}^2}=\frac{27}{16}
\ ,
\eqn{gL0renormb}
\eeq
 and the renormalized amplitude for the ${}^3D_2$ channel
 \begin{equation}
\begin{aligned}
\hat \CA_\text{ren}^{(4)}  &= \lim_{\ell\to 2}  \left(\hat\CA^{(4)} + \delta\hat\CA_{(6a)}\right) 
 &\cr
 &=
  \lim_{\ell\to 2} \left( \hat\CA^{(4)} - \hat k^4\left(\frac{k}{\mu} \right)^{2(\ell-2)}\left(\xi_{d,0}+\frac{27/16}{\ell-2} \right) 
\Delta_{\ell,0}^2\right)& \cr
    &=  \hat k^4\left(-\frac{4 \xi_{d,0}}{225}-\frac{3}{100} \ln \left(\frac{k^2}{\mu^2}\right)-\frac{9 \pi
   ^2}{4900}+ i \pi \frac{10127}{343000}-\frac{199}{2000}-\frac{3 \ln 2}{50}\right)&\hfill (\mjs)\ \cr
     &\equiv 
   \hat k^4\left(-\frac{4 \tilde \xi_{d,0}}{225}-\frac{3}{100} \ln \left(\frac{k^2}{\mu^2}\right)+i \pi\frac{10127 
   }{343000}\right)\ &\hfill (\mjsb).
\eqn{C20}
\end{aligned}
\end{equation}
 
 One can extend the calculation to one higher order by including the graphs in Fig.~\ref{fig:cts}(b), with the calculation being very similar to the one for the ${}^3P_1$ channel in \eq{b63p1}, the result being:
 \beq
 \delta \hat\CA_{(6b)} &=& 
 -2i \frac{\left(1+2(-1)^\ell\right)}{\ell(\ell+1)} \hat k^5\left(\xi_{d,0}+\frac{27/16}{\ell-2} \right)\left(\frac{k}{\mu}\right)^{2(\ell-2)}  \Delta_{\ell,0}^2\ .
\eeq
Adding this to $\hat \CA^{(5)}$ an taking the $\ell\to 2$ limit gives
 \begin{equation}
\begin{aligned}
\hat \CA_\text{ren}^{(5)} \Biggr\vert_{L=2} &=\lim_{\ell\to 2} \left(\hat\CA^{(5)} + \delta\hat\CA_{6(b)}\right)& \cr
 &=
\hat k^5\left[ -\frac{4 i \tilde \xi_{d,0}}{225}-\frac{3}{100} i \ln  \frac{k^2}{\mu^2 } -\frac{9 \pi
   ^3}{171500}+\frac{20703 i \pi ^2}{6002500}-\frac{817949431 \pi }{27731550000}+\frac{i}{32} \right]&\hfill(\mjsb)
   \end{aligned}
   \end{equation}

  \subsection{Renormalization of the   $J=2$ coupled channel amplitudes to $n=2,3$}
  
  I now turn to renormalization of the coupled $L=J\pm 1$ channels, starting with $J=2$.  In general \eq{Seqmix} and \eq{contactpotential} must be modified for the coupled channels to allow for mixing counterterms, namely
   \beq
\left[\left(\partial_\rho^2+ \frac{2}{\rho} \partial_\rho+1\right) - \frac{\ell_a(\ell_a+1)}{\rho^2} \right]u_{\ell_a}(\rho)\hspace{-1.5in}&&  \cr &=& \sum_b\left(  \left(1-2(-1)^j\right) \,\hat k  \frac{v_{ab}}{\rho^3}\,u_{\ell_b}(\rho)  +   \CC^{(j)}_{L_a,L_b}  \hat k^{L_a+L_b+1}  \,\left[\frac{ 1}{\rho^2}\,\left(\partial_\rho^{\ell_b} \right)^\dagger\delta(\rho)\right]\left[\partial_{\rho'}^{\ell_a}u_{\ell_a}(\rho')\right]_{\rho'=0}\right)\ , 
\eqn{contactpotentialmix}
\eeq
Because of the factor of $1/\hat k$ in the definition of the amplitude in \eq{apert}, the contact term proportional to $\CC^{(j)}_{L_a,L_b}$ enters the expansion of the amplitude at order $\hat k^{L_a+L_b}$.    

Expanding about $j=2$ I find the leading amplitude  at $O(\hat k^2)$ to be 
  \beq
 \hat\CA_{J=2}^{(2)} &=&
\frac{\hat  k^2/45}{j-2}\,\begin{pmatrix}
 1 & 0 \\
0& 0 \\
\end{pmatrix}
+ O(1)\ ,
\eqn{P32a2pole}\eeq
and so I introduce the $\CC^{(2)}_{pp} = \sum_n \xi^{(2)}_{pp,n} (j-2)^{-n}$ coupling and fix the simple pole contribution to be
\beq
\xi^{(2)}_{pp,1}= \frac{1/45}{\Delta_{1,0}^2} = \frac{1}{5}\ ,
\eeq
and I find the \MJSb~renormalized amplitude
\beq
\hat \CA_{J=2,\text{ren}^{(2)} }= \hat k^2 \begin{pmatrix}
 -\frac{\tilde \xi^{(2)}_{pp,0}}{9}-\frac{1}{45} \ln \frac{k^2}{\mu^2 } +\frac{13 i \pi }{1050} &\ \ 
   \frac{4 }{225\sqrt{6}}-\frac{19 i \pi }{1575 \sqrt{6}}\  \ \\
 \frac{4}{225\sqrt{6}}-\frac{19 i \pi }{1575 \sqrt{6}} & -\frac{2}{675}-\frac{i \pi }{4725} \\
   \end{pmatrix}
\eeq

Proceeding to the next order I compute the diagrams of Fig.~\ref{fig:cts}(b):
\beq
 \delta \hat\CA_{(6b)} &=& 
 -i  \CC^{(2)}_{pp}\left(2(-1)^j-1\right)  \hat k^3\left(\frac{k}{\mu}\right)^{2(j-2)} 
 \Biggl[
  \Delta_{0,j-1}^2 \begin{pmatrix} 2v_{11} \alpha_{j-1,j-1} & v_{12} \alpha_{j-1,j+1} \\ v_{21} \alpha_{j+1,j-1} & 0\end{pmatrix}
  \Biggr]\ ,
 \eqn{J26b}
 \eeq
Adding this to $\hat \CA^{(3)}$ and taking the $j\to 2$ limit yields the \MJSb~renormalized amplitude
 \beq
  \left[ \hat\CA_{J=2,\text{ren}}^{(3)}\right]_{11} &=& \hat k^3  \left(
   -\frac{i \tilde \xi^{(2)}_{pp,0}}{45}-\frac{1}{225} i \ln \frac{k^2}{\mu ^2}+\frac{53 i \pi
   ^2}{14700}-\frac{50779 \pi }{9261000}-\frac{13 i}{18000} 
    \right)\ ,\cr
   &&\cr
     \left[ \hat\CA_{J=2,\text{ren}}^{(3)}\right]_{12} &=& 
     \hat k^3  \left(
    \frac{i  \tilde \xi^{(2)}_{pp,0}}{45
   \sqrt{6}}+\frac{i \ln \frac{k^2}{\mu ^2} }{225 \sqrt{6}}-\frac{i \pi ^2}{3150
   \sqrt{6}}+\frac{6463 \pi }{1984500 \sqrt{6}}+\frac{71 i}{27000 \sqrt{6}} 
    \right)\ ,\cr &&\cr
     \left[ \hat\CA_{J=2,\text{ren}}^{(3)}\right]_{22} &=& \hat k^3  \left(
     -\frac{103 i}{81000}-\frac{78821 \pi }{275051700}+\frac{i \pi ^2}{39690} 
    \right)
\ .
   \eeq

  %%%%%%%%%%%%%%%%%%
 
 \subsection{Renormalization of the   $J=1$ coupled channel amplitudes to $n=2,3$}

\begin{figure}[t]
\includegraphics[width=8cm]{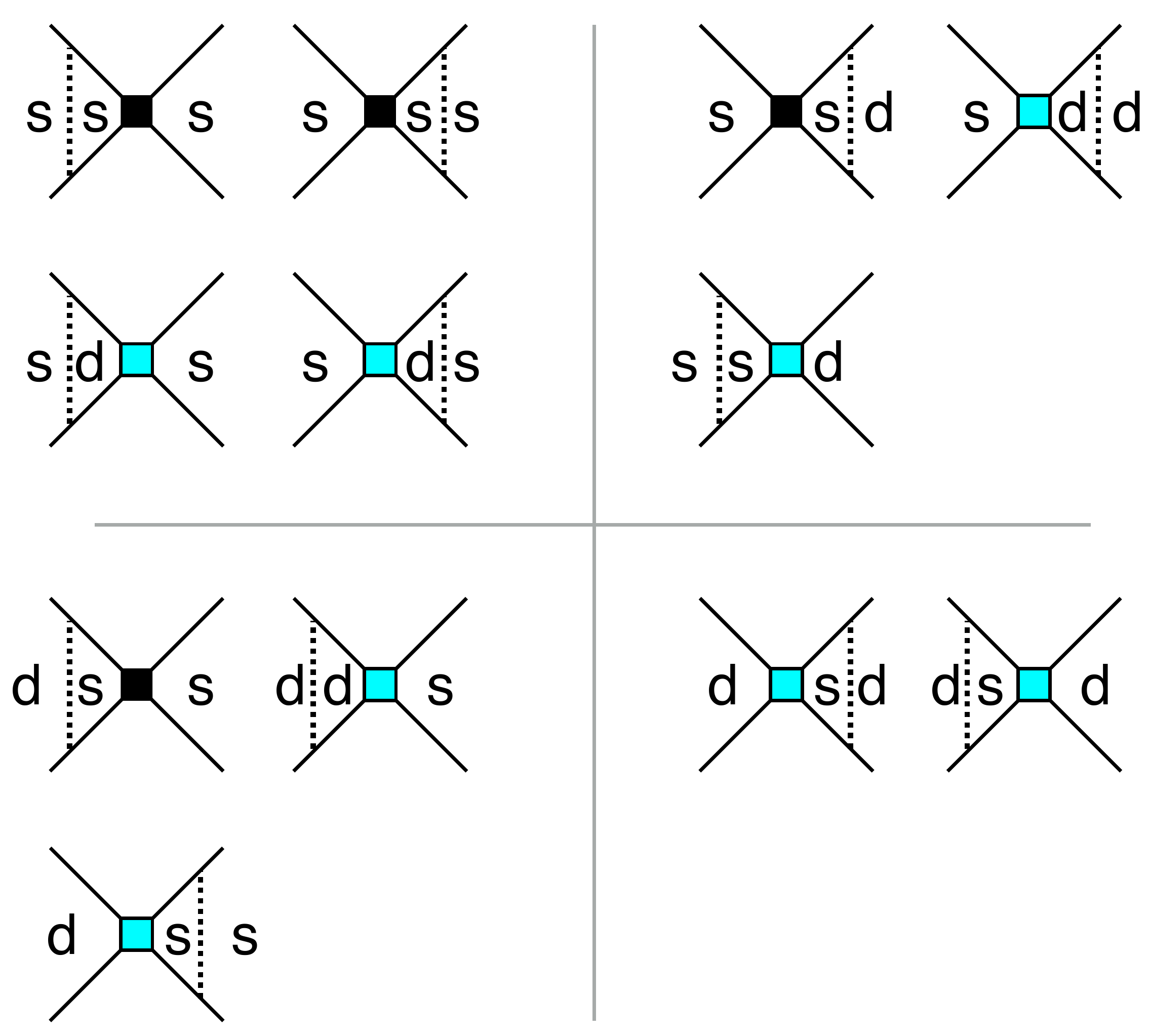}
\caption{ {\it One-loop graphs renormalizing  ${}^3S_1-{}^3D_1$ coupled channel scattering at  $O(\hat k^3)$.  The dark and light square vertices denote the $\CC^{(1)}_{ss}$ and $\CC^{(1)}_{sd}$ contact interactions respectively, for which the pole contributions were fixed at $O(\hat k^2)$ in \eq{J1poles}. The arrangement of the graphs corresponds to the entries in the $2\times 2$ scattering amplitude for the coupled $s-d$ channels. These contributions are computed in \eq{J16b}, and when added to $\hat A^{(3)}$ render it finite at $J=1$.  }}
\label{fig:J1A3}
\end{figure}

Finally I consider   the problem of renormalizing the coupled   ${}^3S_1-{}^3D_1$  partial waves, for which the amplitudes I computed are given in \eq{ampsLJCoupled}.  Expanding those amplitudes about $j=1$ I find them to be finite for $n=0,1$, while exhibiting $(j-1)$ poles at higher orders.  Specifically, one sees that as $j\to 1$ one has
\beq
 \hat\CA_{J=1}^{(2)} &=&
\frac{ 1}{j-1}\left(3 \hat k^2\right)\,\begin{pmatrix}
  2 & -\frac{\sqrt{2}}{5} \\
 -\frac{\sqrt{2}}{5} & 0 \\
\end{pmatrix}
+ O(1)\ ,
\eqn{J1a2pole}\eeq
\beq
 \hat\CA_{J=1}^{(3)} &=& \frac{1}{j-1}\left( -\frac{3i\hat k^3}{5}\right)
 \begin{pmatrix}
 22 & -\frac{13}{\sqrt{2}} \\
 -\frac{13}{\sqrt{2}} & 2 \\
  \end{pmatrix}+ O(1)\ .
\eeq
This indicates that one requires both   $O(\hat k^2)$ $s-s$ and $s-d$  counterterms (proportional to $\CC^{(1)}_{ss}$, and $\CC^{(1)}_{sd}$ respectively)  to renormalize the $n=2$ amplitude, while those same counterterms at one-loop will eliminate the divergence in the imaginary part of the    $n=3$ amplitude.   That one does not need a lower dimension $O(\hat k^0)$  $s-s$ contact interaction  to renormalize the amplitude at $n=0$  appears to arise from the fact that the OPE tensor force does not  have an $s-s$ matrix element at tree level for $J=1$.

For this coupled channel, the analogue of \eq{tree1}, expanded about $j\sim 1$ is
\beq
 \delta \hat\CA_{(6a)} &=& 
 -
\hat k^{2} \left(\frac{k}{\mu} \right)^{2(j-1)}\, \int_0^\infty  d\rho\,\int_0^\infty  d\rho'\,   \delta(\rho)\, \delta(\rho')\,\left[ \CC^{(1)}_{ss}
 j^{(j-1)}_{j-1}(\rho)  j^{(j-1)}_{j-1}(\rho')  \begin{pmatrix}1 & 0\\ 0 & 0\end{pmatrix}+\CC^{(1)}_{sd} j^{(j-1)}_{j-1}(\rho)  j^{(j+1)}_{j+1}(\rho') \begin{pmatrix}0 & 1\\ 1 & 0\end{pmatrix}\right]
\cr &=&
  -\hat k^{2} \left(\frac{k}{\mu} \right)^{2(j-1)}\left[ \CC^{(1)}_{ss}
 \Delta_{j-1,0}\Delta_{j-1,0}\times\begin{pmatrix}1 & 0\\ 0 & 0\end{pmatrix}+\CC^{(1)}_{sd}
 \Delta_{j-1,0}\Delta_{j+1,0}\times\begin{pmatrix}0 & 1\\ 1 & 0\end{pmatrix}\right]\cr &&
   \cr
   &=&
    -\hat k^{2} \left(\frac{k}{\mu} \right)^{2(j-1)}\left[  \CC^{(1)}_{ss}
 \begin{pmatrix}1 & 0\\ 0 & 0\end{pmatrix}+\CC^{(1)}_{sd}
\frac{2}{15}\begin{pmatrix}0 & 1\\ 1 & 0\end{pmatrix}\right]
  \ .
\eqn{treesd}\eeq
These terms must absorb the $1/(j-1)$ poles in \eq{J1a2pole}, 
so I expand 
\beq
\CC^{(1)}_{ss}=  \xi^{(1)}_{ss,0} + \frac{\xi^{(1)}_{ss,1}  }{j-1}\ , \qquad 
 \CC^{(1)}_{sd}=  \xi^{(1)}_{sd,0} + \frac{  \xi^{(1)}_{sd,0}   }{j-1}\ ,  
 \eeq
 and set
 \beq
   \xi^{(1)}_{ss,1}   = 6\ ,\qquad   \xi^{(1)}_{sd,1}  = -\frac{9}{\sqrt{2}}\ .
 \eqn{J1poles}\eeq
 Upon adding the two contributions to the $n=2$ amplitude  and taking the $j\to 1$ limit I find  the renormalized amplitude
 \begin{equation}
\begin{aligned}
  \hat\CA_{J=1,\text{ren}}^{(2)}  &= 
 \hat k^{2} 
\begin{pmatrix}
 -\xi^{(1)}_{ss,0}-6\ln  \frac{k^2}{\mu^2 } -12 \ln 2+\frac{27 i \pi }{5}+\frac{31}{2} & \frac{1}{75} \left(-10
    \xi^{(1)}_{sd,0}+45 \sqrt{2} \ln \frac{4 k^2}{\mu^2 } -183 \sqrt{2}\right)-\frac{81 i \pi }{35 \sqrt{2}}
   \\
 \frac{1}{75} \left(-10   \xi^{(1)}_{sd,0}+45 \sqrt{2} \ln  \frac{4 k^2}{\mu^2 } -183 \sqrt{2}\right)-\frac{81 i
   \pi }{35 \sqrt{2}} & 2-\frac{27 i \pi }{70} \\ \end{pmatrix}
 &\hfill  (\mjs)\ \cr &&\cr
  &=\hat k^{2}\
  \begin{pmatrix}
 -\tilde \xi^{(1)}_{ss,0}  -6   \ln  \frac{k^2}{\mu^2 } +\frac{27}{5} i \pi    & -\frac{2
    \tilde  \xi^{(1)}_{sd,0} }{15}+\frac{3}{5} \sqrt{2}   \ln  \frac{k^2}{\mu^2
   } -\frac{81 i \pi   }{35 \sqrt{2}} \\
 -\frac{2  \tilde  \xi^{(1)}_{sd,0}  }{15}+\frac{3}{5} \sqrt{2}   \ln  \frac{k^2}{\mu^2
   } -\frac{81 i \pi   }{35 \sqrt{2}} & 2  -\frac{27}{70} i \pi    \\
   \end{pmatrix}
&\hfill  (\mjsb)
 \end{aligned}
 \end{equation}

 %%%%%%%%%%%%%%%%%%%
\begin{figure}[t]
\includegraphics[width=14cm]{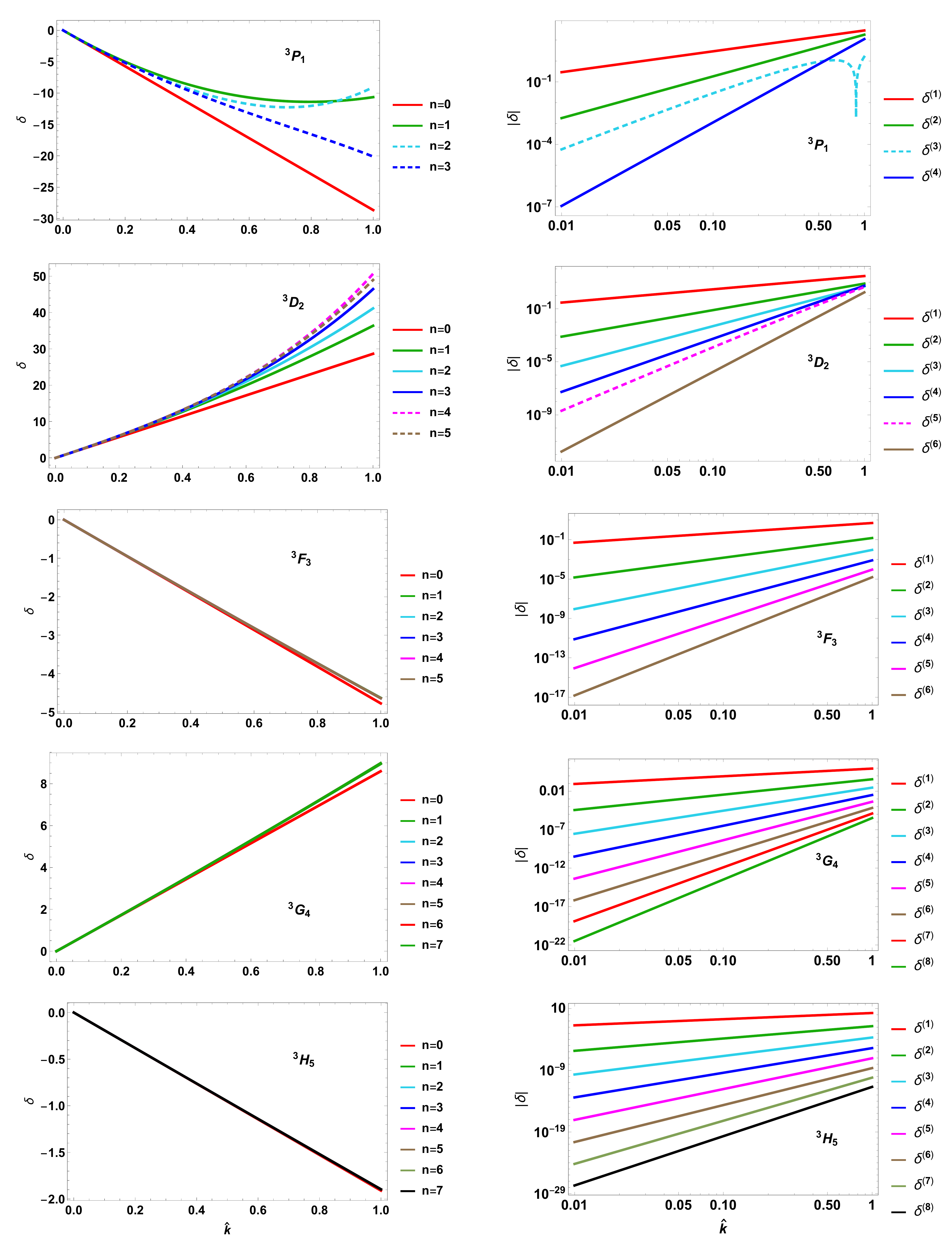}
\caption{ {\it Left column: the cumulative $L=J$ spin-triplet phase shifts in degrees computed from the perturbative expansion of the amplitudes computed to $O(\hat k^n)$, where  $\hat k$ is defined in \eq{khatdef}.  Right column: for the perturbative contributions to the phase shifts (in radians).     Dashed lines indicate quantities dependent on finite counterterms, each of which was set to $-1$ for this plot at renormalization scale $\mu = \Lambda_{NN}$. }}
\label{fig:PhaseShifts1}
\end{figure}
%%%%%%%%%%%%%%%%%%%%%%%%%%%%%%%

 At order $n=3$ one must evaluate graphs of type 6(b) with the same $ss$ and $sd$ contact interactions.  The contributions that renormalize $\hat A^{(3)}$ are  shown in Fig.~\ref{fig:J1A3}
 with the result  
\beq
 \delta \hat\CA_{(6b)} &=& 
 -i \left(2(-1)^j-1\right)  \hat k^3\left(\frac{k}{\mu}\right)^{2(j-1)} 
 \Biggl[
   \CC^{(1)}_{ss}\Delta_{0,j-1}^2 \begin{pmatrix} 2v_{11} \alpha_{j-1,j-1} & v_{12} \alpha_{j-1,j+1} \\ v_{21} \alpha_{j+1,j-1} & 0\end{pmatrix}
  \cr &&\cr &&\qquad +  \CC^{(1)}_{sd}\Delta_{0,j-1}\Delta_{0,j+1} \begin{pmatrix} 2v_{12} \alpha_{j+1,j-1} & v_{11} \alpha_{j-1,j-1}+v_{22} \alpha_{j+1,j+1} \\  v_{11} \alpha_{j-1,j-1}+v_{22} \alpha_{j+1,j+1}& 2 v_{21}  \alpha_{j-1,j+1} \end{pmatrix}\Biggr]\ ,
 \eqn{J16b}
 \eeq
 where $\alpha_{\ell_1,\ell_2}$ is given in \eq{abc} and $v_{ab}$ in \eq{bfepsdef}.  Making the replacements
 \beq
 \CC^{(1)}_{ss}\to \tilde \xi^{(1)}_{ss,0}+ \frac{\xi^{(1)}_{ss,1}}{j-1}\ ,\qquad 
 \CC^{(1)}_{sd} \to \tilde \xi^{(1)}_{sd,0} + \frac{\xi^{(1)}_{sd,1}}{j-1}\ ,
 \eeq
 where $ \xi^{(1)}_{ss,1}$ and $\xi^{(1)}_{sd,1}$ have been determined in \eq{J1poles}, adding to $\hat \CA^{(3)}$ in \eq{ampsLJCoupled}, and taking the $j\to 1$ limit, I arrive at the renormalized \MJSb~amplitude
 \beq
  \left[ \hat\CA_{J=1,\text{ren}}^{(3)}\right]_{11} &=& \hat k^3  
  \left(2 i  \tilde \xi^{(1)}_{ss,0}-\frac{2}{15} i \sqrt{2} \tilde \xi^{(1)}_{sd,0}+\frac{66}{5} i \ln
  \frac{k^2}{\mu^2 } +\frac{459 i \pi ^2}{50}+\frac{46827 \pi }{3500}+\frac{27 i}{8} \right)\ ,\cr
   &&\cr
     \left[ \hat\CA_{J=1,\text{ren}}^{(3)}\right]_{12} &=& \hat k^3 
      \left(-\frac{i  \tilde \xi^{(1)}_{ss,0}}{\sqrt{2}}+\frac{i \tilde \xi^{(1)}_{sd,0}}{5}-\frac{39}{10} i \sqrt{2}
   \ln  \frac{k^2}{\mu^2 } +\frac{27 i \pi ^2}{14 \sqrt{2}}-\frac{7461 \pi }{980
   \sqrt{2}}-\frac{11 i}{8 \sqrt{2}} \right)\ ,\cr &&\cr
     \left[ \hat\CA_{J=1,\text{ren}}^{(3)}\right]_{22} &=& \hat k^3 
      \left(-\frac{2}{15} i \sqrt{2} \tilde \xi^{(1)}_{sd,0}+\frac{6}{5} i \ln  \frac{k^2}{\mu^2
   } +\frac{1107 i \pi ^2}{4900}+\frac{459171 \pi }{343000}-\frac{5 i}{16}\right)
\ .
   \eeq
As expected, the amplitude at this order is finite without further renormalizations, since there could not be a local counterterm at $O(\hat k^3)$.

%%%%%%%%%%%%%%%%%%%%%%%%%%%%%%%

\begin{figure}[t]
\includegraphics[width=18cm]{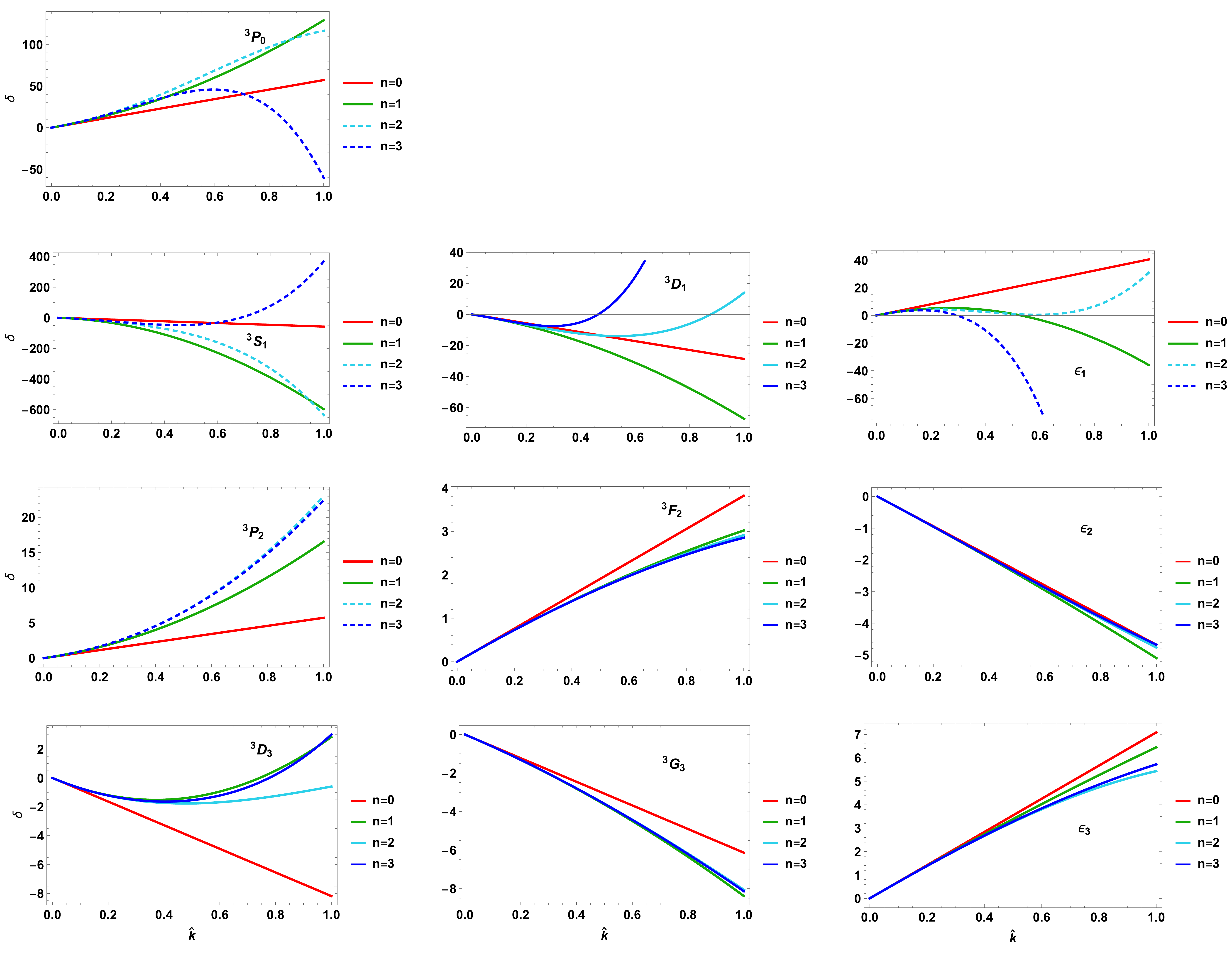}
%{LJcoupledplot.pdf}
\caption{ {\it Phase shifts in degrees versus $\hat k$ for $S=1$, $L=J\pm1$ partial waves.   Dashed lines indicate quantities dependent on finite counterterms, each of which was set to $-1$ for this plot at renormalization scale $\mu = \Lambda_{NN}$. For $J=1$ one  expands around the free fermion limit for the ${}^3S_1$ channel, not the unitary fermion wave function.
%in the next section I consider the expansion around the unitary fermion limit.
}}
\label{fig:PhaseShifts2}
\end{figure}
%%%%%%%%%%%%%%%%%%%%%%%%%%%%%%%

  \section{Phase shifts for low $J$}
\label{sec:lowJ}
 Now that  amplitudes for low $J$ have been renormalized, one can look at the corresponding phase shifts, which are given in Appendix~\ref{sec:rendelta}  in terms of the finite parts of the counterterms.   When I plot the phase shifts, in each case I set the counterterm to $\tilde \xi_{L,0} = -1$, and all phase shifts that depend on the value of the counterterm are marked with a dashed line.  It is worth noticing that while in each  case where the  graphs in Fig.~\ref{fig:cts}(a) and Fig.~\ref{fig:cts}(b) contributed to $\hat\CA^{(n)}$ and  $\hat\CA^{(n+1)}$ respectively, the phase shifts only depend on the counterterms at order $\delta^{(n)}$, but not $\delta^{(n+1)}$.  Thus two orders are dashed for the cumulative phase shift, but only one is dashed on the corresponding Lepage plot.

%%%%%%%%%%%%%%%%%%%%%%%%%%%%%%%
\begin{figure}[t]
\includegraphics[width=18cm]{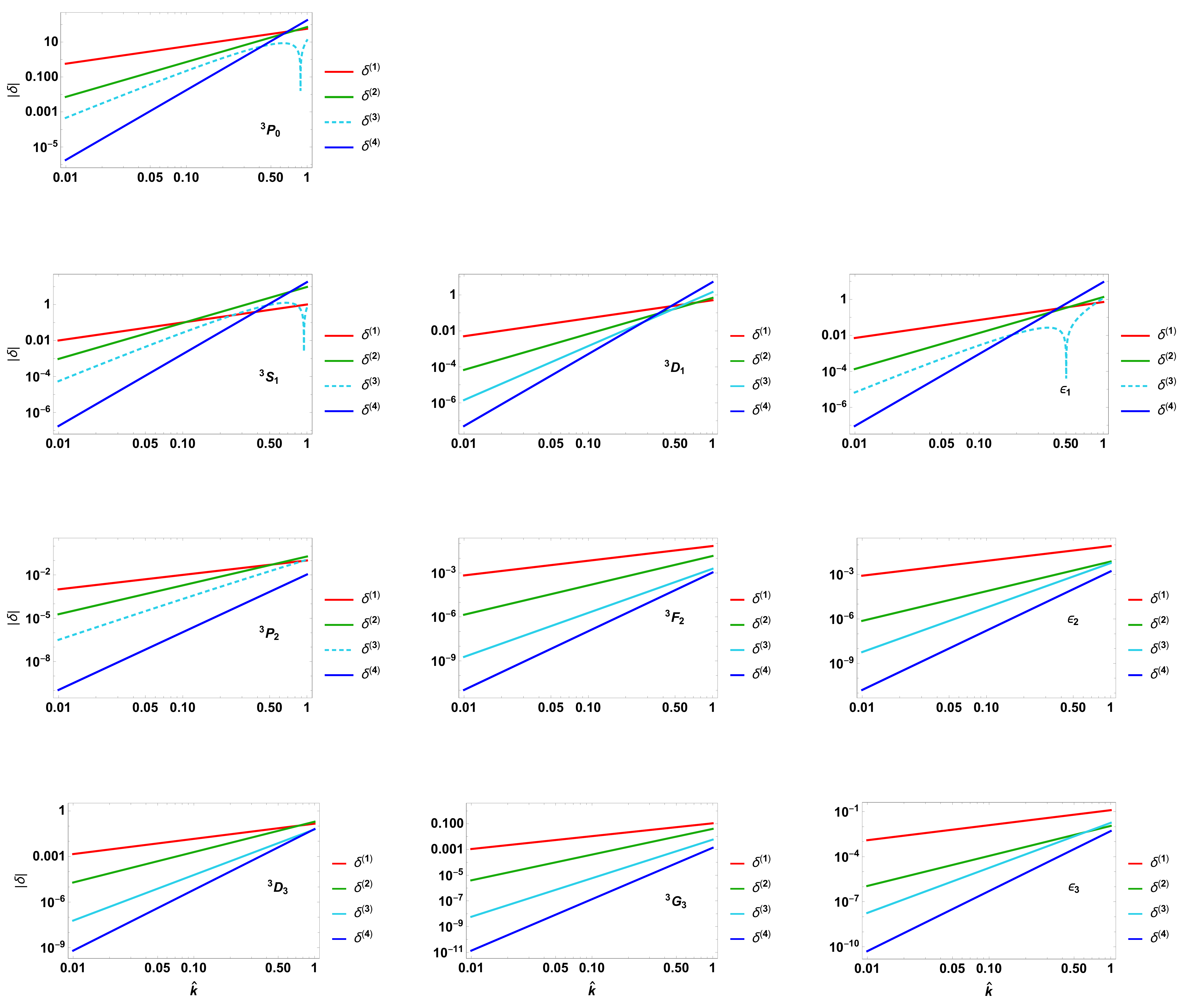}
%{LJcoupledplot.pdf}
\caption{ {\it Lepage log-log plots of $\delta^{(n)}$ phase shifts in the $L=J\pm 1$ channels, indicating the convergence properties of the the perturbative expansion.  }}
\label{fig:Lepage2}
\end{figure}
%%%%%%%%%%%%%%%%%%%%%%%%%%%%%%%

 \subsection{The $L=J$ channels}
 Fig.~\ref{fig:PhaseShifts1} shows the low lying partial waves for the $L=J$ spin triplet channels, and  the corresponding Lepage plots. One see that the perturbative expansion for the $L=3,\ldots$ partial waves converge very rapidly, the same conclusion reached in Refs.~\cite{Nogga:2005hy,Birse:2005um,Wu:2018lai}.  The ${}^3D_2$ partial wave shows adequate convergence in the phase shift plot, which  is well confirmed by the corresponding Lepage plot. The only partial wave that might be problematic in Fig.~\ref{fig:PhaseShifts1} is ${}^3P_1$, yet even for this partial wave the Lepage plot makes it evident that the radius of convergence extends to $\hat k\sim 1$, even though the convergence is slow. Note that if one only computed the tree and one-loop ladder diagrams as in FMS,  the $n=0$ and $n=1$ lines in the cumulative phase shift plots  would give the impression that the perturbative expansion had broken down at very low $\hat k$ for the $p$ and $d$ waves, while the Lepage plots  suggest that the worry is misplaced.

 \subsection{The $L=J\pm1$ channels}
I now turn to the coupled channels, plotting the phase shifts for low angular momentum channels in Fig.~\ref{fig:PhaseShifts2}.  One striking feature in the phase shift plots is that in almost every case there is a large deviation between the leading and subleading contributions (red and green curves) as remarked on by FMS, even more striking than in the $L=J$ channels. Overall, however, for $J\ge 2$, the Lepage  plots in Fig.~\ref{fig:Lepage2} show reasonably good convergence    below $\hat k=1$.
In contrast, one sees evidence of convergence problems in the $J=0$ (${}^3P_0$) and $J=1$ partial waves (${}^3S_1$, ${}^3D_1$, $\epsilon_1$). In the ${}^3P_0$ case the Lepage plot indicates relatively uniform convergence up to $\hat k \sim 0.7$.  In the $J=1$ coupled channels, however, the Lepage plots show the intersection points occurring at lower $\hat k$ with each successive order in the ${}^3D_1$ and $\epsilon_1$  channels, while not showing much pattern in the ${}^3S_1$.

 \section{Discussion and connections with the method of Cavagnero, Gao and Birse}
 
 In this paper I have focused on the observation by Fleming, Mehen and Stewart in Ref.~\cite{Fleming:1999ee} that due to the first two ladder diagrams in Fig.~\ref{fig:ladders},   the KSW expansion did not appear to converge well at NNLO in multiple spin-triplet nucleon-nucleon scattering channels, and that the problem was not a result of finite pion mass, but could be seen in the chiral limit. One feature they noted were powers of $\pi$ in the amplitudes which appeared to reduce the radius of convergence of the expansion.  From their work it was unclear whether the poor convergence they observed could be a general pathology of doing perturbation theory with a $1/r^3$ potential, or whether the coefficient of the potential was simply too strong for perturbation theory to be valid, or whether in fact the perturbative expansion was valid but simply slow to converge.  Pursuing that observation  techniques were developed here to compute those ladder diagrams to relatively high order in the chiral limit in all spin-triplet partial waves simultaneously.  By examining the amplitudes for large angular momentum I was able to show that there was no general pathology with the expansion, at least up to order $n\sim L$. The radius of convergence is affected by the strength of the potential, which alternates with $L$ because of the $\tau_1\cdot\tau_2 =[1+2(-1)^L]$ isospin factor, but it generally grows rapidly with angular momentum.   Furthermore, while factors of $\pi$ appear in the amplitudes exponentiated to powers that increase with order,  those  factors of $\pi$ only enter the expressions for the phase shifts linearly,  and then only  at  even terms in the $\hat k$ expansion, and so they are not implicated in harming the perturbative expansion.  
 
 To analyze the phase shifts at low $L$ beyond order $n=2L$ a method was devised  to regulate and renormalize the scattering amplitudes by analytically continuing angular momentum to non-integer values, subtracting poles at positive integer angular momentum.  Results suggest that there is no problem with the convergence of the ladder diagrams for the $L=J$ channels, even down to $L=J=1$; nor is there an apparent problem in the coupled $L=J\pm1$ channels for $J\ge 2$.  Lack of convergence was evident though in the $J=0$ and coupled $J=1$ channels, although in both cases there is model dependence on finite counterterms starting at two loops.  It is not entirely clear whether my results are relevant for the $J=1$ channels because I am  expanding around the noninteracting solutions, while in the KSW expansion one expands about the unitary fermion limit for the ${}^3S_1$ channel. It should be possible to extend the methods developed here to explore the unitary fermion limit, by perturbing around $\ell=-1$ instead of $\ell=0$ for the ${}^3S_1$ partial wave, although I have not looked at this closely.  However, one has no reason to expect that the unitary fermion limit will improve convergence. The problems seen in the  ${}^3P_0$ partial wave have no such excuses in any case.    
 
\begin{figure}[t]
\includegraphics[width=18cm]{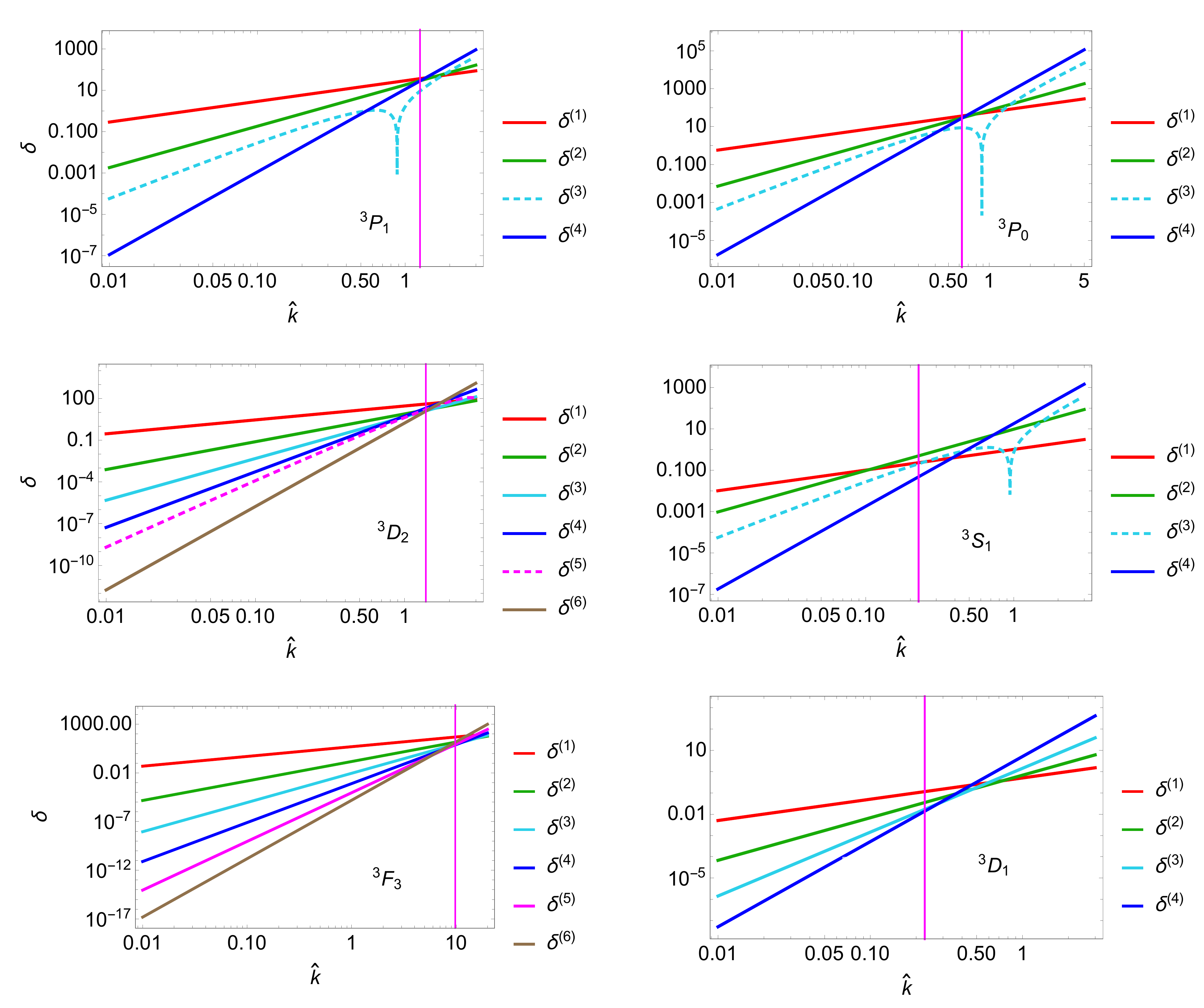}
%{LJcoupledplot.pdf}
\caption{ {\it  Lepage plots for several of the lower partial waves, showing $\hat k_c$ from Ref.~\cite{birse1999renormalisation} as a vertical magenta line. These provide visual confirmation that the branch point in $\nu$, defined in \eq{Cav}, is what determines the radius of convergence of the KSW expansion. }}
\label{fig:birseplot}
\end{figure}

Knowing that the KSW expansion converges for all but the lowest spin-triplet partial waves is a reassuring result, but does not in itself render the expansion useful for nuclear physics due to the dominant role played by low angular momentum scattering. A deeper understanding of the convergence problem can be found in the 2005 paper by Birse \cite{birse1999renormalisation}, building on earlier work by Cavagnero and Gao \cite{cavagnero1994secular,gao1999repulsive}.  That earlier work had shown that the exact angular momentum $L$ solution for a $1/r^3$ potential  could be found in the form of a sum of spherical Bessel functions
\beq
u_\ell = \sum_n\, c_n j_{L+  n+\nu }(k r)\ ,
\eqn{Cav}\eeq 
where $\nu$ is an energy dependent parameter that starts at $\nu=0$ for $\hat k=0$, drops to $\nu=-\half$ at a critical momentum $\hat k_c$ (which depends on $L$), and then splits into complex conjugate pairs for higher $\hat k$.  Since $\nu$ vanishes at $\hat k=0$, a strict expansion in powers of $\hat k$ will be a sum of spherical Bessel functions of integer order, $j_{L+n}$, along with their derivatives with respect to order.  This is what has been presented in this paper, and in fact the coefficient of the $\partial_\nu j_{\ell+\nu}\vert_{\nu=0}$ term agrees with the leading piece Cavagnero-Gao solution when the latter expanded is expanded in powers of $\hat k$ \footnote{M. Birse, private communication.}.  In Ref.~\cite{birse1999renormalisation}, Birse argued that $\hat k_c$ would be  the value of  $\hat k$ where the EFT expansion necessarily broke down, due to the branch point in the solution for the angular momentum shift $\nu(\hat k)$. Using the notation of this paper, the values of $\hat k_c$ he computed for several partial waves are shown in Table~\ref{tab:kc}. These values are then overlaid as vertical magenta lines on the relevant Lepage plots in Fig.~\ref{fig:birseplot}, where the  value for $\hat k_c$ computed for  ${}^3S_1$ is plotted in the ${}^3D_1$ plot. These plots offer striking  confirmation that indeed $\hat k_c$ is the scale which controls the convergence of the EFT expansion.

The framework of Cavagnero, Gao and Birse offers the tantalizing possibility that the effect of the branch point in $\nu$ could be accounted for either analytically or semi-analytically, allowing for systematic improvement of the EFT expansion.  The range of an EFT is normally extended by identifying nonanalyticities which control the convergence of the perturbative expansion in powers of energy, and then adding propagating degrees of freedom to the EFT which reproduce them explicitly.  In this case it is not clear whether there is new physics that might be added to the EFT to account for the branch point in $\nu$ explicitly, or whether there are alternative approaches, such as an expansion about the chiral limit, as proposed in Ref.~\cite{Beane:2001bc}, which might extend the validity of the EFT, but both approaches seem worth exploring.

%%%%%%%%%%%
\begin{table}
\caption{\label{tab:kc} Critical momenta $\hat k_c$ where perturbation theory is expected to break down for the $1/r^3$ potential, from Ref.~\cite{birse1999renormalisation}. }
\begin{ruledtabular}
\begin{tabular}{lccccc}
\   &${}^3P_1$ &$ {}^3D_2$&$ {}^3F_3$ & $ {}^3P_0$  &$ {}^3S_1$  \\
\hline
$\hat k_c$  & 1.26 & 1.39&9.85 & 0.63 & 0.23 \\
\end{tabular}
\end{ruledtabular}
\end{table}
%%%%%%%%%%%%%%

%%%%%%%%%%%%%%%%%%%%%%%%%%%%%%%
%%%%%%%%%%%%%%%%%%%%%%%%%%%%%%%

\vfill\eject
 
 \begin{acknowledgments}
I would like to thank S. Beane, M. Birse, A. Bulgac, M. J. Savage, I. Stewart, and U. Van Kolck for useful discussions.  M. Campbell participated in some of the early stages of this research. This work  was supported by the DOE Grant No. DE-FG02-00ER41132, and by the Thomas L. and Margo G. Wyckoff Endowed Faculty Fellowship.    
\end{acknowledgments}

\appendix
 
 %%%%%%%%%%%%%%%%
 %%%%%  Appendix A, derivation of Bessel formula
 %%%%%%%%%%%%%%%%%
 
   \section{Derivation of Bessel integral formula}
   \label{sec:Bessel}
In this appendix I give the derivation of \eq{jgjint} and \eq{abc}, namely the evaluation of
\beq
\CI_{\ell,m}(\rho)&\equiv& \int_0^\infty \frac{d\rho'}{\rho'} g_{\ell}(\rho,\rho') j_{m}(\rho')
\cr 
&=&
h^{(1)}_{\ell}(\rho)  \int_0^\rho \frac{d\rho'}{\rho'} j_{\ell}(\rho') j_{m}(\rho') + j_{\ell}(\rho) \int_\rho^\infty \frac{d\rho'}{\rho'} h^{(1)}_{\ell}(\rho') j_{m}(\rho') \ .
%
%&=&
% \alpha_{\ell m}\, j_{\ell}(\rho) + \beta_{\ell m}\, j_{m-1}(\rho) + \gamma_{\ellm} \,  j_{m+1}(\rho)\ .
 \eqn{jgjint2}
 \eeq
%with  $\rho$-independent coefficients given by 
%\beq
%\alpha_{\ell m} &=& -\frac{2 e^{i\pi(m-\ell)/2}}{(m-\ell-1)(m-\ell+1)(\ell+m)(2 + \ell+m)}\cr &&\cr
%\beta_{\ell m} &=& \frac{i}{(m-\ell-1)(\ell+m)(1+2m)}\cr&&\cr
%\gamma_{\ell m} &=& \frac{i}{(m-\ell+1)(2 + \ell+m)(1 +2m)}\ .
%\eqn{abc2}
%\eeq
%where

 This can be computed by using the indefinite integral in Eq. (77) of Ref.~\cite{bloomfield2017indefinite}:
\beq
L^{-1}_{\ell m}(x) &\equiv \int dx \, \frac{j_\ell(x) j_m(x)}{x}
= \frac{j_\ell(x) j_m(x)}{\ell + m}
- \frac{x j_{\ell+1}(x) j_m(x)}{(  \ell - m+1)(\ell+m)}
+ \frac{x j_\ell(x) j_{m+1}(x)}{(\ell- m -1)(\ell+m)}
-\frac{2 x^2\left(j_{\ell+1}(x) j_{m+1}(x) + j_\ell(x)j_m(x)\right)}{(\ell+m)(\ell+m+2)(\ell-m+1)(\ell-m-1)}\ ,
\eqn{int1}\eeq
which has the asymptotic limits
 \beq
 L^{-1}_{\ell m}(0) &=& \begin{cases} 0 & \Re[\ell+m]>0 \\ \infty & \Re[\ell+m]< 0\end{cases}\ ,\qquad
 L^{-1}_{\ell m}(\infty)   =-\frac{2 \cos \left(\frac{1}{2} \pi  (\ell-m)\right)}{(\ell-m-1) (\ell-m+1) (\ell+m) (\ell+m+2)}\ .
  \eqn{ILims} \eeq
  By making use of the relation
 \beq
  h^1_\nu(x) = j_\nu(x) + i y_\nu(x) = \frac{e^{-i \pi\nu}j_\nu(x) - i j_{-\nu-1}(x)}{\cos\pi\nu}
  \ ,
  %\qquad  j_\nu(x) =\frac{x}{2\nu+1} \left(j_{\nu+1}(x) + j_{\nu-1}(x)\right)\ .
 \eqn{H1}\eeq
 $\CI_{\ell,m}(\rho)$ can be rewritten as
\beq
\CI_{\ell,m}(\rho) &=& \left(
\frac{e^{-i \pi\ell}j_{\ell}(\rho) - i j_{-\ell-1}(\rho) }{\cos\pi\ell}\right) L^{-1}_{\ell m}(x)\Biggl\vert^\rho_0 
+ j_{\ell}(\rho) \left( \frac{e^{-i \pi\ell}L^{-1}_{\ell m}(x) - i L^{-1}_{-\ell-1,m}(x)}{\cos\pi\ell} \right)\Biggl\vert_\rho^\infty \cr
&=&
-\frac{2 e^{-\frac{1}{2} i \pi  (\ell-m)} j_\ell(r)}{(\ell-m-1) (\ell-m+1) (\ell+m)
   (\ell+m+2)}\cr &&
   +\frac{i j_{m-1}(r)}{(2 m+1) (-\ell+m-1) (\ell+m)}-\frac{i j_{m+1}(r)}{(2 m+1) (\ell-m-1) (\ell+m+2)}\ ,
\eqn{Ival}\eeq
where the final formula was obtained using  \eq{int1} and  \eq{ILims}, assuming that $\Re[\ell+m]>0$, as well as the standard Bessel function identities
 \beq
 j_\nu(z) =\frac{z}{2\nu+1} \left[j_{\nu+1}(z) + j_{\nu-1}(z)\right]\ ,\qquad
  j_{\nu-1}(z)  j_{-\nu-1}(z)  +   j_{\nu}(z) j_{-\nu}(z) = \frac{\cos\pi\nu}{z^2}\ .
  \eeq
The first of these identities allows one to eliminate dependence on $j_m$ in favor of $j_{m\pm 1}$, while the second of the above identities makes it possible for $\CI_{\ell,m}(\rho) $   to be written as a linear polynomial in Bessel functions rather than cubic,  which is what enables a recursive solution of multi-loop diagrams. The above expression \eqn{Ival} can be written as
\beq
\CI_{\ell,m}(\rho) = \alpha_{\ell m} j_\ell(\rho) + \beta_{\ell m} j_{m-1}(\rho) + \gamma_{\ell m}   j_{m+1}(\rho)   \ ,
   \eqn{rslt}
   \eeq
   where
    \beq
\alpha_{\ell m} &=& -\frac{2 i^{m-\ell}}{(m-(\ell+1))(m-(\ell-1))(\ell+m)(2 + \ell+m)}\cr &&\cr
\beta_{\ell m} &=& \frac{i}{(m-(\ell+1))(\ell+m)(1+2m)}\cr&&\cr
\gamma_{\ell m} &=& \frac{i}{(m-(\ell-1))(2 + \ell+m)(1 +2m)}\ .
\eeq
  Note that the imaginary part of the amplitude arises entirely from the $i$ introduced in \eq{H1}; this makes sense, since it is the asymptotic  part of the scattering wave function proportional to the irregular Bessel function which signifies scattering, which in turn is related to the imaginary part of the scattering amplitude through the optical theorem.  Also note that  \eq{rslt} is taken to be the definition of $\CI_{\ell,m}(\rho) $ for all $\ell$, $m$ --- including values for which the integral is divergent. 
  
  %%%%%%%%%%%%%%%
 %  Appendix B
 %%%%%%%%%%%%

   \section{The scattering amplitudes and phase shifts for the spin triplet $L=J$ channels to $O(\hat k^7)$}
   \label{sec:LeqJamps}
   I give here the scattering amplitude $\hat \CA =  (M\Lambda_{NN}/4\pi)\CA$ for the first eight ladder diagrams (i.e., to seven loops) of the series shown in Fig.~\ref{fig:ladders} for the spin-triplet $L=J$ partial waves.  The amplitudes are found to have the general form
   \beq
   \hat\CA^{(n)} = \hat k^n \frac{\left(1+2(-1)^\ell\right)^{n+1}}{\left(\ell(\ell+1)\right)^n}\, \sum_{k=-n}^{n+2}\sum_{p =1}^n\frac{A^{(n)}_{kp}}{(2\ell+k)^p}\ ,
   \eeq
   where the $ A^{(n)}_{kp}$ coefficients (many of which vanish) are independent of $\ell$.  In this form I find the following results from the recursion procedure discussed in \S~\ref{sec:LeqJcomputation}:
\beq
\hat\CA^{(0)}&=& \left(1+2(-1)^\ell\right)  \times\Biggl[-\frac{1}{\ell+1}
+ \frac{1}{\ell}\Biggr]
\cr
&&\cr
&&\cr
\hat\CA^{(1)}&=&\hat k \frac{\left(1+2(-1)^\ell\right)^2}{\left(\ell(\ell+1)\right)}
\times\Biggl[
\frac{3 \pi }{8(2 \ell+3)}
-\frac{i}{\ell+1}
-\frac{3 \pi }{4(2 \ell+1)}
+\frac{i}{\ell}
+\frac{3 \pi }{8(2 \ell-1)}\Biggr]
\cr
&&\cr
&&\cr
\hat\CA^{(2)}&=&\hat k^2 \frac{\left(1+2(-1)^\ell\right)^3}{\left(\ell(\ell+1)\right)^2} 
\times\Biggl[
-\frac{4}{9 (\ell+2)}
+
\frac{3 i \pi }{4(2 \ell+3)}
+
\frac{2}{\ell+1}
-\frac{3 i \pi }{2(2 \ell+1)}
-\frac{2}{\ell}
+
\frac{3 i \pi }{4(2 \ell-1)}
+
\frac{4}{9 (\ell-1)}\Biggr]
\cr
&&\cr
&&\cr
\hat\CA^{(3)}&=&\hat k^3 \frac{\left(1+2(-1)^\ell\right)^4}{\left(\ell(\ell+1)\right)^3} \cr &&\cr &&
\times\Biggl[
\frac{375 \pi }{2048 (2 \ell+5)}
-\frac{8 i}{9 (\ell+2)}
+
\frac{3 (-250+3 i \pi ) \pi }{512 (2 \ell+3)}+\frac{9 (1+3 i \pi ) \pi }{256 (2 \ell+3)^2}+\frac{27 \pi }{128 (2 \ell+3)^3}
+
\frac{3 i}{\ell+1}
\cr
&&\cr
&&\quad
+\frac{2625 \pi }{1024 (2 \ell+1)}-\frac{9 i \pi ^2}{64 (2 \ell+1)^2}-\frac{9 \pi }{32 (2 \ell+1)^3}
-\frac{3 i}{\ell}
+\frac{9 (-1+3 i \pi ) \pi }{256 (1-2 \ell)^2}+\frac{3 (-250-3 i \pi ) \pi }{512 (2 \ell-1)}+\frac{27 \pi }{128 (2 \ell-1)^3}
\cr
&&\cr
&&\quad
+\frac{8 i}{9 (\ell-1)}
+\frac{375 \pi }{2048 (2 \ell-3)}
\Biggr]
\cr
&&\cr
&&\cr
\hat\CA^{(4)}&=&\hat k^4 \frac{\left(1+2(-1)^\ell\right)^5}{\left(\ell(\ell+1)\right)^4} \cr &&\cr &&
\times\Biggl[
-\frac{4}{25 (\ell+3)}
+\frac{375 i \pi }{1024 (2 \ell+5)}
+\frac{16}{45 (\ell+2)^2}+\frac{332+48 i \pi }{135 (\ell+2)}
-\frac{9 \pi  (5 \pi +502 i)}{1280 (2 \ell+3)}+\frac{9 (-3 \pi +i) \pi }{128 (2 \ell+3)^2}+\frac{27 i \pi }{64 (2 \ell+3)^3}
\cr
&&\cr
&&\quad
-\frac{6}{\ell+1}
+
\frac{22601 i \pi }{4608 (2 \ell+1)}+\frac{9 \pi ^2}{32 (2 \ell+1)^2}-\frac{9 i \pi }{16 (2 \ell+1)^3}
+
\frac{6}{\ell}
-\frac{9 \pi  (3 \pi +i)}{128 (1-2 \ell)^2}+\frac{9 \pi  (5 \pi -502 i)}{1280 (2 \ell-1)}+\frac{27 i \pi }{64 (2 \ell-1)^3}
\cr
&&\cr
&&\quad
+
\frac{16}{45 (\ell-1)^2}+\frac{4 i (12 \pi +83 i)}{135 (\ell-1)}
+
\frac{375 i \pi }{1024 (2 \ell-3)}
+
\frac{4}{25 (\ell-2)}\Biggr]
\cr
&&\cr
&&\cr
\hat\CA^{(5)}&=&\hat k^5 \frac{\left(1+2(-1)^\ell\right)^6}{\left(\ell(\ell+1)\right)^5} 
\cr &&\cr &&
\times\Biggl[\frac{2100875 \pi }{42467328 (2 \ell+7)}
-\frac{8 i}{25 (\ell+3)}
-\frac{5625 \pi }{32768 (2 \ell+5)^2}+\frac{375 (-199-30 i \pi ) \pi }{131072 (2 \ell+5)}
+\frac{448 i}{405 (\ell+2)^2}-\frac{8 (168 \pi -697 i)}{1215 (\ell+2)}
\cr
&&\cr
&&\quad
+\frac{3 \pi  \left(6050131+31080 i \pi +9600 \pi ^2\right)}{2621440 (2 \ell+3)}+\frac{9 \pi  \left(259-2373 i \pi +30 \pi ^2\right)}{32768 (2 \ell+3)^2}-\frac{27 \pi  \left(791+30 i \pi +12 \pi ^2\right)}{16384 (2 \ell+3)^3}+\frac{81 i \pi  (6 \pi +5 i)}{4096 (2 \ell+3)^4}
\cr
&&\cr
&&\quad
+\frac{243 \pi }{1024 (2 \ell+3)^5}
-\frac{10 i}{\ell+1}
-\frac{\pi  \left(14833861+38880 \pi ^2\right)}{1769472 (2 \ell+1)}+\frac{6219 i \pi ^2}{8192 (2 \ell+1)^2}+\frac{9 \pi  \left(691+8 \pi ^2\right)}{4096 (2 \ell+1)^3}
-\frac{27 i \pi ^2}{256 (2 \ell+1)^4}\cr
&&\cr
&&\quad
-\frac{27 \pi }{128 (2 \ell+1)^5}
+\frac{10 i}{\ell}
-\frac{9 \pi  \left(259+2373 i \pi +30 \pi ^2\right)}{32768 (1-2\ell)^2}+\frac{81 (5+6 i \pi ) \pi }{4096 (2\ell-1)^4}+\frac{3 \pi  \left(6050131-31080 i \pi +9600 \pi ^2\right)}{2621440 (2 \ell-1)}
\cr
&&\cr
&&\quad
-\frac{27 \pi  \left(791-30 i \pi +12 \pi ^2\right)}{16384 (2 \ell-1)^3}+\frac{243 \pi }{1024 (2 \ell-1)^5}
+\frac{448 i}{405 (\ell-1)^2}-\frac{8 (168 \pi +697 i)}{1215 (\ell-1)}
+\frac{5625 \pi }{32768 (2\ell-3)^2}
\cr
&&\cr
&&\quad
+\frac{375 i \pi  (30 \pi +199 i)}{131072 (2 \ell-3)}
+\frac{8 i}{25 (\ell-2)}
+\frac{2100875 \pi }{42467328 (2 \ell-5)}\Biggr]
\cr
&&\cr
&&\cr
\hat\CA^{(6)}&=&\hat k^6 \frac{\left(1+2(-1)^\ell\right)^7}{\left(\ell(\ell+1)\right)^6} \cr &&\cr &&
\times\Biggl[
-\frac{1024}{30625 (\ell+4)}
+
\frac{2100875 i \pi }{21233664 (2 \ell+7)}
+
\frac{48}{875 (\ell+3)^2}+\frac{4 (449+60 i \pi )}{4375 (\ell+3)}
+
\frac{375 \pi  (210 \pi -113 i)}{458752 (2 \ell+5)}-\frac{5625 i \pi }{16384 (2 \ell+5)^2}
\cr
&&\cr
&&\quad
+
\frac{4 \left(-2644619-1083720 i \pi +37800 \pi ^2\right)}{1063125 (\ell+2)}+\frac{32 (-9031-630 i \pi )}{70875 (\ell+2)^2}-\frac{64}{225 (\ell+2)^3}
\cr
&&\cr
&&\quad
+
\frac{3 i \pi  \left(1224996797+30813720 i \pi +1680000 \pi ^2\right)}{229376000 (2 \ell+3)}
+\frac{9 \pi  \left(36683 i+67005 \pi +750 i \pi ^2\right)}{409600 (2 \ell+3)^2}-\frac{81 i \pi  \left(1489+50 i \pi +20 \pi ^2\right)}{40960 (2 \ell+3)^3}
\cr
&&\cr
&&\quad
-\frac{81 \pi  (6 \pi +5 i)}{2048 (2 \ell+3)^4}+\frac{243 i \pi }{512 (2 \ell+3)^5}
+
\frac{20}{\ell+1}
-\frac{i \pi  \left(1765710641+4860000 \pi ^2\right)}{110592000 (2 \ell+1)}-\frac{18145 \pi ^2}{12288 (2 \ell+1)^2}
\cr
&&\cr
&&\quad
+\frac{i \pi  \left(18145+216 \pi ^2\right)}{6144 (2 \ell+1)^3}+\frac{27 \pi ^2}{128 (2 \ell+1)^4}-\frac{27 i \pi }{64 (2 \ell+1)^5}
-\frac{20}{\ell}
-\frac{9 i \pi  \left(36683+67005 i \pi +750 \pi ^2\right)}{409600 (1-2\ell)^2}-\frac{81 \pi  (6 \pi -5 i)}{2048 (2\ell-1)^4}
\cr
&&\cr
&&\quad
+\frac{3 \pi  \left(1224996797 i+30813720 \pi +1680000 i \pi ^2\right)}{229376000 (2 \ell-1)}-\frac{81 i \pi  \left(1489-50 i \pi +20 \pi ^2\right)}{40960 (2 \ell-1)^3}+\frac{243 i \pi }{512 (2 \ell-1)^5}
\cr
&&\cr
&&\quad
-\frac{4 \left(-2644619+1083720 i \pi +37800 \pi ^2\right)}{1063125 (\ell-1)}+\frac{32 i (630 \pi +9031 i)}{70875 (\ell-1)^2}+\frac{64}{225 (\ell-1)^3}
+
\frac{5625 i \pi }{16384 (2\ell-3)^2}
\cr
&&\cr
&&\quad
-\frac{375 \pi  (210 \pi +113 i)}{458752 (2 \ell-3)}
+
\frac{48}{875 (\ell-2)^2}+\frac{4 i (60 \pi +449 i)}{4375 (\ell-2)}
+
\frac{2100875 i \pi }{21233664 (2 \ell-5)}
+
\frac{1024}{30625 (\ell-3)}
\cr
&&\cr
&&\cr
\hat\CA^{(7)}&=&\hat k^7 \frac{\left(1+2(-1)^\ell\right)^8}{\left(\ell(\ell+1)\right)^7} \cr &&\cr &&
\times\Biggl[\frac{110270727 \pi }{13421772800 (2 \ell+9)}
-\frac{2048 i}{30625 (\ell+4)}
-\frac{73530625 \pi }{2717908992 (2 \ell+7)^2}-\frac{2100875 i \pi  (105 \pi -902 i)}{16307453952 (2 \ell+7)}
+
\frac{96 i}{875 (\ell+3)^2}
\cr
&&\cr
&&\quad
-\frac{24 (20 \pi -33 i)}{4375 (\ell+3)}
-\frac{375 \pi  \left(1310557-586005 i \pi +25200 \pi ^2\right)}{469762048 (2 \ell+5)}+\frac{5625 (5581+1230 i \pi ) \pi }{33554432 (2 \ell+5)^2}+\frac{3459375 \pi }{8388608 (2 \ell+5)^3}
\cr
&&\cr
&&\quad
+
\frac{8 i \left(-798123-417240 i \pi +26600 \pi ^2\right)}{354375 (\ell+2)}
+\frac{1216 (70 \pi -549 i)}{70875 (\ell+2)^2}-\frac{2432 i}{2025 (\ell+2)^3}
\cr
&&\cr
&&\quad
+
\frac{3 \pi  \left(-612941353594-32597125365 i \pi -1006656000 \pi ^2-1386000 i \pi ^3\right)}{58720256000 (2 \ell+3)}
\cr
&&\cr
&&\quad
+\frac{9 i \pi  \left(310448813 i+321860880 \pi +10120200 i \pi ^2+140400 \pi ^3\right)}{838860800 (2 \ell+3)^2}
+\frac{81 \pi  \left(894058+84225 i \pi +21110 \pi ^2+360 i \pi ^3\right)}{10485760 (2 \ell+3)^3}
\cr
&&\cr
&&\quad
+\frac{243 \pi  \left(5615-4144 i \pi +144 \pi ^2-12 i \pi ^3\right)}{1048576 (2 \ell+3)^4}
-\frac{243 \pi  \left(518+45 i \pi +9 \pi ^2\right)}{65536 (2 \ell+3)^5}
+\frac{10935 i \pi  (\pi +2 i)}{65536 (2 \ell+3)^6}+\frac{10935 \pi }{32768 (2 \ell+3)^7}
\cr
&&\cr
&&\quad
+
\frac{35 i}{\ell+1}
+
\frac{\pi  \left(225862173824671+1166738256000 \pi ^2\right)}{8153726976000 (2 \ell+1)}-\frac{i \pi ^2 \left(120581933+124416 \pi ^2\right)}{37748736 (2 \ell+1)^2}
\cr
&&\cr
&&\quad
-\frac{\pi  \left(120581933+2634768 \pi ^2\right)}{18874368 (2 \ell+1)^3}+\frac{27 i \pi ^2 \left(1937+4 \pi ^2\right)}{65536 (2 \ell+1)^4}
+\frac{27 \pi  \left(1937+24 \pi ^2\right)}{32768 (2 \ell+1)^5}-\frac{405 i \pi ^2}{4096 (2 \ell+1)^6}-\frac{405 \pi }{2048 (2 \ell+1)^7}
\cr
&&\cr
&&\quad
-\frac{35 i}{\ell}
-\frac{243 i \pi  \left(-5615 i+4144 \pi -144 i \pi ^2+12 \pi ^3\right)}{1048576 (2\ell-1)^4}
+\frac{9 \pi  \left(310448813+321860880 i \pi +10120200 \pi ^2+140400 i \pi ^3\right)}{838860800 (1-2\ell)^2}
\cr
&&\cr
&&\quad
+\frac{10935 (2+i \pi ) \pi }{65536 (2\ell-1)^6}
+\frac{3 i \pi  \left(612941353594 i+32597125365 \pi +1006656000 i \pi ^2+1386000 \pi ^3\right)}{58720256000 (2 \ell-1)}
\cr
&&\cr
&&\quad
+\frac{81 \pi  \left(894058-84225 i \pi +21110 \pi ^2-360 i \pi ^3\right)}{10485760 (2 \ell-1)^3}
-\frac{243 \pi  \left(518-45 i \pi +9 \pi ^2\right)}{65536 (2 \ell-1)^5}+\frac{10935 \pi }{32768 (2 \ell-1)^7}
\cr
&&\cr
&&\quad
-\frac{8 i \left(-798123+417240 i \pi +26600 \pi ^2\right)}{354375 (\ell-1)}-\frac{1216 (70 \pi +549 i)}{70875 (\ell-1)^2}+\frac{2432 i}{2025 (\ell-1)^3}
+
\frac{5625 i \pi  (1230 \pi +5581 i)}{33554432 (2\ell-3)^2}
\cr
&&\cr
&&\quad
-\frac{375 \pi  \left(1310557+586005 i \pi +25200 \pi ^2\right)}{469762048 (2 \ell-3)}+\frac{3459375 \pi }{8388608 (2 \ell-3)^3}
+
\frac{96 i}{875 (\ell-2)^2}-\frac{24 (20 \pi +33 i)}{4375 (\ell-2)}
\cr
&&\cr
&&\quad
+
\frac{73530625 \pi }{2717908992 (2\ell-5)^2}+\frac{2100875 i \pi  (105 \pi +902 i)}{16307453952 (2 \ell-5)}
+
\frac{2048 i}{30625 (\ell-3)}
+
\frac{110270727 \pi }{13421772800 (2 \ell-7)}
\Biggr]
\eqn{ampsLeqJ}\eeq

  From these amplitudes one derives the following phase shifts, whose reality indicates that the above scattering amplitudes are consistent with unitarity:
   \beq
\delta^{(1)}&=&\hat k \frac{ \left(1+2(-1)^\ell\right)} {\ell(\ell+1)}\cr &&\cr
\delta^{(2)}&=& \hat k^2\frac{ \left(1+2(-1)^\ell\right)^2}{ \ell^2(\ell+1)^2}   \Biggl[\pi  \Bigl(\frac{3}{16 (2 \ell+1)}+\frac{9}{32 (2 \ell+3)}+\frac{9}{32 (2 \ell-1)}\Bigr) \Biggr]\cr &&\cr
\delta^{(3)}&=&\hat k^3 \frac{ \left(1+2(-1)^\ell\right)^3}{ \ell^3(\ell+1)^3}\Biggl[ \frac{8}{9 (\ell-1)}-\frac{8}{9 (\ell+2)}\Biggr]
\cr &&\cr
\delta^{(4)}&=& \hat k^4 \frac{ \left(1+2(-1)^\ell\right)^4}{ \ell^4(\ell+1)^4}\Biggl[ \pi  \Bigl(-\frac{531}{1024 (2 \ell-1)}-\frac{1377}{4096 (2 \ell+1)}-\frac{531}{1024 (2 \ell+3)}+\frac{5625}{8192 (2 \ell+5)}+\frac{189}{1024 (2 \ell-1)^2}
\cr &&\qquad\qquad\qquad\qquad
-\frac{189}{1024 (2 \ell+3)^2}+\frac{81}{512 (2
   \ell-1)^3}+\frac{9}{128 (2 \ell+1)^3}+\frac{81}{512 (2 \ell+3)^3}+\frac{5625}{8192 (2 \ell-3)}\Bigr)\Biggr]
\cr &&\cr
\delta^{(5)}&=& \hat k^5 \frac{ \left(1+2(-1)^\ell\right)^5}{ \ell^5(\ell+1)^5} \Biggl[-\frac{56}{27 (\ell-1)}+\frac{56}{27 (\ell+2)}-\frac{24}{25 (\ell+3)}+\frac{32}{45 (\ell-1)^2}+\frac{32}{45 (\ell+2)^2}+\frac{24}{25 (\ell-2)} \Biggr]
\cr &&\cr
\delta^{(6)}&=& \hat k^6  \frac{ \left(1+2(-1)^\ell\right)^6}{ \ell^6(\ell+1)^6} \Biggl[\pi  \Bigl(-\frac{64}{81 (\ell+2)}+\frac{73530625}{169869312 (2 \ell-5)}-\frac{219375}{524288 (2 \ell-3)}+\frac{2456271}{2097152 (2 \ell-1)}
\cr&&\qquad\qquad\qquad\qquad
+\frac{5585365}{7077888 (2 \ell+1)}+\frac{2456271}{2097152 (2
   \ell+3)}-\frac{219375}{524288 (2 \ell+5)}+\frac{73530625}{169869312 (2 \ell+7)}
   \cr&&\qquad\qquad\qquad\qquad
+\frac{84375}{131072 (2 \ell-3)^2}-\frac{126225}{131072 (2 \ell-1)^2}+\frac{126225}{131072 (2 \ell+3)^2}-\frac{84375}{131072 (2
   \ell+5)^2}
   \cr&&\qquad\qquad\qquad\qquad
-\frac{32967}{65536 (2 \ell-1)^3}-\frac{4779}{16384 (2 \ell+1)^3}-\frac{32967}{65536 (2 \ell+3)^3}+\frac{5103}{16384 (2 \ell-1)^4}-\frac{5103}{16384 (2 \ell+3)^4}
\cr&&\qquad\qquad\qquad\qquad
+\frac{729}{4096 (2 \ell-1)^5}+\frac{27}{512 (2
   \ell+1)^5}+\frac{729}{4096 (2 \ell+3)^5}-\frac{64}{81 (\ell-1)}\Bigr)\Biggr]
\cr &&\cr
\delta^{(7)}&=&\hat k^7 \frac{ \left(1+2(-1)^\ell\right)^7}{ \ell^7(\ell+1)^7} \Biggl[ -\frac{168}{625 (\ell-2)}+\frac{36424}{16875 (\ell-1)}-\frac{36424}{16875 (\ell+2)}+\frac{168}{625 (\ell+3)}-\frac{12288}{30625 (\ell+4)}
\cr&&\qquad\qquad\qquad\qquad
+\frac{288}{875 (\ell-2)^2}-\frac{33856}{7875 (\ell-1)^2}-\frac{33856}{7875
   (\ell+2)^2}+\frac{288}{875 (\ell+3)^2}+\frac{128}{225 (\ell-1)^3}
   \cr&&\qquad\qquad\qquad\qquad
-\frac{128}{225 (\ell+2)^3}+\frac{12288}{30625 (\ell-3)} \Biggr]
\cr &&\cr
\delta^{(8)}&=& \hat k^8 \frac{ \left(1+2(-1)^\ell\right)^8}{ \ell^8(\ell+1)^8} \Biggl[\pi \Bigl(\frac{16384}{6075 (\ell+2)}+\frac{6947055801}{53687091200 (2 \ell-7)}-\frac{2279449375}{32614907904 (2 \ell-5)}-\frac{751786875}{268435456 (2 \ell-3)}
\cr&&\qquad\qquad\qquad\qquad
-\frac{5453519139}{3355443200 (2
   \ell-1)}-\frac{178732992665}{86973087744 (2 \ell+1)}-\frac{5453519139}{3355443200 (2 \ell+3)}-\frac{751786875}{268435456 (2 \ell+5)}
   \cr&&\qquad\qquad\qquad\qquad
-\frac{2279449375}{32614907904 (2 \ell+7)}+\frac{6947055801}{53687091200 (2s
   \ell+9)}+\frac{256}{405 (\ell+2)^2}+\frac{2573571875}{10871635968 (2 \ell-5)^2}
   \cr&&\qquad\qquad\qquad\qquad
-\frac{187396875}{134217728 (2 \ell-3)^2}+\frac{2936049363}{671088640 (2 \ell-1)^2}-\frac{2936049363}{671088640 (2
   \ell+3)^2}
   \cr&&\qquad\qquad\qquad\qquad
+\frac{187396875}{134217728 (2 \ell+5)^2}-\frac{2573571875}{10871635968 (2 \ell+7)^2}+\frac{51890625}{33554432 (2 \ell-3)^3}+\frac{3929391}{4194304 (2 \ell-1)^3}
\cr&&\qquad\qquad\qquad\qquad
+\frac{77958125}{75497472 (2
   \ell+1)^3}+\frac{3929391}{4194304 (2 \ell+3)^3}+\frac{51890625}{33554432 (2 \ell+5)^3}-\frac{9186615}{4194304 (2 \ell-1)^4}
   \cr&&\qquad\qquad\qquad\qquad
+\frac{9186615}{4194304 (2 \ell+3)^4}-\frac{10935}{16384 (2 \ell-1)^5}-\frac{44955}{131072 (2
   \ell+1)^5}-\frac{10935}{16384 (2 \ell+3)^5}
   \cr&&\qquad\qquad\qquad\qquad
+\frac{76545}{131072 (2 \ell-1)^6}-\frac{76545}{131072 (2 \ell+3)^6}+\frac{32805}{131072 (2 \ell-1)^7}+\frac{405}{8192 (2 \ell+1)^7}
\cr&&\qquad\qquad\qquad\qquad
+\frac{32805}{131072 (2
   \ell+3)^7}+\frac{16384}{6075 (\ell-1)}-\frac{256}{405 (\ell-1)^2}\Bigr)\Biggr]
  \eqn{LeqJphase} \eeq
There are many interesting repeating patterns in the above formulas for both the amplitudes and the phase shifts, which suggests that it might be possible to find a closed expression for the phase shifts to all orders in the perturbative expansion, but I do not pursue  this idea here.

%%%%%%%%%%%%%%%%%%%%%%%%%%%%%%%%
%%%  Appendix C
%%%%%%%%%%%

   \section{The scattering amplitudes and phases  for the spin triplet $L=J\pm1$ channels   to $O(\hat k^3)$}
   \label{sec:LJcoupledJg2}

        By solving the  recursion relation \eq{Arecur} up to $n=3$ for $j\ge 1$  I arrive at the following results for the coupled scattering amplitudes (with $\CA_{12} = \CA_{21}$).  (For the special case  of $J=0$, the uncoupled ${}^3P_0$ channel is given by the  ${}^3P_1$ results for the amplitudes in \eq{ampsLeqJ} modified by multiplying $\CA^{(n)}$  by a factor of $(-2)^{n+1}$ at each order, accounting for the different coefficient of the $1/r^3$ potential in the Schr\"odinger equation).

 \beq
\hat\CA^{(0)}_{11} &=& \left(2(-1)^j-1\right)  \Biggl[  \frac{1}{j}-\frac{2}{2 j+1} \Biggr]
\cr  &&\cr
\hat\CA^{(0)}_{12} &=& \left(2(-1)^j-1\right)  \sqrt{j (j+1)} \Biggl[  -\frac{1}{j+1}+\frac{4}{2 j+1}-\frac{1}{j} \Biggr]
\cr &&\cr
\hat\CA^{(0)}_{22} &=& \left(2(-1)^j-1\right)  \Biggl[   \frac{2}{2 j+1}-\frac{1}{j+1}  \Biggr]
\cr  &&\cr &&\cr
\hat\CA^{(1)}_{11} &=& \hat k\,\left(2(-1)^j-1\right)^2 \Biggl[ \pi  \left(\frac{1}{8 (2 j-3)}+\frac{15}{8 (2 j+1)}+\frac{3}{2 (2 j+1)^2}-\frac{1}{j}\right)+i \left(\frac{1}{j^2}-\frac{1}{j+1}+\frac{8}{2 j+1}-\frac{3}{j}\right)\Biggr]\cr  &&\cr
\hat\CA^{(1)}_{12} &=& \hat k\, \left(2(-1)^j-1\right)^2 \sqrt{j (j+1)}
\cr &&\cr &&
\times
 \Biggl[\pi  \left(-\frac{1}{j+1}-\frac{1}{4 (2 j-1)}+\frac{1}{4 (2 j+3)}-\frac{3}{(2 j+1)^2}+\frac{1}{j}\right)
+i \left(-\frac{1}{j^2}+\frac{4}{j+1}-\frac{16}{2
   j+1}+\frac{1}{(j+1)^2}+\frac{4}{j}\right)\Biggr]
\cr &&\cr
\hat\CA^{(1)}_{22} &=& \hat k\, \left(2(-1)^j-1\right)^2  \Biggl[ \pi  \left(\frac{15}{8 (2 j+1)}+\frac{1}{8 (2 j+5)}-\frac{3}{2 (2 j+1)^2}-\frac{1}{j+1}\right)+i \left(\frac{3}{j+1}-\frac{8}{2 j+1}+\frac{1}{(j+1)^2}+\frac{1}{j}\right)\Biggr]
\cr  &&\cr &&\cr
\hat\CA^{(2)}_{11}&=&\hat k^2\left(2(-1)^j-1\right)^3
 \cr &&\cr&&\times \Biggl[\left(
-\frac{2}{j^3}+\frac{26}{3 j^2}-\frac{2}{9 (j-1)}-\frac{23}{j}-\frac{92}{9 (j+1)}+\frac{3008}{45 (2 j+1)}-\frac{2}{(j+1)^2}+\frac{1}{45 (j-2)}   \right)\cr &&
   \qquad\qquad
  +i \pi  \left(-\frac{2}{j^2}-\frac{2}{j+1}+\frac{1}{24 (2 j-3)}+\frac{1}{4 (2 j-1)}-\frac{111}{8 (2 j+1)}+\frac{1}{4 (2 j+3)}-\frac{27}{2 (2 j+1)^2}+\frac{26}{3 j}\right)
   \Biggr]
 \ ,\cr &&\cr&&\cr
 \hat\CA^{(2)}_{12}   &=& \hat k^2 \left(2(-1)^j-1\right)^3 \sqrt{j (j+1)}
\cr &&\cr &&
\times
 \Biggl[\left(\frac{2}{j^3}-\frac{52}{5 j^2}+\frac{167}{5 j}+\frac{167}{5 (j+1)}+\frac{1}{45 (j+2)}-\frac{6016}{45 (2 j+1)}+\frac{52}{5 (j+1)^2}+\frac{2}{(j+1)^3}+\frac{1}{45 (j-1)}\right)
 \cr &&\cr &&  \quad
 +i \pi  \left(\frac{2}{j^2}+\frac{52}{5 (j+1)}-\frac{1}{120 (2 j-3)}-\frac{1}{3 (2 j-1)}+\frac{1}{3 (2 j+3)}+\frac{1}{120 (2 j+5)}+\frac{2}{(j+1)^2}+\frac{27}{(2 j+1)^2}-\frac{52}{5 j}\right) \Biggr]
    \ ,\cr &&\cr
  \hat\CA^{(2)}_{22}&=& \hat k^2\left(2(-1)^j-1\right)^3\cr &&\cr&&\times 
   \Biggl[\left( -\frac{2}{j^2}+\frac{23}{j+1}+\frac{2}{9 (j+2)}-\frac{1}{45 (j+3)}-\frac{3008}{45 (2 j+1)}+\frac{26}{3 (j+1)^2}+\frac{2}{(j+1)^3}+\frac{92}{9 j}\right)
 \cr &&\cr &&  \quad
 +i \pi  \left(\frac{26}{3 (j+1)}+\frac{1}{4 (2 j-1)}-\frac{111}{8 (2 j+1)}+\frac{1}{4 (2 j+3)}+\frac{1}{24 (2 j+5)}+\frac{2}{(j+1)^2}+\frac{27}{2 (2 j+1)^2}-\frac{2}{j}\right) \Biggr] \,
   %%%%%
   \cr &&\cr&&\cr
 %%%%%  %
 \hat\CA^{(3)}_{11}&=&  \hat k^3 \left(2(-1)^j-1\right)^4
 \cr &&\times
 \Biggl[ 
 i \left(-\frac{3}{j^4}+\frac{49}{3 j^3}-\frac{778}{15 j^2}-\frac{22}{135 (j-1)}+\frac{683}{5 j}+\frac{3148}{45 (j+1)}+\frac{2}{135 (j+2)}-\frac{278656}{675 (2 j+1)}+\frac{94}{5
   (j+1)^2}\right. \cr &&\cr &&
   \qquad\qquad
   \left.+\frac{3}{(j+1)^3}+\frac{1}{225 (j-2)}\right)
 \cr && \cr &&\quad
   +\pi  \left(\frac{3}{j^3}-\frac{49}{3 j^2}-\frac{94}{5 (j+1)}+\frac{3}{2560 (2 j-5)}+\frac{31}{5760 (2 j-3)}+\frac{1135}{768 (2 j-1)}-\frac{8703}{128 (2 j+1)}+\frac{1735}{4608 (2 j+3)}\right.
   \cr &&\cr &&
   \qquad\qquad
-\frac{1}{480 (2
   j+5)}-\frac{3}{(j+1)^2}-\frac{5}{96 (2 j-3)^2}-\frac{23}{64 (2 j-1)^2}-\frac{1059}{16 (2 j+1)^2}+\frac{23}{64 (2 j+3)^2}+\frac{1}{32 (2 j-3)^3}
    \cr &&\cr &&
   \qquad\quad
   \left.+\frac{3}{32 (2 j-1)^3}+\frac{387}{32 (2 j+1)^3}+\frac{3}{32 (2
   j+3)^3}+\frac{45}{4 (2 j+1)^4}+\frac{778}{15 j}\right)
 \cr && \cr &&\qquad
+i \pi ^2 \left(\frac{1}{j^2}-\frac{1}{j+1}-\frac{5}{192 (2 j-3)}-\frac{23}{128 (2 j-1)}+\frac{855}{64 (2 j+1)}+\frac{23}{128 (2 j+3)}+\frac{1}{64 (2 j-3)^2}\right.
 \cr &&\cr &&
   \qquad\quad
   \left.
+\frac{3}{64 (2 j-1)^2}+\frac{387}{64 (2
   j+1)^2}+\frac{3}{64 (2 j+3)^2}+\frac{45}{8 (2 j+1)^3}-\frac{17}{3 j}\right)
    \Biggr]
  \ ,\cr &&\cr
  \hat\CA^{(3)}_{12} &=& \hat k^3 \left(2(-1)^j-1\right)^4 \sqrt{j (j+1)}
  \cr && \cr &&\times
  \Biggl[ 
  i \left(\frac{3}{j^4}-\frac{286}{15 j^3}+\frac{3176}{45 j^2}+\frac{13}{270 (j-1)}-\frac{27872}{135 j}-\frac{27872}{135 (j+1)}+\frac{13}{270 (j+2)}-\frac{1}{1350 (j+3)}+\frac{557312}{675 (2 j+1)}\right.
 \cr &&\cr &&
   \qquad\quad
   \left.
-\frac{3176}{45
   (j+1)^2}-\frac{286}{15 (j+1)^3}-\frac{3}{(j+1)^4}-\frac{1}{1350 (j-2)}\right)
 \cr && \cr &&\quad
   +
   \pi  \left(-\frac{3}{j^3}+\frac{286}{15 j^2}+\frac{3176}{45 (j+1)}+\frac{13}{1920 (2 j-3)}-\frac{11}{9 (2 j-1)}+\frac{11}{9 (2 j+3)}-\frac{13}{1920 (2 j+5)}+\frac{286}{15 (j+1)^2}\right.
 \cr &&\cr &&
   \qquad\quad
-\frac{1}{960 (2
   j-3)^2}+\frac{5}{12 (2 j-1)^2}+\frac{4281}{32 (2 j+1)^2}+\frac{5}{12 (2 j+3)^2}-\frac{1}{960 (2 j+5)^2}+\frac{3}{(j+1)^3} \cr &&\cr &&
   \qquad\quad
   \left.
-\frac{45}{2 (2 j+1)^4}-\frac{3176}{45 j}\right)
 \cr && \cr &&\quad
   +
i \pi ^2 \left(-\frac{1}{j^2}+\frac{32}{5 (j+1)}-\frac{1}{1920 (2 j-3)}+\frac{5}{24 (2 j-1)}-\frac{1665}{64 (2 j+1)}+\frac{5}{24 (2 j+3)}-\frac{1}{1920 (2 j+5)}\right.
 \cr &&\cr &&
   \qquad\qquad
   \left.
+\frac{1}{(j+1)^2}-\frac{45}{4 (2 j+1)^3}+\frac{32}{5
   j}\right)
  \Biggr] \,
   %%%%%
   \cr &&\cr&&\cr
 %%%%%  %
 \hat\CA^{(3)}_{22}&=&  \hat k^3 \left(2(-1)^j-1\right)^4
 \cr &&\times
 \Biggl[ i \left(-\frac{3}{j^3}+\frac{94}{5 j^2}-\frac{3148}{45 j}-\frac{683}{5 (j+1)}+\frac{22}{135 (j+2)}-\frac{1}{225 (j+3)}+\frac{278656}{675 (2 j+1)}-\frac{778}{15 (j+1)^2}\right. \cr &&\cr &&
   \qquad\qquad
   \left. -\frac{49}{3
   (j+1)^3}-\frac{3}{(j+1)^4}-\frac{2}{135 (j-1)} 
   \right)
 \cr && \cr &&\quad
   +\pi\left(\frac{3}{j^2}+\frac{778}{15 (j+1)}-\frac{1}{480 (2 j-3)}+\frac{1735}{4608 (2 j-1)}-\frac{8703}{128 (2 j+1)}+\frac{1135}{768 (2 j+3)}+\frac{31}{5760 (2 j+5)}
   \right.
   \cr &&\cr &&
   \qquad\qquad
   +\frac{3}{2560 (2 j+7)}+\frac{49}{3
   (j+1)^2}-\frac{23}{64 (2 j-1)^2}+\frac{1059}{16 (2 j+1)^2}+\frac{23}{64 (2 j+3)^2}+\frac{5}{96 (2 j+5)^2}+\frac{3}{(j+1)^3}
    \cr &&\cr &&
   \qquad\qquad
   \left.+\frac{3}{32 (2 j-1)^3}+\frac{387}{32 (2 j+1)^3}+\frac{3}{32 (2 j+3)^3}+\frac{1}{32 (2
   j+5)^3}-\frac{45}{4 (2 j+1)^4}-\frac{94}{5 j}  \right)
 \cr && \cr &&\qquad
+i \pi ^2 \left(\frac{17}{3 (j+1)}-\frac{23}{128 (2 j-1)}-\frac{855}{64 (2 j+1)}+\frac{23}{128 (2 j+3)}+\frac{5}{192 (2 j+5)}+\frac{1}{(j+1)^2}+\frac{3}{64 (2 j-1)^2}  \right.
 \cr &&\cr &&
   \qquad\qquad
   \left.
+\frac{387}{64 (2 j+1)^2}+\frac{3}{64 (2
   j+3)^2}+\frac{1}{64 (2 j+5)^2}-\frac{45}{8 (2 j+1)^3}+\frac{1}{j}\right) 
    \Biggr]
  \ .
 \eqn{ampsLJCoupled} \eeq

   The above scattering amplitudes can be solved for the phase shifts for   $\delta_\pm$ corresponding to angular momentum $\ell=j\pm1$, and the mixing phase $\epsilon$ by expanding
  \beq
  \delta_\pm = \sum_{n=1}^\infty \delta_{\pm}^{(n)} \ ,\qquad  \epsilon = \sum_{n=1}^\infty \epsilon^{(n)}
\eeq
where $\delta_\pm^{(n)}$ and $\epsilon^{(n)}$ are all $O(\hat k^n)$, and then expanding both sides of the equation
 \beq
S = 1 + \frac{i M k}{2\pi}\CA = 1 + 2 i \hat k \hat\CA =
\begin{pmatrix}
e^{2i\delta_-}\cos2\epsilon & i e^{i(  \delta_-+  \delta_+)}\sin2  \epsilon\\ i e^{i( \delta_- +  \delta_+)}\sin2 \epsilon &  e^{2i  \delta_+}\cos2  \epsilon\end{pmatrix}
\eeq
in powers of $\hat k$.  Using the above expressions derived for $\CA^{(n)}$ up to $n=3$ I find the following expressions for $\delta_\pm^{(n)}$ and $\epsilon^{(n)}$ for $n=1,\ldots,4$, for $j\ge 2$:
\beq
\delta_-^{(1)} &=&\hat k \left(1-2(-1)^j\right)\left[\frac{2}{2 j+1}-\frac{1}{j}
\right]\ ,\cr
&&\cr
\delta_-^{(2)} &=&\pi\, \hat k^2 \left(1-2(-1)^j\right)^2\left[  \frac{1}{8 (2 j-3)}+\frac{15}{8 (2 j+1)}+\frac{3}{2 (2 j+1)^2}-\frac{1}{j}
\right]\ ,\cr
&&\cr
\delta_-^{(3)} &=&\hat k^3 \left(1-2(-1)^j\right)^3\left[\frac{4}{3 j^3}-\frac{20}{3 j^2}+\frac{2}{9 (j-1)}+\frac{21}{j}+\frac{110}{9
   (j+1)}-\frac{3008}{45 (2 j+1)}+\frac{2}{(j+1)^2}-\frac{32}{3 (2 j+1)^3}-\frac{1}{45 (j-2)}
\right]\ ,\cr
&&\cr
\delta_-^{(4)} &=& \pi\,\hat k^4 \left(1-2(-1)^j\right)^4\left[  \frac{1}{j^3}-\frac{23}{3 j^2}-\frac{24}{j+1}+\frac{3}{2560 (2 j-5)}+\frac{19}{1152 (2 j-3)}+\frac{1519}{768
   (2 j-1)}\right.
   \cr
   &&
   \qquad\qquad\qquad\qquad -\frac{1599}{128 (2 j+1)}+\frac{2503}{4608 (2 j+3)}-\frac{1}{480 (2 j+5)}-\frac{3}{(j+1)^2}-\frac{5}{96 (2 j-3)^2}-\frac{23}{64 (2 j-1)^2}
  \cr
   &&
   \qquad\qquad\qquad\qquad -\frac{1059}{16 (2 j+1)^2}+\frac{23}{64 (2 j+3)^2}+\frac{1}{32 (2
   j-3)^3}+\frac{3}{32 (2 j-1)^3}+\frac{387}{32 (2 j+1)^3}+\frac{3}{32 (2 j+3)^3}
   \cr 
   &&
      \qquad\qquad\qquad\qquad 
\left.
   -\frac{51}{4 (2 j+1)^4}+\frac{1304}{45 j}
 \right]\ ,  
\eqn{deltamJneq1}
\eeq

\beq
\delta_+^{(1)} &=&\hat k \left(1-2(-1)^j\right)\left[  \frac{1}{j+1}-\frac{2}{2 j+1}
\right]\ ,\cr
&&\cr
\delta_+^{(2)} &=&\pi\, \hat k^2 \left(1-2(-1)^j\right)^2\left[  \frac{15}{8 (2 j+1)}+\frac{1}{8 (2 j+5)}-\frac{3}{2 (2 j+1)^2}-\frac{1}{j+1} 
\right]\ ,\cr
&&\cr
\delta_+^{(3)} &=&\hat k^3 \left(1-2(-1)^j\right)^3\left[  \frac{2}{j^2}-\frac{21}{j+1}-\frac{2}{9 (j+2)}+\frac{1}{45 (j+3)}+\frac{3008}{45 (2
   j+1)}-\frac{20}{3 (j+1)^2}-\frac{4}{3 (j+1)^3}+\frac{32}{3 (2 j+1)^3}-\frac{110}{9 j}
\right]\ ,\cr
&&\cr
\delta_+^{(4)} &=& \pi\,\hat k^4 \left(1-2(-1)^j\right)^4\left[  \frac{3}{j^2}+\frac{1304}{45 (j+1)}-\frac{1}{480 (2 j-3)}+\frac{2503}{4608 (2 j-1)}-\frac{1599}{128 (2
   j+1)}+\frac{1519}{768 (2 j+3)}
\right.
   \cr
   &&
   \qquad\qquad\qquad\qquad +\frac{19}{1152 (2 j+5)}+\frac{3}{2560 (2 j+7)}+\frac{23}{3 (j+1)^2}-\frac{23}{64 (2 j-1)^2}+\frac{1059}{16 (2 j+1)^2}+\frac{23}{64 (2 j+3)^2}
  \cr
   &&
   \qquad\qquad\qquad\qquad+\frac{5}{96 (2
   j+5)^2}+\frac{1}{(j+1)^3}+\frac{3}{32 (2 j-1)^3}+\frac{387}{32 (2 j+1)^3}+\frac{3}{32 (2 j+3)^3}+\frac{1}{32 (2 j+5)^3}
   \cr
   &&
  \qquad\qquad\qquad\qquad  \left. +\frac{51}{4 (2 j+1)^4}-\frac{24}{j}\right]\ ,  
\eqn{deltapJneq1}
\eeq

\beq
\epsilon^{(1)} &=&\hat k \left(1-2(-1)^j\right) \sqrt{j (j+1)}\,\left[  \frac{1}{j+1}-\frac{4}{2 j+1}+\frac{1}{j}
\right]\ ,\cr
&&\cr
\epsilon^{(2)} &=&\pi\, \hat k^2 \left(1-2(-1)^j\right)^2\sqrt{j (j+1)}\,\left[  -\frac{1}{j+1}-\frac{1}{4 (2 j-1)}+\frac{1}{4 (2 j+3)}-\frac{3}{(2 j+1)^2}+\frac{1}{j}
\right]\ ,\cr
&&\cr
\epsilon^{(3)} &=&\hat k^3 \left(1-2(-1)^j\right)^3\sqrt{j (j+1)}\,\left[   -\frac{3}{2
   j^3}+\frac{257}{30 j^2}-\frac{461}{15 j}-\frac{461}{15 (j+1)}-\frac{1}{45 (j+2)}+\frac{5536}{45 (2 j+1)}-\frac{257}{30 (j+1)^2}
   \right.
   \cr &&
   \qquad   \qquad   \qquad   \qquad \qquad   \qquad \quad  
 \left.  -\frac{3}{2 (j+1)^3}+\frac{32}{3 (2 j+1)^3}-\frac{1}{45 (j-1)}
\right]\ ,\cr
&&\cr
\epsilon^{(4)} &=& \pi\,\hat k^4 \left(1-2(-1)^j\right)^4\sqrt{j (j+1)}\,\left[ -\frac{3}{2 j^3}+\frac{23}{2 j^2}+\frac{10699}{225 (j+1)}+\frac{131}{28800 (2 j-3)}-\frac{29}{18 (2 j-1)}+\frac{29}{18 (2 j+3)} \right.
   \cr
   &&
   \qquad   \qquad   \qquad   \qquad \qquad   \qquad \quad  
   -\frac{131}{28800 (2 j+5)}+\frac{23}{2 (j+1)^2}-\frac{1}{960 (2
   j-3)^2}
     \cr
   &&
  \qquad   \qquad   \qquad   \qquad \qquad   \qquad \quad  
  +\frac{5}{12 (2 j-1)^2}+\frac{3321}{32 (2 j+1)^2}+\frac{5}{12 (2 j+3)^2}-\frac{1}{960 (2 j+5)^2}+\frac{3}{2 (j+1)^3}
   \cr
   &&
  \qquad   \qquad   \qquad   \qquad \qquad   \qquad \quad  
 \left.  +\frac{3}{2 (2 j+1)^4}-\frac{10699}{225 j}
  \right]\ ,  
\eqn{epsJneq1}
\eeq

 %%%%%%%%%%%%%%%
 %  Appendix D
 %%%%%%%%%%%%

   \section{The renormalized phase shifts: ${}^3P_1$, ${}^3D_2$, ${}^3S_1$, $\epsilon_1$, ${}^3P_0$, ${}^3P_2$}
   \label{sec:rendelta}
   
I give here results for the renormalized phase shifts (in radians)  computed from the renormalized amplitudes in \S~\ref{sec:renormalization}.

\subsection{ ${}^3P_1$ to $O(\hat k^3)$}

\beq
\delta^{(1)}  =  -\half \,\hat k    \ ,\qquad
\delta^{(2)}  =    \frac{\pi}{10}\,\hat k^2 \ , \qquad
\delta^{(3)}  =    -\frac{1}{36}\left(3 + 4 \tilde\xi_{p,0} -8\ln\hat k\right)\,\hat k^3 \ ,\qquad
\delta^{(4)} &=& -\frac {1291\pi}{21000}\,\hat k^4    \ .
\eeq

\subsection{ ${}^3D_2$ to $O(\hat k^5)$}

\beq
\delta^{(1)}  &=&   \half  \hat k  \ ,\qquad
\delta^{(2)}  =    \frac{3\pi}{70} \hat k^2\ , \qquad
\delta^{(3)}  =      \frac{1}{12} \hat k^3 \ ,\qquad
\delta^{(4)}  =     \frac{10127\pi}{343000} \hat k^4 \ , \cr &&\cr
\delta^{(5)} & =&    \frac{1}{ 176400}\left(-3136 \tilde\xi_{d,0} + 27 (245 + 12 \pi^2 + 784 \ln 2) - 
  10584 \ln 4 \hat k\right)  \hat k^5  \ ,\qquad
\delta^{(6)}  =      - \frac{260003581\pi}{27731550000}\hat k^6 \ .
\eeq

\subsection{ ${}^3S_1$ and $\epsilon_1$  to $O(\hat k^3)$}

\beq
\delta^{(1)}  =      -\hat k\ ,\qquad
\delta^{(2)}  =     -3\pi  \,\hat k^2\ , \qquad
\delta^{(3)}  =     -\frac{1}{3}\left(5+3\tilde\xi_{ss,0} - 36\ln\hat k\right)  \,\hat k^3 \ ,\qquad
\delta^{(4)}  =   \frac{19527\pi}{3500}    \,\hat k^4 \ .
\eeq

 \beq
\epsilon^{(1)}  =   \frac{1}{\sqrt{2} }  \,\hat k  \ ,\qquad
\epsilon^{(2)}  =     -\frac{3\pi}{5\sqrt{2}}  \,\hat k^2\ , \qquad
\epsilon^{(3)}  =     \frac{1}{240}\left(-32\tilde\xi_{sd,0} +\sqrt{2}\left(175+288\ln\hat k\right)\right)\,\hat k^3  \ ,\qquad
\epsilon^{(4)}  =    \frac{7971\pi}{1960\sqrt{2}}   \,\hat k^4 \ .
\eeq

\subsection{ ${}^3P_0$ to $O(\,\hat k^3)$}

\beq
\delta^{(1)}  =      \hat k \ ,\qquad
\delta^{(2)}  =   \frac{2\pi}{5}    \,\hat k^2\ , \qquad
\delta^{(3)}  =    
\frac{2}{9}\left(3+4\tilde\xi_{p,0} - 8\ln\hat k  \right) \,\hat k^3 \ ,\qquad
\delta^{(4)}  =  -\frac{2582\pi}{2625}     \,\hat k^4 \ .
\eeq

\subsection{ ${}^3P_2$ to $O(\,\hat k^3)$}

\beq
\delta^{(1)}  =    \frac{1}{10}   \,\hat k \ ,\qquad
\delta^{(2)}  =   \frac{3\pi}{50}   \,\hat k^2 \ , \qquad
\delta^{(3)}  =     -\frac{1}{4500}\left(-9+500\tilde\xi_{p,0} +200\ln\hat k\right) \,\hat k^3 \ ,\qquad
\delta^{(4)}  =  -\frac{157757\pi}{46305000}     \,\hat k^4\ .
\eeq
 
 %%%%%%%%%%%%%%%%%%%%%%
\bibliography{Kaplan}

%merlin.mbs apsrev4-1.bst 2010-07-25 4.21a (PWD, AO, DPC) hacked
%Control: key (0)
%Control: author (8) initials jnrlst
%Control: editor formatted (1) identically to author
%Control: production of article title (-1) disabled
%Control: page (0) single
%Control: year (1) truncated
%Control: production of eprint (0) enabled
\begin{thebibliography}{31}%
\makeatletter
\providecommand \@ifxundefined [1]{%
 \@ifx{#1\undefined}
}%
\providecommand \@ifnum [1]{%
 \ifnum #1\expandafter \@firstoftwo
 \else \expandafter \@secondoftwo
 \fi
}%
\providecommand \@ifx [1]{%
 \ifx #1\expandafter \@firstoftwo
 \else \expandafter \@secondoftwo
 \fi
}%
\providecommand \natexlab [1]{#1}%
\providecommand \enquote  [1]{``#1''}%
\providecommand \bibnamefont  [1]{#1}%
\providecommand \bibfnamefont [1]{#1}%
\providecommand \citenamefont [1]{#1}%
\providecommand \href@noop [0]{\@secondoftwo}%
\providecommand \href [0]{\begingroup \@sanitize@url \@href}%
\providecommand \@href[1]{\@@startlink{#1}\@@href}%
\providecommand \@@href[1]{\endgroup#1\@@endlink}%
\providecommand \@sanitize@url [0]{\catcode `\\12\catcode `\$12\catcode
  `\&12\catcode `\#12\catcode `\^12\catcode `\_12\catcode `\%12\relax}%
\providecommand \@@startlink[1]{}%
\providecommand \@@endlink[0]{}%
\providecommand \url  [0]{\begingroup\@sanitize@url \@url }%
\providecommand \@url [1]{\endgroup\@href {#1}{\urlprefix }}%
\providecommand \urlprefix  [0]{URL }%
\providecommand \Eprint [0]{\href }%
\providecommand \doibase [0]{http://dx.doi.org/}%
\providecommand \selectlanguage [0]{\@gobble}%
\providecommand \bibinfo  [0]{\@secondoftwo}%
\providecommand \bibfield  [0]{\@secondoftwo}%
\providecommand \translation [1]{[#1]}%
\providecommand \BibitemOpen [0]{}%
\providecommand \bibitemStop [0]{}%
\providecommand \bibitemNoStop [0]{.\EOS\space}%
\providecommand \EOS [0]{\spacefactor3000\relax}%
\providecommand \BibitemShut  [1]{\csname bibitem#1\endcsname}%
\let\auto@bib@innerbib\@empty
%</preamble>
\bibitem [{\citenamefont {Fleming}\ \emph {et~al.}(2000)\citenamefont
  {Fleming}, \citenamefont {Mehen},\ and\ \citenamefont
  {Stewart}}]{Fleming:1999ee}%
  \BibitemOpen
  \bibfield  {author} {\bibinfo {author} {\bibfnamefont {S.}~\bibnamefont
  {Fleming}}, \bibinfo {author} {\bibfnamefont {T.}~\bibnamefont {Mehen}}, \
  and\ \bibinfo {author} {\bibfnamefont {I.~W.}\ \bibnamefont {Stewart}},\
  }\href {\doibase 10.1016/S0375-9474(00)00221-9} {\bibfield  {journal}
  {\bibinfo  {journal} {Nucl.Phys.}\ }\textbf {\bibinfo {volume} {A677}},\
  \bibinfo {pages} {313} (\bibinfo {year} {2000})},\ \Eprint
  {http://arxiv.org/abs/nucl-th/9911001} {arXiv:nucl-th/9911001 [nucl-th]}
  \BibitemShut {NoStop}%
%%CITATION = NUCL-TH/9911001;%%
\bibitem [{\citenamefont {Birse}(2005)}]{birse2005renormalization}%
  \BibitemOpen
  \bibfield  {author} {\bibinfo {author} {\bibfnamefont {M.~C.}\ \bibnamefont
  {Birse}},\ }\href@noop {} {\bibfield  {journal} {\bibinfo  {journal} {Journal
  of Physics A: Mathematical and General}\ }\textbf {\bibinfo {volume} {39}},\
  \bibinfo {pages} {L49} (\bibinfo {year} {2005})}\BibitemShut {NoStop}%
\bibitem [{\citenamefont {Weinberg}(1990)}]{Weinberg:1990rz}%
  \BibitemOpen
  \bibfield  {author} {\bibinfo {author} {\bibfnamefont {S.}~\bibnamefont
  {Weinberg}},\ }\href {\doibase 10.1016/0370-2693(90)90938-3} {\bibfield
  {journal} {\bibinfo  {journal} {Phys.Lett.}\ }\textbf {\bibinfo {volume}
  {B251}},\ \bibinfo {pages} {288} (\bibinfo {year} {1990})}\BibitemShut
  {NoStop}%
%%CITATION = PHLTA,B251,288;%%
\bibitem [{\citenamefont {Weinberg}(1991)}]{Weinberg:1991um}%
  \BibitemOpen
  \bibfield  {author} {\bibinfo {author} {\bibfnamefont {S.}~\bibnamefont
  {Weinberg}},\ }\href {\doibase 10.1016/0550-3213(91)90231-L} {\bibfield
  {journal} {\bibinfo  {journal} {Nucl.Phys.}\ }\textbf {\bibinfo {volume}
  {B363}},\ \bibinfo {pages} {3} (\bibinfo {year} {1991})}\BibitemShut
  {NoStop}%
%%CITATION = NUPHA,B363,3;%%
\bibitem [{\citenamefont {Ordonez}\ and\ \citenamefont {van
  Kolck}(1992)}]{Ordonez:1992xp}%
  \BibitemOpen
  \bibfield  {author} {\bibinfo {author} {\bibfnamefont {C.}~\bibnamefont
  {Ordonez}}\ and\ \bibinfo {author} {\bibfnamefont {U.}~\bibnamefont {van
  Kolck}},\ }\href {\doibase 10.1016/0370-2693(92)91404-W} {\bibfield
  {journal} {\bibinfo  {journal} {Phys.Lett.}\ }\textbf {\bibinfo {volume}
  {B291}},\ \bibinfo {pages} {459} (\bibinfo {year} {1992})}\BibitemShut
  {NoStop}%
%%CITATION = PHLTA,B291,459;%%
\bibitem [{\citenamefont {Machleidt}\ and\ \citenamefont
  {Entem}(2011)}]{machleidt2011chiral}%
  \BibitemOpen
  \bibfield  {author} {\bibinfo {author} {\bibfnamefont {R.}~\bibnamefont
  {Machleidt}}\ and\ \bibinfo {author} {\bibfnamefont {D.~R.}\ \bibnamefont
  {Entem}},\ }\href@noop {} {\bibfield  {journal} {\bibinfo  {journal} {Physics
  Reports}\ }\textbf {\bibinfo {volume} {503}},\ \bibinfo {pages} {1} (\bibinfo
  {year} {2011})}\BibitemShut {NoStop}%
\bibitem [{\citenamefont {Epelbaum}\ \emph {et~al.}(2009)\citenamefont
  {Epelbaum}, \citenamefont {Hammer},\ and\ \citenamefont
  {Mei{\ss}ner}}]{epelbaum2009modern}%
  \BibitemOpen
  \bibfield  {author} {\bibinfo {author} {\bibfnamefont {E.}~\bibnamefont
  {Epelbaum}}, \bibinfo {author} {\bibfnamefont {H.-W.}\ \bibnamefont
  {Hammer}}, \ and\ \bibinfo {author} {\bibfnamefont {U.-G.}\ \bibnamefont
  {Mei{\ss}ner}},\ }\href@noop {} {\bibfield  {journal} {\bibinfo  {journal}
  {Reviews of Modern Physics}\ }\textbf {\bibinfo {volume} {81}},\ \bibinfo
  {pages} {1773} (\bibinfo {year} {2009})}\BibitemShut {NoStop}%
\bibitem [{\citenamefont {Bedaque}\ \emph
  {et~al.}(1999{\natexlab{a}})\citenamefont {Bedaque}, \citenamefont {Hammer},\
  and\ \citenamefont {Van~Kolck}}]{bedaque1999three}%
  \BibitemOpen
  \bibfield  {author} {\bibinfo {author} {\bibfnamefont {P.~F.}\ \bibnamefont
  {Bedaque}}, \bibinfo {author} {\bibfnamefont {H.-W.}\ \bibnamefont {Hammer}},
  \ and\ \bibinfo {author} {\bibfnamefont {U.}~\bibnamefont {Van~Kolck}},\
  }\href@noop {} {\bibfield  {journal} {\bibinfo  {journal} {Nuclear Physics
  A}\ }\textbf {\bibinfo {volume} {646}},\ \bibinfo {pages} {444} (\bibinfo
  {year} {1999}{\natexlab{a}})}\BibitemShut {NoStop}%
\bibitem [{\citenamefont {Nogga}\ \emph {et~al.}(2005)\citenamefont {Nogga},
  \citenamefont {Timmermans},\ and\ \citenamefont {van Kolck}}]{Nogga:2005hy}%
  \BibitemOpen
  \bibfield  {author} {\bibinfo {author} {\bibfnamefont {A.}~\bibnamefont
  {Nogga}}, \bibinfo {author} {\bibfnamefont {R.~G.~E.}\ \bibnamefont
  {Timmermans}}, \ and\ \bibinfo {author} {\bibfnamefont {U.}~\bibnamefont {van
  Kolck}},\ }\href {\doibase 10.1103/PhysRevC.72.054006} {\bibfield  {journal}
  {\bibinfo  {journal} {Phys. Rev.}\ }\textbf {\bibinfo {volume} {C72}},\
  \bibinfo {pages} {054006} (\bibinfo {year} {2005})},\ \Eprint
  {http://arxiv.org/abs/nucl-th/0506005} {arXiv:nucl-th/0506005 [nucl-th]}
  \BibitemShut {NoStop}%
%%CITATION = NUCL-TH/0506005;%%
\bibitem [{\citenamefont {Kaplan}\ \emph {et~al.}(1996)\citenamefont {Kaplan},
  \citenamefont {Savage},\ and\ \citenamefont {Wise}}]{Kaplan:1996xu}%
  \BibitemOpen
  \bibfield  {author} {\bibinfo {author} {\bibfnamefont {D.~B.}\ \bibnamefont
  {Kaplan}}, \bibinfo {author} {\bibfnamefont {M.~J.}\ \bibnamefont {Savage}},
  \ and\ \bibinfo {author} {\bibfnamefont {M.~B.}\ \bibnamefont {Wise}},\
  }\href {\doibase 10.1016/0550-3213(96)00357-4} {\bibfield  {journal}
  {\bibinfo  {journal} {Nucl. Phys.}\ }\textbf {\bibinfo {volume} {B478}},\
  \bibinfo {pages} {629} (\bibinfo {year} {1996})},\ \Eprint
  {http://arxiv.org/abs/nucl-th/9605002} {arXiv:nucl-th/9605002 [nucl-th]}
  \BibitemShut {NoStop}%
%%CITATION = NUCL-TH/9605002;%%
\bibitem [{\citenamefont {Kaplan}\ \emph
  {et~al.}(1998{\natexlab{a}})\citenamefont {Kaplan}, \citenamefont {Savage},\
  and\ \citenamefont {Wise}}]{Kaplan:1998tg}%
  \BibitemOpen
  \bibfield  {author} {\bibinfo {author} {\bibfnamefont {D.~B.}\ \bibnamefont
  {Kaplan}}, \bibinfo {author} {\bibfnamefont {M.~J.}\ \bibnamefont {Savage}},
  \ and\ \bibinfo {author} {\bibfnamefont {M.~B.}\ \bibnamefont {Wise}},\
  }\href {\doibase 10.1016/S0370-2693(98)00210-X} {\bibfield  {journal}
  {\bibinfo  {journal} {Phys.Lett.}\ }\textbf {\bibinfo {volume} {B424}},\
  \bibinfo {pages} {390} (\bibinfo {year} {1998}{\natexlab{a}})},\ \Eprint
  {http://arxiv.org/abs/nucl-th/9801034} {arXiv:nucl-th/9801034 [nucl-th]}
  \BibitemShut {NoStop}%
%%CITATION = NUCL-TH/9801034;%%
\bibitem [{\citenamefont {Kaplan}\ \emph
  {et~al.}(1998{\natexlab{b}})\citenamefont {Kaplan}, \citenamefont {Savage},\
  and\ \citenamefont {Wise}}]{Kaplan:1998we}%
  \BibitemOpen
  \bibfield  {author} {\bibinfo {author} {\bibfnamefont {D.~B.}\ \bibnamefont
  {Kaplan}}, \bibinfo {author} {\bibfnamefont {M.~J.}\ \bibnamefont {Savage}},
  \ and\ \bibinfo {author} {\bibfnamefont {M.~B.}\ \bibnamefont {Wise}},\
  }\href {\doibase 10.1016/S0550-3213(98)00440-4} {\bibfield  {journal}
  {\bibinfo  {journal} {Nucl.Phys.}\ }\textbf {\bibinfo {volume} {B534}},\
  \bibinfo {pages} {329} (\bibinfo {year} {1998}{\natexlab{b}})},\ \Eprint
  {http://arxiv.org/abs/nucl-th/9802075} {arXiv:nucl-th/9802075 [nucl-th]}
  \BibitemShut {NoStop}%
%%CITATION = NUCL-TH/9802075;%%
\bibitem [{\citenamefont {Bedaque}\ \emph
  {et~al.}(1999{\natexlab{b}})\citenamefont {Bedaque}, \citenamefont {Hammer},\
  and\ \citenamefont {van Kolck}}]{Bedaque:1998kg}%
  \BibitemOpen
  \bibfield  {author} {\bibinfo {author} {\bibfnamefont {P.~F.}\ \bibnamefont
  {Bedaque}}, \bibinfo {author} {\bibfnamefont {H.}~\bibnamefont {Hammer}}, \
  and\ \bibinfo {author} {\bibfnamefont {U.}~\bibnamefont {van Kolck}},\ }\href
  {\doibase 10.1103/PhysRevLett.82.463} {\bibfield  {journal} {\bibinfo
  {journal} {Phys.Rev.Lett.}\ }\textbf {\bibinfo {volume} {82}},\ \bibinfo
  {pages} {463} (\bibinfo {year} {1999}{\natexlab{b}})},\ \Eprint
  {http://arxiv.org/abs/nucl-th/9809025} {arXiv:nucl-th/9809025 [nucl-th]}
  \BibitemShut {NoStop}%
%%CITATION = NUCL-TH/9809025;%%
\bibitem [{\citenamefont {Birse}\ \emph {et~al.}(1999)\citenamefont {Birse},
  \citenamefont {McGovern},\ and\ \citenamefont
  {Richardson}}]{birse1999renormalisation}%
  \BibitemOpen
  \bibfield  {author} {\bibinfo {author} {\bibfnamefont {M.~C.}\ \bibnamefont
  {Birse}}, \bibinfo {author} {\bibfnamefont {J.~A.}\ \bibnamefont {McGovern}},
  \ and\ \bibinfo {author} {\bibfnamefont {K.~G.}\ \bibnamefont {Richardson}},\
  }\href@noop {} {\bibfield  {journal} {\bibinfo  {journal} {Physics Letters
  B}\ }\textbf {\bibinfo {volume} {464}},\ \bibinfo {pages} {169} (\bibinfo
  {year} {1999})}\BibitemShut {NoStop}%
\bibitem [{\citenamefont {Bedaque}\ \emph {et~al.}(2003)\citenamefont
  {Bedaque}, \citenamefont {Rupak}, \citenamefont {Griesshammer},\ and\
  \citenamefont {Hammer}}]{Bedaque:2002yg}%
  \BibitemOpen
  \bibfield  {author} {\bibinfo {author} {\bibfnamefont {P.~F.}\ \bibnamefont
  {Bedaque}}, \bibinfo {author} {\bibfnamefont {G.}~\bibnamefont {Rupak}},
  \bibinfo {author} {\bibfnamefont {H.~W.}\ \bibnamefont {Griesshammer}}, \
  and\ \bibinfo {author} {\bibfnamefont {H.-W.}\ \bibnamefont {Hammer}},\
  }\href {\doibase 10.1016/S0375-9474(02)01402-1} {\bibfield  {journal}
  {\bibinfo  {journal} {Nucl.Phys.}\ }\textbf {\bibinfo {volume} {A714}},\
  \bibinfo {pages} {589} (\bibinfo {year} {2003})},\ \Eprint
  {http://arxiv.org/abs/nucl-th/0207034} {arXiv:nucl-th/0207034 [nucl-th]}
  \BibitemShut {NoStop}%
%%CITATION = NUCL-TH/0207034;%%
\bibitem [{\citenamefont {Barford}\ and\ \citenamefont
  {Birse}(2003)}]{barford2003renormalization}%
  \BibitemOpen
  \bibfield  {author} {\bibinfo {author} {\bibfnamefont {T.}~\bibnamefont
  {Barford}}\ and\ \bibinfo {author} {\bibfnamefont {M.~C.}\ \bibnamefont
  {Birse}},\ }\href@noop {} {\bibfield  {journal} {\bibinfo  {journal}
  {Physical Review C}\ }\textbf {\bibinfo {volume} {67}},\ \bibinfo {pages}
  {064006} (\bibinfo {year} {2003})}\BibitemShut {NoStop}%
\bibitem [{\citenamefont {Griesshammer}(2005)}]{Griesshammer:2005ga}%
  \BibitemOpen
  \bibfield  {author} {\bibinfo {author} {\bibfnamefont {H.~W.}\ \bibnamefont
  {Griesshammer}},\ }\href {\doibase 10.1016/j.nuclphysa.2005.05.202}
  {\bibfield  {journal} {\bibinfo  {journal} {Nucl.Phys.}\ }\textbf {\bibinfo
  {volume} {A760}},\ \bibinfo {pages} {110} (\bibinfo {year} {2005})},\ \Eprint
  {http://arxiv.org/abs/nucl-th/0502039} {arXiv:nucl-th/0502039 [nucl-th]}
  \BibitemShut {NoStop}%
%%CITATION = NUCL-TH/0502039;%%
\bibitem [{\citenamefont {Nishida}\ and\ \citenamefont
  {Son}(2007)}]{Nishida:2007pJ}%
  \BibitemOpen
  \bibfield  {author} {\bibinfo {author} {\bibfnamefont {Y.}~\bibnamefont
  {Nishida}}\ and\ \bibinfo {author} {\bibfnamefont {D.~T.}\ \bibnamefont
  {Son}},\ }\href {\doibase 10.1103/PhysRevD.76.086004} {\bibfield  {journal}
  {\bibinfo  {journal} {Phys.Rev.}\ }\textbf {\bibinfo {volume} {D76}},\
  \bibinfo {pages} {086004} (\bibinfo {year} {2007})},\ \Eprint
  {http://arxiv.org/abs/0706.3746} {arXiv:0706.3746 [hep-th]} \BibitemShut
  {NoStop}%
%%CITATION = ARXIV:0706.3746;%%
\bibitem [{\citenamefont {Chen}\ \emph {et~al.}(1999)\citenamefont {Chen},
  \citenamefont {Rupak},\ and\ \citenamefont {Savage}}]{chen1999nucleon}%
  \BibitemOpen
  \bibfield  {author} {\bibinfo {author} {\bibfnamefont {J.-W.}\ \bibnamefont
  {Chen}}, \bibinfo {author} {\bibfnamefont {G.}~\bibnamefont {Rupak}}, \ and\
  \bibinfo {author} {\bibfnamefont {M.~J.}\ \bibnamefont {Savage}},\
  }\href@noop {} {\bibfield  {journal} {\bibinfo  {journal} {Nuclear Physics
  A}\ }\textbf {\bibinfo {volume} {653}},\ \bibinfo {pages} {386} (\bibinfo
  {year} {1999})}\BibitemShut {NoStop}%
\bibitem [{\citenamefont {Rupak}(2000)}]{rupak2000precision}%
  \BibitemOpen
  \bibfield  {author} {\bibinfo {author} {\bibfnamefont {G.}~\bibnamefont
  {Rupak}},\ }\href@noop {} {\bibfield  {journal} {\bibinfo  {journal} {Nuclear
  Physics A}\ }\textbf {\bibinfo {volume} {678}},\ \bibinfo {pages} {405}
  (\bibinfo {year} {2000})}\BibitemShut {NoStop}%
\bibitem [{\citenamefont {Cirigliano}\ \emph {et~al.}(2018)\citenamefont
  {Cirigliano}, \citenamefont {Dekens}, \citenamefont {Mereghetti},\ and\
  \citenamefont {Walker-Loud}}]{cirigliano2018neutrinoless}%
  \BibitemOpen
  \bibfield  {author} {\bibinfo {author} {\bibfnamefont {V.}~\bibnamefont
  {Cirigliano}}, \bibinfo {author} {\bibfnamefont {W.}~\bibnamefont {Dekens}},
  \bibinfo {author} {\bibfnamefont {E.}~\bibnamefont {Mereghetti}}, \ and\
  \bibinfo {author} {\bibfnamefont {A.}~\bibnamefont {Walker-Loud}},\
  }\href@noop {} {\bibfield  {journal} {\bibinfo  {journal} {Physical Review
  C}\ }\textbf {\bibinfo {volume} {97}},\ \bibinfo {pages} {065501} (\bibinfo
  {year} {2018})}\BibitemShut {NoStop}%
\bibitem [{\citenamefont {Beane}\ \emph {et~al.}(2002)\citenamefont {Beane},
  \citenamefont {Bedaque}, \citenamefont {Savage},\ and\ \citenamefont {van
  Kolck}}]{Beane:2001bc}%
  \BibitemOpen
  \bibfield  {author} {\bibinfo {author} {\bibfnamefont {S.~R.}\ \bibnamefont
  {Beane}}, \bibinfo {author} {\bibfnamefont {P.~F.}\ \bibnamefont {Bedaque}},
  \bibinfo {author} {\bibfnamefont {M.~J.}\ \bibnamefont {Savage}}, \ and\
  \bibinfo {author} {\bibfnamefont {U.}~\bibnamefont {van Kolck}},\ }\href
  {\doibase 10.1016/S0375-9474(01)01324-0} {\bibfield  {journal} {\bibinfo
  {journal} {Nucl. Phys.}\ }\textbf {\bibinfo {volume} {A700}},\ \bibinfo
  {pages} {377} (\bibinfo {year} {2002})},\ \Eprint
  {http://arxiv.org/abs/nucl-th/0104030} {arXiv:nucl-th/0104030 [nucl-th]}
  \BibitemShut {NoStop}%
%%CITATION = NUCL-TH/0104030;%%
\bibitem [{\citenamefont {Beane}\ \emph {et~al.}(2009)\citenamefont {Beane},
  \citenamefont {Kaplan},\ and\ \citenamefont {Vuorinen}}]{Beane:2008bt}%
  \BibitemOpen
  \bibfield  {author} {\bibinfo {author} {\bibfnamefont {S.~R.}\ \bibnamefont
  {Beane}}, \bibinfo {author} {\bibfnamefont {D.~B.}\ \bibnamefont {Kaplan}}, \
  and\ \bibinfo {author} {\bibfnamefont {A.}~\bibnamefont {Vuorinen}},\ }\href
  {\doibase 10.1103/PhysRevC.80.011001} {\bibfield  {journal} {\bibinfo
  {journal} {Phys. Rev.}\ }\textbf {\bibinfo {volume} {C80}},\ \bibinfo {pages}
  {011001} (\bibinfo {year} {2009})},\ \Eprint {http://arxiv.org/abs/0812.3938}
  {arXiv:0812.3938 [nucl-th]} \BibitemShut {NoStop}%
%%CITATION = ARXIV:0812.3938;%%
\bibitem [{\citenamefont {Birse}(2006)}]{Birse:2005um}%
  \BibitemOpen
  \bibfield  {author} {\bibinfo {author} {\bibfnamefont {M.~C.}\ \bibnamefont
  {Birse}},\ }\href {\doibase 10.1103/PhysRevC.74.014003} {\bibfield  {journal}
  {\bibinfo  {journal} {Phys. Rev.}\ }\textbf {\bibinfo {volume} {C74}},\
  \bibinfo {pages} {014003} (\bibinfo {year} {2006})},\ \Eprint
  {http://arxiv.org/abs/nucl-th/0507077} {arXiv:nucl-th/0507077 [nucl-th]}
  \BibitemShut {NoStop}%
%%CITATION = NUCL-TH/0507077;%%
\bibitem [{\citenamefont {Pav\'on~Valderrama}\ \emph
  {et~al.}(2017)\citenamefont {Pav\'on~Valderrama}, \citenamefont
  {S\'anchez~S\'anchez}, \citenamefont {Yang}, \citenamefont {Long},
  \citenamefont {Carbonell},\ and\ \citenamefont {van
  Kolck}}]{PavonValderrama:2016lqn}%
  \BibitemOpen
  \bibfield  {author} {\bibinfo {author} {\bibfnamefont {M.}~\bibnamefont
  {Pav\'on~Valderrama}}, \bibinfo {author} {\bibfnamefont {M.}~\bibnamefont
  {S\'anchez~S\'anchez}}, \bibinfo {author} {\bibfnamefont {C.~J.}\
  \bibnamefont {Yang}}, \bibinfo {author} {\bibfnamefont {B.}~\bibnamefont
  {Long}}, \bibinfo {author} {\bibfnamefont {J.}~\bibnamefont {Carbonell}}, \
  and\ \bibinfo {author} {\bibfnamefont {U.}~\bibnamefont {van Kolck}},\ }\href
  {\doibase 10.1103/PhysRevC.95.054001} {\bibfield  {journal} {\bibinfo
  {journal} {Phys. Rev.}\ }\textbf {\bibinfo {volume} {C95}},\ \bibinfo {pages}
  {054001} (\bibinfo {year} {2017})},\ \Eprint
  {http://arxiv.org/abs/1611.10175} {arXiv:1611.10175 [nucl-th]} \BibitemShut
  {NoStop}%
%%CITATION = ARXIV:1611.10175;%%
\bibitem [{\citenamefont {Wu}\ and\ \citenamefont {Long}(2019)}]{Wu:2018lai}%
  \BibitemOpen
  \bibfield  {author} {\bibinfo {author} {\bibfnamefont {S.}~\bibnamefont
  {Wu}}\ and\ \bibinfo {author} {\bibfnamefont {B.}~\bibnamefont {Long}},\
  }\href {\doibase 10.1103/PhysRevC.99.024003} {\bibfield  {journal} {\bibinfo
  {journal} {Phys. Rev.}\ }\textbf {\bibinfo {volume} {C99}},\ \bibinfo {pages}
  {024003} (\bibinfo {year} {2019})},\ \Eprint
  {http://arxiv.org/abs/1807.04407} {arXiv:1807.04407 [nucl-th]} \BibitemShut
  {NoStop}%
%%CITATION = ARXIV:1807.04407;%%
\bibitem [{\citenamefont {Cavagnero}(1994)}]{cavagnero1994secular}%
  \BibitemOpen
  \bibfield  {author} {\bibinfo {author} {\bibfnamefont {M.}~\bibnamefont
  {Cavagnero}},\ }\href@noop {} {\bibfield  {journal} {\bibinfo  {journal}
  {Physical Review A}\ }\textbf {\bibinfo {volume} {50}},\ \bibinfo {pages}
  {2841} (\bibinfo {year} {1994})}\BibitemShut {NoStop}%
\bibitem [{\citenamefont {Gao}(1999)}]{gao1999repulsive}%
  \BibitemOpen
  \bibfield  {author} {\bibinfo {author} {\bibfnamefont {B.}~\bibnamefont
  {Gao}},\ }\href@noop {} {\bibfield  {journal} {\bibinfo  {journal} {Physical
  Review A}\ }\textbf {\bibinfo {volume} {59}},\ \bibinfo {pages} {2778}
  (\bibinfo {year} {1999})}\BibitemShut {NoStop}%
\bibitem [{\citenamefont {Wu}\ and\ \citenamefont
  {Ohmura}(1962)}]{wu2014quantum}%
  \BibitemOpen
  \bibfield  {author} {\bibinfo {author} {\bibfnamefont {T.-Y.}\ \bibnamefont
  {Wu}}\ and\ \bibinfo {author} {\bibfnamefont {T.}~\bibnamefont {Ohmura}},\
  }\href@noop {} {\emph {\bibinfo {title} {Quantum theory of scattering}}},\
  Prentice-Hall international series in physics\ (\bibinfo  {publisher}
  {Prentice-Hall},\ \bibinfo {address} {Englewood Cliffs, N.J.},\ \bibinfo
  {year} {1962})\BibitemShut {NoStop}%
\bibitem [{\citenamefont {Stapp}\ \emph {et~al.}(1957)\citenamefont {Stapp},
  \citenamefont {Ypsilantis},\ and\ \citenamefont
  {Metropolis}}]{stapp1957phase}%
  \BibitemOpen
  \bibfield  {author} {\bibinfo {author} {\bibfnamefont {H.~P.}\ \bibnamefont
  {Stapp}}, \bibinfo {author} {\bibfnamefont {T.~J.}\ \bibnamefont
  {Ypsilantis}}, \ and\ \bibinfo {author} {\bibfnamefont {N.}~\bibnamefont
  {Metropolis}},\ }\href@noop {} {\bibfield  {journal} {\bibinfo  {journal}
  {Physical Review}\ }\textbf {\bibinfo {volume} {105}},\ \bibinfo {pages}
  {302} (\bibinfo {year} {1957})}\BibitemShut {NoStop}%
\bibitem [{\citenamefont {Bloomfield}\ \emph {et~al.}()\citenamefont
  {Bloomfield}, \citenamefont {Face},\ and\ \citenamefont
  {Moss}}]{bloomfield2017indefinite}%
  \BibitemOpen
  \bibfield  {author} {\bibinfo {author} {\bibfnamefont {J.~K.}\ \bibnamefont
  {Bloomfield}}, \bibinfo {author} {\bibfnamefont {S.~H.}\ \bibnamefont
  {Face}}, \ and\ \bibinfo {author} {\bibfnamefont {Z.}~\bibnamefont {Moss}},\
  }\href@noop {} {\bibfield  {journal} {\bibinfo  {journal} {arXiv preprint
  arXiv:1703.06428}\ }}\Eprint {http://arxiv.org/abs/1703.06428}
  {arXiv:1703.06428} \BibitemShut {NoStop}%
\end{thebibliography}%
\end{document}